\documentclass[acmsmall]{acmart}
\makeatletter\@ifundefined{KV@Gin@artifact}{\define@key{Gin}{artifact}[]{}}{}\makeatother

\usepackage[utf8]{inputenc}
\usepackage[T1]{fontenc}
\usepackage{xspace}

\usepackage{amsmath,amsfonts,amsthm}
\usepackage{mathtools}
\usepackage{hyperref}
\usepackage{stmaryrd}
\usepackage{etoolbox}
\usepackage{float}
\usepackage{bm}
\usepackage{graphicx}
\usepackage{textcomp}
\usepackage{cleveref}
\usepackage{tabto}
\usepackage{thm-restate}
\usepackage{blkarray}
\usepackage{setspace}
\usepackage{multirow}
\usepackage{caption}
\usepackage{subcaption}
\usepackage{listings}
\usepackage{mathdots}
\usepackage{wrapfig}

\usepackage[dvipsnames]{xcolor}
\usepackage{alltt}

\usepackage{thmtools}

\theoremstyle{acmplain}

\declaretheoremstyle[
  spaceabove=.5\baselineskip\@plus.2\baselineskip\@minus.2\baselineskip,
  spacebelow=.5\baselineskip\@plus.2\baselineskip\@minus.2\baselineskip,
  bodyfont=\normalfont,
  headindent=\parindent,
  headfont=\scshape,
  notefont=\scshape,
  qed=$\triangleleft$
]{examplestyle}

\declaretheorem[
  style=examplestyle,
  name=Example,
  numberwithin=section
]{example}

\def\BibTeX{{\rm B\kern-.05em{\sc i\kern-.025em b}\kern-.08em
T\kern-.1667em\lower.7ex\hbox{E}\kern-.125emX}}

\usepackage{tikz}
\usetikzlibrary{
  decorations.pathreplacing,
  calc,
  arrows,
  positioning,
  backgrounds,
  automata,
  snakes,
  shapes
}
\usepackage{pgfplots}

\usepackage{svg}

\usepackage{colortbl}

\usepackage{algorithm}
\usepackage[noend]{algpseudocode}

\algnewcommand\algorithmiccontext{\textbf{Context:}}
\algnewcommand\Context{\item[\algorithmiccontext]}
\algnewcommand\algorithmicbranchoutput{\textbf{Branch Output:}}
\algnewcommand\BranchOutput{\item[\algorithmicbranchoutput]}
\algnewcommand\algorithmicndbranchoutput{\textbf{Output of each branch ($\beta$):}}
\algnewcommand\NDBranchOutput{\item[\algorithmicndbranchoutput]}
\algnewcommand\algorithmicglobaloutput{\textbf{Global Output:}}
\algnewcommand\GlobalOutput{\item[\algorithmicglobaloutput]}
\algnewcommand\algorithmicglobalspec{\textbf{Ensuring:}}
\algnewcommand\GlobalSpec{\item[\algorithmicglobalspec]}
\algnewcommand\algorithmicswitch{\textbf{switch}}
\algnewcommand\algorithmiccase{\textbf{case}}
\algnewcommand\algorithmicforeach{\textbf{foreach}}
\algnewcommand\algorithmicrep{\textbf{repeat}}
\algnewcommand\algorithmicnondet{\textbf{nondet}}
\algnewcommand\algorithmicor{\textbf{or}}
\algnewcommand\algorithmicassert{\textbf{assert}}
\algnewcommand\algorithmiclet{\textbf{let}}
\algnewcommand\Assert[1]{\State \algorithmicassert(#1)}
\algdef{SE}[SWITCH]{Switch}{EndSwitch}[1]{\algorithmicswitch\ #1\ \algorithmicdo}{\algorithmicend\ \algorithmicswitch}%
\algtext*{EndSwitch}%
\algdef{SE}[CASE]{Case}{EndCase}[1]{\algorithmiccase\ #1\ \textbf{:}}{\algorithmicend\ \algorithmiccase}%
\algtext*{EndCase}%
\algdef{S}[FOR]{ForEach}[1]{\algorithmicforeach\ #1\ \algorithmicdo}%
\algdef{S}[FOR]{Rep}[1]{\algorithmicrep\ #1\ \algorithmicdo}%
\algdef{SE}[NONDET]{Nondet}{EndNondet}{\algorithmicnondet\ }{\algorithmicend\ \algorithmicnondet}%
\algtext*{EndNondet}%
\algdef{SE}[Or]{Or}{EndOr}{\algorithmicor\ }{\algorithmicend\ \algorithmicor}%
\algtext*{EndOr}%
\algdef{SE}[LET]{Let}{EndLet}{\algorithmiclet\ }{\algorithmicend\ \algorithmiclet}%
\algtext*{EndLet}%

\AtBeginEnvironment{algorithm}{\setstretch{1.00}}

\newcounter{stepnum}


\makeatletter
\@ifpackageloaded{stix}{%
}{%
  \DeclareFontEncoding{LS2}{}{\noaccents@}
  \DeclareFontSubstitution{LS2}{stix}{m}{n}
  \DeclareSymbolFont{stix@largesymbols}{LS2}{stixex}{m}{n}
  \SetSymbolFont{stix@largesymbols}{bold}{LS2}{stixex}{b}{n}
  \DeclareMathDelimiter{\lBrace}{\mathopen} {stix@largesymbols}{"E8}%
  {stix@largesymbols}{"0E}
  \DeclareMathDelimiter{\rBrace}{\mathclose}{stix@largesymbols}{"E9}%
  {stix@largesymbols}{"0F}
}
\makeatother

\newtheorem{fact}{Fact}[section]
\newtheorem{remark}{Remark}[section]
\crefname{fact}{fact}{facts}
\Crefname{fact}{Fact}{Facts}
\crefname{remark}{remark}{remarks}
\Crefname{remark}{Remark}{Remarks}

\newtheorem*{thmstar}{Theorem}
\crefname{thmstar}{theorem}{theorems}
\Crefname{thmstar}{Theorem}{Theorems}

\newif\iffullversion
\fullversiontrue
\input{macros}

\acmJournal{PACMPL}

\begin{document}

\title{Parametric Recognizability over the Integers}

\author{Jakub Kaszycki}
\orcid{0009-0000-1155-3354}
\affiliation{
    \institution{University of Warsaw}
    \city{Warsaw}
    \country{Poland}
}
\email{j.kaszycki@uw.edu.pl}
\affiliation{
    \institution{IMDEA Software Institute}
    \city{Madrid}
    \country{Spain}
}

\author{Alessio Mansutti}
\orcid{0000-0002-1104-7299}
\affiliation{
    \institution{IMDEA Software Institute}
    \city{Madrid}
    \country{Spain}
}
\email{alessio.mansutti@imdea.org}

\author{Mikhail Starchak}
\orcid{0000-0002-2288-9483}
\affiliation{
    \institution{Max Planck Institute for Software Systems (MPI-SWS)}
    \city{Saarbr\"ucken}
    \country{Germany}
}
\email{starchak@mpi-sws.org}

\begin{abstract}

    B\"uchi arithmetic on a fixed integer $k\geq 2$ is the first-order theory of the structure $(\Z; 0, 1, +, V_k, \leq)$,
    where the binary relation~$V_k(x,y)$ is true if and only if $y$ is the greatest power of $k$ dividing $x$. By the
    B\"uchi-Bruy\`ere theorem, this theory completely describes the sets of integers that are recognizable by finite
    automata when encoded in the $k$-ary positional number system.

    Motivated by recent advances in symbolic automata, in this paper we study expressivity and decidability problems for B\"uchi
    arithmetic with a \emph{parametric positional encoding}, where $k$ is treated as a variable rather
    than a fixed constant. We show that this \emph{parametric} B\"uchi arithmetic provides a logical characterization of a
    class of \emph{parametric symbolic automata} over Linear Integer Arithmetic (LIA), thus extending the B\"uchi-Bruy\`ere theorem
    to the parametric setting.
    Building on this characterization, we give an algorithm showing that
    for every first-order sentence~$\phi_t$ (parametric in
    $t$) in the language of $(\Z; 0, 1, +, V_t, \leq)$,
    the set of bases $k \geq 2$ where $\phi_k$ is true (in $\Z$) is ultimately periodic.

    At a technical level, the original B\"uchi-Bruy\`ere theorem translates finite automata to the $\exists^*\forall^*$
    fragment of B\"uchi arithmetic, while our proof translates parametric symbolic automata to the $\exists^*\forall^*\exists^*$
    fragment of parametric B\"uchi arithmetic. We trace this discrepancy in the number of quantifier alternations to the
    use of the map $x \mapsto k \cdot x$ multiplying a natural number by the fixed base~$k$, whose parametric counterpart
    $x \mapsto t \cdot x$ is not definable in parametric B\"uchi arithmetic. Indeed, we show that given a $\exists^*\forall^*$
    sentence~$\phi_t$ in the language of ${(\Z; 0, 1, +, x \mapsto t \cdot x, V_t, \leq)}$,
    the existence of a base $k\geq 2$ such that $\phi_k$ is true is an undecidable problem.

    Finally, we provide a prototype implementation of both parametric and standard symbolic automata,
    and experiments on problems that ask to decide whether a formula is satisfied \emph{for all (or for some) bases $k \geq 2$} in an input segment $[\bmin..\bmax]$. These experiments indicate that parametric symbolic automata are more efficient when the range of bases is large, or so is $\bmin$.
\end{abstract}

\ccsdesc[500]{Theory of computation~Automated reasoning}

\keywords{Symbolic automata; B\"uchi arithmetic; Decidability; SMT-solving}

\setcopyright{cc}
\setcctype{by}
\acmYear{2027} \acmVolume{11} \acmNumber{POPL} \acmArticle{45}
\acmMonth{1} \acmDOI{10.1145/3871747}

\maketitle

\section{Introduction}\label{sec:introduction}

Establishing connections between computational models and arithmetic theories is a classic topic in theoretical computer science.
The focus of this paper is related to ``limitations'' of a fundamental theorem of this kind, established by B\"uchi and Bruy\`ere
in~\cite{Buechi1960,Bruyere85}. For a fixed integer $k \geq 2$, consider the first-order theory of the
structure~$(\N; 0, 1, +, V_k, \leq)$, which extends Linear Integer Arithmetic
(LIA, also known as Presburger arithmetic) with a binary predicate $V_k$ interpreted as follows: $V_k(x,y)$ is true whenever
$y$ is the largest power of $k$ dividing $x$. For example, $V_2(14,2)$ and $V_2(15,1)$ hold, while \( V_k(0,y) \) does not hold for any \( y \).
The B\"uchi-Bruy\`ere theorem tells us that the expressiveness of this arithmetic theory
coincides to that of regular languages:

\begin{theorem}[B\"uchi-Bruy\`ere theorem~\cite{BruyereHMV94}]\label{theorem:bb}
    Fix an integer~$k \geq 2$. Then, every set $\mathcal{S} \subseteq \N^n$ is recognizable by a finite automaton
    over the alphabet~$\{0,...,k-1\}^n$ if and only if $\mathcal{S}$~is first-order definable in the structure $(\N; 0, 1, +, V_k, \leq)$.
\end{theorem}

Here, non-negative integers can be encoded in the most- or least-significant digit first notation (\emph{msdf}
and \emph{lsdf}, respectively) as words $a_{\ell - 1}\,\dots\, a_{1} \, a_{0}$ in the language
$\{0,1,\dots,k-1\}^+$. In the case of \emph{msdf}, this word defines the integer $a_{\ell - 1}\cdot k^{\ell - 1}
    + \dots + a_1\cdot k + a_0$. We say that $S\subseteq\N$ is \emph{recognizable} by an automaton~$\mathcal{A}$
if it accepts exactly \emph{all} words representing elements of $S$.
By reading $k$-ary expansions of integers synchronously, the definition is generalized on dimensions $n\geq 1$
(see formal definition in~\Cref{subsec:encodings}). A set $\mathcal{S}\subseteq\N^n$ is (first-order) \emph{definable} in the structure
$(\N; 0, 1, +, V_k, \leq)$ if there is a formula $\phi$ in the first-order language of this structure with $n$
free variables such that $\vec a\in \mathcal{S}$ if and only if $\phi(\vec a)$ is true.
Then,~\Cref{theorem:bb} states that \emph{recognizability} by finite automata corresponds to \emph{definability} in $(\N; 0, 1, +, V_k, \leq)$.
This result can be naturally extended to work over the integers~\cite{Boigelot21}.
In this case, integers are represented in the $k$'s complement encoding, i.e., as words
$a_{\ell}\,\dots \, a_{0}$ in the language $\{0,k-1\}\cdot\{0,1,\dots,k-1\}^*$.
When $a_\ell = 0$, this word defines $a_{\ell - 1}\cdot k^{\ell - 1}
    + \dots + a_1\cdot k + a_0$.
Otherwise, if $a_\ell = k-1$, it encodes $-k^{\ell} + a_{\ell - 1}\cdot k^{\ell - 1}
    + \dots + a_1\cdot k + a_0$.
For example, `$9877$' represents in the $10$'s complement encoding the integer
$-123 = -1000 + 877$.

\begin{example}
    The two finite automata below recognize binary (\Cref{fig:example-v2-1mod3}) and decimal (\Cref{fig:example-v10-1mod3}) words representing
    the integers $x$
    that satisfy the formula $V_k(x,x)\land \exists y : x = 3\cdot y + 1$, i.e., powers of $k$ (with
    $k=2$ and $k=10$, respectively) congruent to $1$ modulo $3$.

    \vspace{-0.7em}
    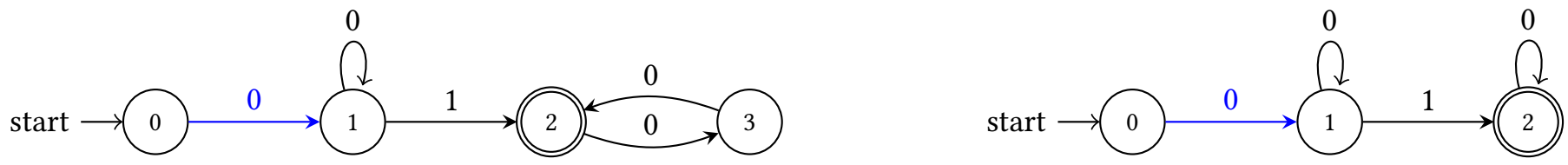
\begin{figure}[h]
        \begin{subfigure}{.5\textwidth}
            \centering
            \begin{tikzpicture}[node distance = 0.7cm]
                \scriptsize
                \node (0) [state, initial left, scale=0.8] {$0$};
                \node (1) [state, right=1cm of 0, scale=0.8] {$1$};
                \node (2) [state, accepting, right=1cm of 1, scale=0.8] {$2$};
                \node (3) [state, right=1cm of 2, scale=0.8] {$3$};
                \path[-stealth]
                (0) edge[above, blue] node{$0$} (1)
                (1) edge[loop above] node{$0$} (1)
                (1) edge[above] node{$1$} (2)
                (2) edge[above, bend right=20] node{$0$} (3)
                (3) edge[above, bend right=20] node{$0$} (2);
            \end{tikzpicture}
            \caption{for $V_2(x,x)\land \exists y : x = 3\cdot y + 1$}
            \label{fig:example-v2-1mod3}
        \end{subfigure}
        \begin{subfigure}{.45\textwidth}
            \centering
            \begin{tikzpicture}[node distance = 1cm]
                \scriptsize
                \node (0) [state, initial left, scale=0.8] {$0$};
                \node (1) [state, right=1cm of 0, scale=0.8] {$1$};
                \node (2) [state, accepting, right=1cm of 1, scale=0.8] {$2$};
                \path[-stealth]
                (0) edge[above, blue] node{$0$} (1)
                (1) edge[loop above] node{$0$} (1)
                (1) edge[above] node{$1$} (2)
                (2) edge[loop above] node{$0$} (2);
            \end{tikzpicture}
            \caption{for $V_{10}(x,x)\land \exists y : x = 3\cdot y + 1$}
            \label{fig:example-v10-1mod3}
        \end{subfigure}
        \caption{Automata in base $2$ and base $10$ in signed msdf encoding.}
        \label{fig:example-v2-or-10-1mod3}
    \end{figure}
    \vspace{-0.7em}
    \noindent
    For example, the automaton from \Cref{fig:example-v2-1mod3} accepts words `$01$', `$0100$', and `$010000$'
    representing the integers $1$, $4$, and $16$, but does not accept any word with odd number of zeros after `$1$'.
    The automaton from \Cref{fig:example-v10-1mod3} accepts all powers of $10$: they are always congruent to $1$ modulo $3$.
    The first transition (in {\color{blue}blue}) decides the sign of the integer.
    Transitions to sink states are omitted.
\end{example}

\subsection{Applications of B\"uchi-Bruy\`ere theorem}

The proof of~\Cref{theorem:bb} is constructive, and shows the decidability of the first-order theory of the structure $(\Z; 0, 1, +, V_k, \leq)$, which we henceforth call \emph{B\"uchi arithmetic (on base $k$)}.
This result gave raise to the study of \emph{automatic structures} (the structures that can be represented by~automata)~\cite{Hodgson76,KhoussainovN95,Gradel20},
and of \emph{numeration systems} admitting automatic representations~\cite{Shallit94,BruyereH97,HieronymiT18}.

In the context of automatic structures, B\"uchi arithmetic is bi-interpretable with the weak second-order theory of one successor (WS1S):
alongside the translation from B\"uchi arithmetic to automata,
in his pioneering paper~\cite{Buechi1960} B\"uchi establishes translations between automata and WS1S in both directions.
WS1S has found its applications in formal verification and is supported by several solvers~\cite{ElgaardKM98,FiedorHJLV17,DAntoniV17,FiedorHLV19}.
These applications extend to bit-vector analysis,
since basic bitwise operations can easily be represented by finite automata and in B\"uchi arithmetic; see~\cite{SchueleS06,BhatSHG25}.
B\"uchi arithmetic is also related to the linear theory of $p$-adic numbers with divisibility $(\Q_p; 0, 1, +, |_p, =)$ \cite{Hodgson76}, where $p$ is a fixed prime number;
we touch upon this~connection in \Cref{subsec:limitations}.

In the context of numeration systems, the B\"uchi-Bruy\`ere theorem has found important applications in word combinatorics, which studies periodicity and the (non-)existence of patterns in
sequences of (infinite) words~\cite{AlloucheS03}. Automata-based theorem provers such as \mbox{Walnut}~\cite{Mousavi16,Shallit22} and Pecan~\cite{OeiMSH21,HieronymiMOSSS24}
are actively used to prove new results in combinatorics on words (now in the hundreds), and to re-prove old ones---sometimes spotting and correcting errors; see, e.g.,~\cite[Section~9.11.6]{Shallit22}.

Another consequence of \Cref{theorem:bb} is that LIA and the classic integer linear
programming feasibility problem (ILP) can be decided via automata theory.
In the absence of quantifier alternation (e.g., in ILP), this connection is often impractical:
as shown in~\cite{AmatGM25}, automata-based tools are usually outperformed by algebraic methods on the QF\_LIA benchmarks
of the \mbox{SMT-LIB} repository~\cite{BarrettFT16}.
Automata can, however, be the best option on benchmarks with large coefficients and sparse solution sets~\cite{GaneshBD02}.
For LIA formulas featuring quantifier alternations, the automata-based approach is more competitive.
This is in part due to a line of improvements on the representation of
linear equalities and inequalities via finite automata,
starting with~\cite{WolperB95} and continuing through~\cite{WolperB00,Boigelot21} to~\cite{HabermehlHHHL24}.
The latter paper shows that a careful combination of
algebraic and automata-based methods can efficiently handle LIA formulas with one quantifier alternation.

\subsection{Some limitations of B\"uchi-Bruy\`ere theorem}\label{subsec:limitations}

In practice, compiling a formula from B\"uchi arithmetic (or LIA) down to automata comes
with a limitation: the
number of transitions between any two states can
be of the order of $k^{n}$, where $n$ is the number of variables in the formula and $k$ is the base of the encoding. This blow-up is usually
mitigated by means of binary decision diagrams: ordered (OBDDs) as in~\cite{WolperB00} or multi-terminal (MTBDDs) as in~\cite{HabermehlHHHL24}.
However, this solution is only effective for small values of~$k$
(and in fact, the tools from both mentioned papers fix $k=2$).

Our work takes as its starting point an alternative to binary decision diagrams, which is inspired by the algorithms for constructing automata for linear
equations and inequalities from~\cite[Algorithms~1~and~2]{Boigelot21}.
An inspection of these algorithms reveals that $k$ influences the control flow only through LIA
formulas: every branch condition involving $k$ is a linear arithmetic constraint.
This observation naturally leads us to a class of \emph{symbolic automata}, in the sense of~\cite{DantoniV21}, where
transitions are labelled directly by LIA formulas representing these branch conditions, and where each variable represents a digit base $k\geq 2$.

\begin{example}\label{ex:adders}
    \Cref{fig:example-adder-10} shows the result of applying~\cite[Algorithm~1]{Boigelot21} to the equation $(x+y=z)$, with base $k=10$.
    The figure hides the fact that there are more than $70$ transitions between each pair of states $(p,q) \in \{0,1\} \times \{0,1\}$.
    The automaton contains $3$ states and~$330$~transitions.
    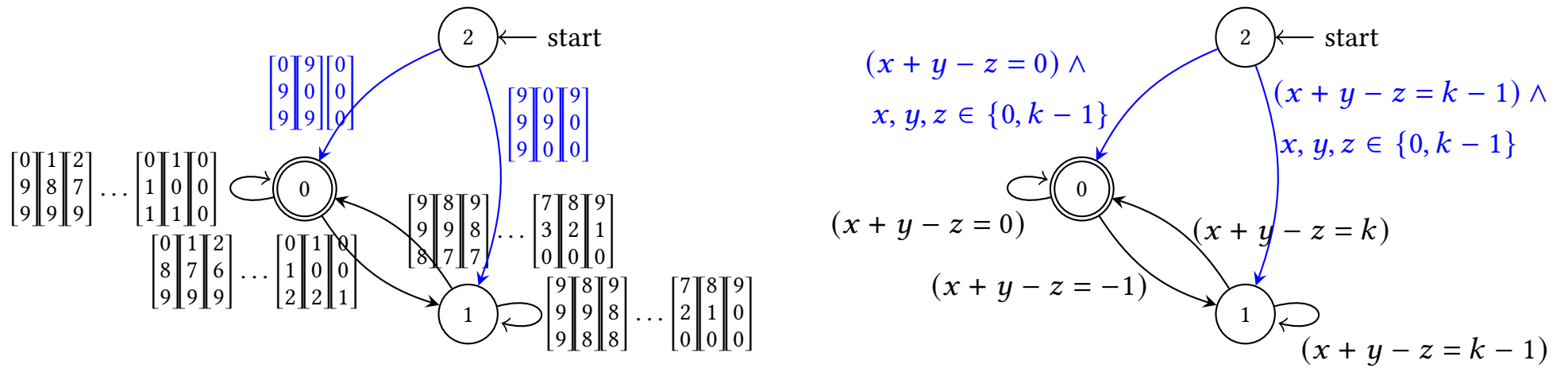
\begin{figure}[h]
        \vspace{-0.7em}
        \begin{subfigure}{.5\textwidth}
            \centering
            \begin{tikzpicture}[node distance = 2cm]
                \scriptsize
                \node (2) [state,initial right,scale=0.8] {$2$};
                \node (0) [state,accepting, below left=0.9cm and 1.0cm of 2,scale=0.8] {$0$};
                \node (1) [state, below=1.8cm of 2,scale=0.8] {$1$};
                \path[-stealth]
                (1) edge[loop right] node{\scalebox{0.8}{$\!\!\begin{bmatrix}9\\9\\9\end{bmatrix}\!\!\!\begin{bmatrix}8\\9\\8\end{bmatrix}\!\!\!
                                \begin{bmatrix}9\\8\\8\end{bmatrix}\dots\begin{bmatrix}7\\2\\0\end{bmatrix}\!\!\!\begin{bmatrix}8\\1\\0\end{bmatrix}\!\!\!
                                \begin{bmatrix}9\\0\\0\end{bmatrix}$}} (1)
                (1) edge[right=0.2, bend right=20] node{\scalebox{0.8}{$\!\!\begin{bmatrix}9\\9\\8\end{bmatrix}\!\!\!\begin{bmatrix}8\\9\\7\end{bmatrix}\!\!\!
                                \begin{bmatrix}9\\8\\7\end{bmatrix}\dots\begin{bmatrix}7\\3\\0\end{bmatrix}\!\!\!\begin{bmatrix}8\\2\\0\end{bmatrix}\!\!\!
                                \begin{bmatrix}9\\1\\0\end{bmatrix}$}} (0)
                (0) edge[left=0.2, bend right=20] node{\scalebox{0.8}{$\!\!\begin{bmatrix}0\\8\\9\end{bmatrix}\!\!\!\begin{bmatrix}1\\7\\9\end{bmatrix}\!\!\!
                                \begin{bmatrix}2\\6\\9\end{bmatrix}\dots\begin{bmatrix}0\\1\\2\end{bmatrix}\!\!\!\begin{bmatrix}1\\0\\2\end{bmatrix}\!\!\!
                                \begin{bmatrix}0\\0\\1\end{bmatrix}$}} (1)
                (0) edge[loop left] node{\scalebox{0.8}{$\!\!\begin{bmatrix}0\\9\\9\end{bmatrix}\!\!\!\begin{bmatrix}1\\8\\9\end{bmatrix}\!\!\!
                                \begin{bmatrix}2\\7\\9\end{bmatrix}\dots\begin{bmatrix}0\\1\\1\end{bmatrix}\!\!\!\begin{bmatrix}1\\0\\1\end{bmatrix}\!\!\!
                                \begin{bmatrix}0\\0\\0\end{bmatrix}$}} (0)
                (2) edge[above=0.2, bend left=20, blue] node{\scalebox{0.8}{$\hspace{1cm}\begin{bmatrix}9\\9\\9\end{bmatrix}\!\!\! \begin{bmatrix}0\\9\\0\end{bmatrix}\!\!\! \begin{bmatrix}9\\0\\0\end{bmatrix}$}} (1)
                (2) edge[left=0.1, bend right=20, blue] node{\scalebox{0.8}{$\begin{bmatrix}0\\9\\9\end{bmatrix}\!\!\! \begin{bmatrix}9\\0\\9\end{bmatrix}\!\! \begin{bmatrix}0\\0\\0\end{bmatrix}$}} (0);
                \path ([yshift=-0.45cm]1.center);
            \end{tikzpicture}
            \caption{Automaton for $x+y=z$ base $10$}
            \label{fig:example-adder-10}
        \end{subfigure}
        \begin{subfigure}{.45\textwidth}
            \centering
            \begin{tikzpicture}[node distance = 2cm]
                \scriptsize
                \node (2) [state,initial right,scale=0.8] {$2$};
                \node (0) [state, accepting, below left=0.9cm and 1.0cm of 2,scale=0.8] {$0$};
                \node (1) [state, below=1.8cm of 2,scale=0.8] {$1$};
                \path[-stealth]
                (1) edge[loop right] node[below right,xshift=-0.25cm,yshift=-0.18cm,inner ysep=0pt]{$(x+y-z=k-1)$} (1)
                (1) edge[right=0.2, bend right=20] node{$(x+y-z=k)$} (0)
                (0) edge[left=0.2, bend right=20] node[xshift=0.1cm,yshift=-0.13cm]{$(x+y-z=-1)$} (1)
                (0) edge[loop left] node[below left,xshift=0.25cm,yshift=-0.18cm,inner ysep=0pt]{$(x+y-z=0)$} (0)
                (2) edge[above=0.2, bend left=20, blue] node{
                        $\hspace{2.2cm}\begin{aligned}
                                (x + y & - z = k-1) \,\land \\
                                x,y,z  & \in \{0, k-1\}
                            \end{aligned}$
                    } (1)
                (2) edge[left=0.1, bend right=20, blue] node[xshift=-0.2cm]{
                        $\begin{aligned}
                                (x + y & - z = 0) \,\land \\
                                x,y,z  & \in \{0, k-1\}
                            \end{aligned}$
                    } (0);
                \path ([yshift=-0.45cm]1.center);
            \end{tikzpicture}
            \caption{Symbolic automaton for $x+y=z$ base $k\geq 2$}
            \label{fig:example-adder-par}
        \end{subfigure}
        \caption{Automata for addition in signed msdf encoding.}
        \vspace{-0.7em}
    \end{figure}

    \noindent
    \Cref{fig:example-adder-par} demonstrates a symbolic version for the same equation $(x+y=z)$.
    The variables of formulas for the transitions belong to the segment $[0..k-1] = [0..9]$ (we omit these constraints for presentational simplicity).
    Both automata have the following accepting run
    \begin{align*}
          & \:\scalebox{0.8}{$\begin{bmatrix}0\\9\\0\end{bmatrix}$}\hspace{0.33cm}
        \scalebox{0.8}{$\begin{bmatrix}8\\9\\8\end{bmatrix}$}\hspace{0.33cm}
        \scalebox{0.8}{$\begin{bmatrix}7\\2\\0\end{bmatrix}$}\hspace{0.33cm}
        \scalebox{0.8}{$\begin{bmatrix}9\\1\\0\end{bmatrix}$}\hspace{0.33cm}
        \scalebox{0.8}{$\begin{bmatrix}0\\1\\2\end{bmatrix}$}\hspace{0.33cm}
        \scalebox{0.8}{$\begin{bmatrix}9\\9\\8\end{bmatrix}$}\hspace{0.33cm}
        \scalebox{0.8}{$\begin{bmatrix}0\\1\\1\end{bmatrix}$}
        \quad
        \begin{array}{ll}
             & \gets x \,= \,879 090 \,=\, \hphantom{-1^{6} }0 + 879 090 \\ &\gets y \,= -78 809 \,=\, -10^{6} + 921 191 \\ &\gets z \:= \,800 281 \,=\, \hphantom{-1^{6} }0 + 800 281
        \end{array}
        \\
        2 & \rightarrow 1 \rightarrow 1 \rightarrow 1\rightarrow 0\rightarrow 1 \rightarrow 0\rightarrow 0 \qedhere
    \end{align*}
\end{example}

Observe that in~\Cref{fig:example-adder-par} the base $k$ never occurs as a coefficient of a variable. It is then possible to see the symbolic automaton as a parametric device featuring linear formulas in which the base $k$ is an integer variable interpreted as a natural number greater than or equal to $2$, rather than a fixed integer. This is sometimes desirable in applications.
A classic example is given by the decidability of the existential ($\exists^*$) fragment of LIA with divisibility constraints~\cite{Beltyukov76,Lipshitz78}, which found applications in analysing programs with lists~\cite{BozgaI05}, one-counter and timed automata~\cite{Haase12, BundalaO17, PerezR22}, and in solving word equations~\cite{LinM21}.
The algorithms from~\cite{Beltyukov76,Lipshitz78} require checking satisfiability of an $\exists^*$ formula $\Phi_p$ in the language of $(\Q_p; 0, 1, +, |_p, =)$ \emph{for every prime~$p$} not exceeding some constant.
For a fixed prime $p$,
the formula $\Phi_p$ can be translated in polynomial time into an $\exists^*$ formula of B\"uchi arithmetic~\cite{GuepinH019}.
Motivated by this connection,
\cite{HaaseM21} shows that for $\exists^*$ sentences in B\"uchi arithmetic considered \emph{parametrically in the base}, both the problems of satisfiability for
\emph{some} base $k \geq 2$ and for \emph{all} bases $k \geq 2$ are decidable.
The approach in~\cite{HaaseM21} does not handle quantifier alternation (it is known from~\cite{HaaseR2021,Starchak24} that existential B\"uchi arithmetic is less expressive than full B\"uchi arithmetic) but circumvents the fact that~\Cref{theorem:bb} is limited to a fixed base for the numeration system.

\subsection{Our contributions}\label{subsec:contributions}

In this paper, we establish a parametric version of the B\"uchi-Bruy\`ere theorem,
which provides a unified framework for problems in which a given B\"uchi arithmetic formula
must be solved either for a single large base, or for multiple bases.
The correspondence between arithmetic theories and automatic structures that we establish
is between \emph{parametric B\"uchi arithmetic} and \emph{parametric symbolic automata over LIA}.
The former is a decidable syntactic fragment of the
first-order theory of the structure $(\Z; 0, 1, +, V_t, \leq)$, consisting of B\"uchi arithmetic formulas where the base $t$ is not
a fixed integer constant but a free integer variable (the~\emph{parameter}).
The latter are objects akin to the symbolic automaton in~\Cref{fig:example-adder-par}, but with the base $k$ replaced by the parameter $t$.
We leave the formal definitions to~\Cref{sec:preliminaries},
but intuitively these objects represent functions $\mathcal{F} \colon \N_{\geq 2} \!\to \powerset(\Z^n)$ that map a base $k\geq 2$ to a set
$\mathcal{F}(k)\subseteq\Z^n$;
instead of a set of integer tuples, as in \Cref{theorem:bb}.

\begin{theorem}[Parametric Recognizability]\label{theorem:bb-par}
    A function $\mathcal{F} \colon \N_{\geq 2} \!\to \powerset(\Z^n)$ is recognizable~by a parametric symbolic automaton over LIA if and only if
    $\mathcal{F}$ is definable in parametric B\"uchi arithmetic.
\end{theorem}

Consider a formula $\phi_t$ of parametric B\"uchi arithmetic whose only free variable is the parameter~$t$
(i.e., $\phi_t$ is a \emph{sentence} of this logic; then $n=0$ and $\Z^{0}=\{\emptyset\}$).
This formula represents the function ${\mathcal{F} \colon \N_{\geq 2} \to \{\emptyset,\{\emptyset\}\}}$ defined as follows.
For every $k \geq 2$, let $\phi_k$ be the sentence of B\"uchi arithmetic obtained by replacing $t$ with~$k$.
Then, $\mathcal{F}(k) = \{\emptyset\}$ if $\phi_k$ is true in the structure $(\Z; 0, 1, +, V_k, \leq)$,
and $\mathcal{F}(k) = \emptyset$ otherwise.
By~\Cref{theorem:bb-par}, there is a parametric symbolic automaton $\mathcal{A}$ representing~$\mathcal{F}$.
Starting from~$\mathcal{A}$, we give an algorithm (\Cref{algo:realizable-bases}) that computes the set of all bases $k$
such that $\mathcal{F}(k) = \{\emptyset\}$.
This set is ultimately periodic, and the algorithm returns a finite representation of it.

\begin{restatable}{theorem}{TheoremUltimatelyPeriodic}\label{theorem:ultimately-periodic}
    For every sentence $\phi_t$ in parametric B\"uchi arithmetic, the set of bases $k \geq 2$ for which $\phi_k$ is a true
    sentence in the structure $(\Z; 0, 1, +, V_k, \leq)$ is an ultimately periodic set.
\end{restatable}
This resolves a problem left open in \cite{HaaseM21}: both ``\emph{existence}'' and ``\emph{universality}'' problems are decidable for the
\emph{full} B\"uchi arithmetic, not only the $\exists^*$ fragment.

In contrast to \Cref{theorem:bb}, where $\exists^*\forall^*$ formulas
of B\"uchi arithmetic are sufficient to describe computations of finite automata~\cite{Villemaire92,HaaseR2021}, the left-to-right direction of~\Cref{theorem:bb-par}
requires $\exists^*\forall^*\exists^*$ formulas. We trace back this increase in quantifier alternation to the fact that, while the map $(x\mapsto k\cdot x)$ comes for free when the base is a fixed integer (it is expressible in LIA),
$(x\mapsto t\cdot x)$ is undefinable in the parametric case. In \Cref{sec:multiplication-effects} we analyse the effects
of inclusion of multiplication by the parameter $t$ into parametric B\"uchi arithmetic.

\begin{theorem}\label{theorem:mul}
    The existence problem for the $\exists^*\forall^*$ fragment of parametric B\"uchi arithmetic extended with $(x\mapsto t\cdot x)$
    is undecidable. Therefore, $(x\mapsto t\cdot x)$ is not definable in parametric B\"uchi arithmetic.
\end{theorem}

This theorem is proved by encoding computations of counter machines via
the bitwise minimum operation base $t$, similarly to~\cite{JonesM84,Starchak23}.

In \Cref{sec:tool}, we turn to the applicability of our theoretical results.
First, we give an improved algorithm for translating linear inequalities into parametric
symbolic automata. We then show how parametric Büchi arithmetic can express two
arithmetical problems from the literature as instances of the \textsc{Bounded Existence}
and \textsc{Bounded Universality} problems, i.e., where the base ranges over a finite
interval $[\bmin..\bmax]$. We conclude the paper with an experimental comparison, on these bounded
problems, of parametric versus non-parametric symbolic automata.

\iffullversion
\textit{Full proofs appear in the appendix.}
\else
\textit{Full proofs appear in the appendix of the full version of this paper~\cite{KaszyckiMS26full}.} \textcolor{red}{\textbf{TODO}}%
\fi

\section{Preliminaries}\label{sec:preliminaries}

The naturals~$\N$ include zero, and we write $\N_{\geq c}$ for the natural numbers greater than or equal to $c \in \N$.
We usually use $a,b,c,\dots$ for integers. The powerset of a set~$S$ is denoted~$\mathcal{P}(S)$. The set of all
(total) functions from $A$ to $B$ is denoted~$[A \to B]$.
For a function $f \colon A \to B$ and a pair $(a,b) \in A \times B$,
$f[a \mapsto b] \colon A \to B$ is the function identical to $f$ on all inputs except that $f[a \mapsto b](a) = b$.

\subsection{Logic notation}\label{subsec:Logic}

We assume familiarity with the basic definitions of first-order logic and only briefly recall some notions. A
\emph{signature} is a set of constant, function, and predicate symbols, together with their arities (which are usually
clear from the context and omitted). Given a signature $\sigma$, we write $\Formulas_\sigma$ for the set of all
first-order formulas $\phi,\psi,\dots$ over $\sigma$, using variables from a fixed countable set~$X$. Formulas without
free variables are called \emph{sentences}. Given $m \geq 1$, the~\emph{$\exists^*\forall^*\dots Q_\ell^*$ fragment}
of $\Formulas_\sigma$ is understood as the set of all formulas of $\Formulas_\sigma$ that, once rewritten in their prenex normal form, are of
the form $Q_1 \vec x_1 Q_2 \vec x_2 \dots Q_\ell \vec x_\ell \psi$ for some $\ell \leq m$, where $Q_i \in
   \{{\exists},{\forall}\}$ stands for $\exists$ exactly when $i$ is odd, \ each $\vec x_i$ is a vector of variables from
$X$, and $\psi$ is a quantifier-free formula. The $\exists^*$ fragment (of~$\Formulas_\sigma$) is sometimes called the
\emph{existential} fragment.
Given real numbers $a,b \in \R$, we write $[a..b]$ for the set of integers $\{n \in \Z : a \leq n \leq b\}$, and $[a,b]$ for the real interval $\{x \in \R : a \leq x \leq b\}$.

A \emph{$\sigma$-structure} $\amodel \coloneqq (M; I)$ is composed of a non-empty set $M$ (i.e., the universe) and of
an interpretation $I$ mapping each constant/function/predicate symbol in $\sigma$ to an element/function/relation over
$M$. The \emph{satisfaction relation} $\_\models_{\amodel} \_ \colon [X \to M] \to \Formulas_\sigma \to \{\top, \bot\}$
is defined as usual for first order logic. For $n\in\N_{\geq 1}$, we say that a set $S\subseteq M^n$ is \emph{definable in the structure \amodel}
if there exist $\phi\in\Formulas_\sigma$ with $n$ free variables $y_1,...,y_n\subseteq X$ such that $S=\{(\mu(y_1),\dots,\mu(y_n)) : \mu\models_{\amodel} \phi \}$.
Since the satisfaction of a formula only depends on the values assigned to its free variables, for a formula $\phi$ with free variables among~$x_1,\dots,x_n$, and $a_1,\dots,a_n\in M$, we will sometimes write $[x_1\mapsto a_1,\dots,x_n\mapsto a_n]\models_{\amodel} \phi$
to denote that $\mu\models_{\amodel} \phi$ holds for all (equivalently, for some) valuation $\mu\in[X\to M]$ such that $\mu(x_i)=a_i$ for every $i\in[1..n]$.
We will also sometimes use $\models_{\amodel}$ as a unary relation on
$\Formulas_\sigma$, writing $\models_{\amodel} \phi$ whenever $\mu \models_{\amodel} \phi$ holds for every $\mu \in [X \to M]$.
The \emph{first-order theory of the structure $\amodel$} is the set $\theory(\amodel) \coloneqq \{\phi \in \Formulas_\sigma : \text{$\phi$ is a sentence and $\models_{\amodel} \phi$}\}$.%

\subsection{Parametric B\"uchi arithmetic}\label{subsec:PBA}
The formulas of \emph{parametric B\"uchi arithmetic} are built from the signature $\PBA \coloneqq \{0, 1, +, V_t,
   \leq\}$, where $0$, $1$, are constant symbols, $+$ is a binary function symbol, and $V_t$ and $\leq$ are binary
predicate symbols. We refer to the symbol $t \notin X$ as the \emph{parameter}. Without loss of generality, we assume
these first-order formulas to be built from the grammar
\[
   \phi \Coloneqq
   \top
   \ \mid \    \bot
   \ \mid \    {\textstyle\sum_{i=1}^n a_i \cdot x_i} \leq c
   \ \mid \    V_t(x,y)
   \ \mid \    \lnot \phi
   \ \mid \    \phi \land \phi
   \ \mid \    \exists x\,\phi\,,
\]
where $x,y,x_1,\dots,x_n$ are variables from $X$, and $a_1,\dots,a_n,c$ are integers encoded in binary. We use
standard abbreviations for first-order logic, defining $\lor$, $\implies$, $\iff$, $\forall$, $<$, and $=$ from
$\lnot$, $\land$, $\exists$, $\leq$.

Given~$k \in \N_{\geq 2}$, consider the $\PBA$-structure $\kBAmodel_k \coloneqq (\Z;I)$ interpreting the symbol $V_t$ using the
set $\{(n,d) \in \Z \times \N : \text{$d$ is the largest power of $k$ dividing $n$}
   \}$, and $0,1,+$ and $\leq$ as the standard constants and operators over the integers. We call this \emph{the structure
   of B\"uchi arithmetic on base $k$}, and write $\models_k$ as a shorthand for the satisfaction relation
$\models_{\kBAmodel_k}$. Observe that, for every $n,d \in \Z$, we have $V_t(n,d)$ if and only if $V_t(-n,d)$.
The \emph{semantics}~$\sem{\,\_\,}_{\,-\,} \colon \pbaformulas \to \N_{\geq 2} \to \mathcal{P}[X \to \Z]$ of parametric
B\"uchi arithmetic is defined as the function that given a formula $\phi \in \pbaformulas$ and $k \in \N_{\geq 2}$
returns the set $\sem{\phi}_k \coloneqq \{ \mu : \mu \models_k \phi \}$ of the solutions that $\phi$ has in B\"uchi
arithmetic on base $k$.
We write $\realbases(\phi)$ for the set of \emph{realizable bases} of $\phi$, i.e., the set $\{k \in \N_{\geq 2} :
   \sem{\phi}_k \neq \emptyset\}$.

\smallskip
The \textsc{Existence} and \textsc{Universality}
problems for parametric B\"uchi arithmetic ask, given a formula
$\phi \in \pbaformulas$, whether $\realbases(\phi)$ is non-empty or
equal to $\N_{\geq 2}$,
respectively.

\begin{remark}\label{remark:semantic-differences}
   The semantics $\sem{\,\_\,}_{\,-\,}$ given here is slightly different
   from the one considered in~\Cref{subsec:contributions},
   which had the type $\Formulas_\PBA \to \N_{\geq 2} \to \powerset(\Z^n)$.
   Taking $\powerset[X \to \Z]$ instead of $\powerset(\Z^n)$
   is entirely benign and allows us to avoid reordering variables in proofs and in various constructions.
   Formally, given a formula $\phi(x_1,\dots,x_n) \in \Formulas_\PBA$ with free variables $x_1,\dots,x_n$, the semantics in~\Cref{subsec:contributions}
   is obtained from $\sem{\phi}_{\,-}$ as the map $(k \in \N_{\geq 2}) \mapsto \{ (\mu(x_1),\dots,\mu(x_n)) \mid \mu \in \sem{\phi}_{k} \}$.
\end{remark}

\subsection{Presburger arithmetic with digits}\label{subsec:PrA}
We define \emph{Presburger arithmetic with digits}
as the set of first-order formulas over the signature $\PAt \coloneqq \{0, 1, +, t,
   \leq\}$. Given $k \in \N_{\geq 2}$, we denote by $\kPAmodel_k \coloneqq (\Z;J)$ the structure obtained from
$\kBAmodel_k$ by the replacement of the predicate $V_t$ with the constant $t$ interpreted by the integer~$k$.
The satisfaction relation $\models_k$ and semantics~$\sem{\,\_\,}_{\,-}$ for Presburger arithmetic with digits are then
defined analogously.%

We use the name ``Presburger arithmetic with digits''
   instead of ``parametric Presburger arithmetic'' to not confuse this logic with the one defined
   in~\cite{BogartGNW19,MansuttiS25}, which features a parametric multiplication function $x \mapsto t \cdot x$. We will
   discuss the effects of adding this function to parametric B\"uchi arithmetic in~\Cref{sec:multiplication-effects}.

The qualifier ``with digits'' refers to the fact that, in this logic, the constant symbol $t$ can be used to quantify over
the digits $0,\dots,t-1$ of numbers encoded in base $t$: this is done with \emph{bounded quantifiers} of
the form $(\exists x < t) \, \phi \coloneqq \exists x (0 \leq x \leq t -1 \land \phi)$. We define the \emph{digit
   semantics} $\digsem{\,\_\,}_{\,-} \colon \Formulas_{\PAt} \to (k \colon \N_{\geq 2}) \to \mathcal{P}[X \to [0..k-1]]$
of Presburger arithmetic with digits to be the function given by~$\digsem{\phi}_k \coloneqq \sem{\phi}_k \cap [X \to
      [0..k-1]]$, for every $\phi \in \Formulas_{\PAt}$ and $k \in \N_{\geq 2}$; i.e., $\digsem{\phi}_k$ consists of all
valuations~$\mu$ mapping every variable to a digit base $k$, such that $k,\mu \models \phi$.

\subsection{Symbolic automata}\label{subsec:symbolic-automata}
For simplicity, we consider symbolic automata over first-order signatures, instead of Boolean algebras as
in~\cite{DantoniV21}. Let $\sigma$ be a signature.

\begin{definition}[NSA]
   A \emph{non-deterministic symbolic automaton
      (over~$\sigma$)} is a tuple $(\Phi, Q, q_0, F, \delta)$ where:
   $\Phi$ is a finite set of first-order formulas from~$\Formulas_\sigma$ called \emph{labels},
   $Q$ is a finite set of \emph{states},
   $q_0 \in Q$ is the \emph{initial} state,
   $F\subseteq Q$ is the set of \emph{accepting} states, and
   $\delta: Q\times Q\to \Phi$ is the \emph{transition} function.
\end{definition}

Consider a $\sigma$-structure $\amodel$ with universe~$M$.
A \emph{finite word} over $\amodel$ is a sequence $\vec w =
   w_1w_2\ldots w_n$ of valuations (called \emph{letters}) of variables over $\amodel$,
that is, $\vec w \in [X \to M]^*$. We say that $\vec w$ \emph{reaches state~$q$ from state $p$}
whenever there is a sequence of states (i.e., a \emph{path}) $p=q_1,\ldots, q_{n+1}=q$ from $Q$ such that $w_i \models_{\amodel} \delta(q_i, q_{i+1})$
for every $i \in \{1,\dots,n\}$. A word $\vec w$
is \emph{accepted by an NSA} $\Aut \coloneqq (\Phi, Q, q_0, F, \delta)$
whenever $\vec w$ reaches some state $q \in F$ from the initial state $q_0$.
We write $\Lang_{\amodel}(\Aut)$ (or $\Lang(\Aut)$ when $\amodel$ is clear from the context) for the language of
$\Aut$, i.e., the set of all finite words accepted by $\Aut$.

\begin{example}
   The reader has already seen a symbolic automaton over the formulas
   of LIA in \Cref{fig:example-adder-par}. To make it formally defined, replace $k$ with
   any fixed integer, and add the constraints $(0\leq u\land u\leq k-1)$ for $u\in\{x,y,z\}$
   to every outcoming transition of states $0$~and~$1$.
\end{example}

A symbolic automaton $\Aut$ is \emph{deterministic (for~$\amodel$)}---in short, a \emph{DSA}---whenever, for every $q
   \in Q$ and every $w \colon X \to M$, there is at most one $q' \in Q$ such that $w \models_{\amodel} \delta(q, q')$.
The structure~$\amodel$ is \emph{effective} whenever
there is an algorithm to decide, for a given $\phi \in \Formulas_\sigma$, whether $w \models_{\amodel} \phi$ for some $w \in [X \to M]$.
We recall standard closure properties of symbolic automata:

\begin{restatable}[{\cite{DantoniV21}}]{fact}{FactSFAClosure}\label{fact:sfa-closure}
   Let $\sigma$ be a signature, and let $\amodel$ be an effective $\sigma$-structure. Consider symbolic automata over $\sigma$.
   There are algorithms with the following specifications:
   \begin{description}
      \item[Determinization:] For an NSA $\Aut$, return a DSA $\Aut'$ such that $\Lang(\Aut') = \Lang(\Aut)$.
      \item[Intersection:] For NSAs $\Aut_1$ and $\Aut_2$, return an NSA $\Aut'$ such that $\Lang(\Aut') = \Lang(\Aut_1) \cap
               \Lang(\Aut_2)$.
      \item[Complementation:] For an NSA $\Aut$, return an NSA $\Aut'$ such that $\Lang(\Aut') = [X \to M]^* \setminus \Lang(\Aut)$.
   \end{description}
\end{restatable}
In~\cite{DantoniV21}, effectiveness also requires the Boolean operations on labels to be computable; for formulas, these operations are trivial.
Moreover, in the case of formulas, effectiveness is only used to remove unsatisfiable transitions.
Without it, the algorithms of \Cref{fact:sfa-closure} can still be run and remain correct,
since they only manipulate labels syntactically.

\subsection{Number encodings}\label{subsec:encodings}

To encode integers as words that can be read by a (symbolic) automaton, we adopt the encoding
of~\cite{Boigelot21}: a \emph{most-significant-digit-first representation in $k$'s complement (in short, $k$-msdf)}, for
$k\in\N_{\geq 2}$. As explained in~\Cref{sec:introduction}, in this representation words belong to the regular language
$\CEk \coloneqq \{0,k-1\} \cdot \{0,1,\dots,k-1\}^*$. The leftmost (most-significant) digit, constrained to $0$ or
$k-1$, is the \emph{sign digit}.

For $\ell \geq 0$, the word $\vec w \coloneqq a_{\ell}\,a_{\ell-1}\dots a_0 \in \CEk$ represents the integer
\[
   \intrep{\vec w}_k \coloneqq
   \begin{cases}
      \hphantom{\ -k^{\ell} + \:\:} \textstyle\sum_{i=0}^{\ell-1} a_i\cdot k^{i} & \text{if $a_\ell = 0$,}   \\
      \ -k^{\ell} + \textstyle\sum_{i=0}^{\ell-1} a_i\cdot k^{i}                 & \text{if $a_\ell = k-1$.}
   \end{cases}
\]

A key property of this representation is that padding words by their sign digits does not change the encoded number,
since $-k^\ell = -k^{\ell +1} + (k-1)\cdot k^{\ell}$:

\begin{fact}\label{fact:msdf-padding}
   For every $\vec w \coloneqq a_{\ell}\,a_{\ell-1}\dots a_0 \in \CEk$,
   we have $\intrep{a_\ell\,\vec w}_k = \intrep{\vec w}_k$.
\end{fact}
This makes it simple to encode vectors of integers. Consider words $\vec w_1,\dots,\vec w_n$ from
$\CEk$, with $\vec w_i \coloneqq a_{\ell_i}^{(i)} \dots a_{0}^{(i)}$, and let $m \coloneqq \max(\ell_1,\dots,\ell_n)$. Then, the \emph{convolution} $\vec w_1 \otimes \dots \otimes \vec w_n
   \coloneqq c_m \dots c_0$, where for every $i \in [0..m]$ the symbol $c_i\in \CEk^n$ is defined as $c_i \coloneqq
   (a_{\min(\ell_1,i)}^{(1)},\dots,a_{\min(\ell_n,i)}^{(n)})$, encodes the $n$-tuple of integers $\intrep{\vec w_1 \otimes
      \dots \otimes \vec w_n}_k \coloneqq (\intrep{w_1},\dots,\intrep{w_n})$.

A second property concerns the computation of the additive inverse $-n$ of a non-zero integer~$n$. Directly from the
definition of $\intrep{ \_ }_k$, one can verify that:
\begin{fact}\label{fact:msdf-inverses}
   If $\vec w \coloneqq a_\ell \dots a_0 \in \CEk \setminus \{0\}^*$, then $-\intrep{\vec w}_k = \intrep{(k-1 -a_\ell) (k- 1 - a_{\ell-1}) \dots (k- 1 - a_0)}_k + 1$.
\end{fact}
\noindent
For example, in
the decimal encoding we have $-\intrep{0700}_{10} = -700 = -1000 + 299 + 1 = \intrep{9299}_{10} + 1$.

We will soon need to consider words whose letters are maps from variables to $[0..k-1]$, rather than elements in $\CEk$.
Formally, we define the set $\Mk \coloneqq [X \to \{0,k-1\}] \cdot [X \to [0..k-1]]^*$, i.e., the set of non-empty words $\vec w \coloneqq w_\ell \dots w_0$ over the alphabet $[X \to [0..k-1]]$, such that ${w_\ell \in [X \to \{0,k-1\}]}$. As done for words in~$\CEk$, we will refer to $w_\ell$ as the \emph{sign digit}.
Note that, for every $x \in X$,
the word $\vec w(x) \coloneqq w_\ell(x) \dots w_0(x)$ belongs to $\CEk$.
Then, we can transform any word $\vec w \in \Mk$ into a map
$\intrep{\vec w}_k \colon X \to \Z$ by applying $\intrep{\cdot}_k$ variable-wise,
defining $\intrep{\vec w}_k(x) \coloneqq \intrep{w_\ell(x)\dots w_0(x)}_k$
for every $x \in X$.
In fact, for every $k \in \N_{\geq 2}$, every map $\mu \colon X \to \Z$ and every finite set $Y \subseteq X$,
there is $\vec w \in \Mk$ such that $\intrep{\vec w}_k(y) = \mu(y)$ for every $y \in Y$.

\section{A Parametric B\"uchi-Bruy\`ere theorem}\label{sec:parametric-BB}

In this section, we establish~\Cref{theorem:bb-par}.
We first discuss the semantics of symbolic automata used
in this theorem (\Cref{subsec:msdf-automata}),
to then prove the right-to-left direction in \Cref{subsec:from-formulas-to-automata}.
The left-to-right direction is proved in \Cref{subsec:from-automata-to-formulas}.
Related results are given in~\Cref{subsec:corollaries}.
For simplicity of the exposition, the translation into symbolic automata provided in~\Cref{subsec:from-formulas-to-automata} is the simplest possible rather than the most efficient one.
We leave the discussion of optimizations to~\Cref{sec:tool}, where we also discuss the implementation of the translation and an experimental evaluation.

\subsection{Parametric symbolic automata}\label{subsec:msdf-automata}

Consider the class~$\AutPAt$ of all non-deterministic symbolic automata with labels that are formulas from Presburger arithmetic with digits.
We will now define a map
\[
    \LangMSD \colon \AutPAt \to (\N_{\geq 2} \to \powerset[X \to \Z])
\]
giving these automata semantics akin to that of parametric B\"uchi arithmetic.
Roughly speaking, $\LangMSD$ will associate to $\Aut\in \AutPAt$ a function that,
for a given base $k \in \N_{\geq 2}$, returns the set of assignments
$X \to \Z$ whose $k$-msdf encodings are accepted by the symbolic automaton
obtained when replacing $t$ by~$k$. That is, through the semantics $\LangMSD$ a single automaton represents
an infinite \emph{family} of symbolic automata.
When considering this semantics, we will call an automaton from~$\AutPAt$ a~\emph{parametric symbolic automaton}.

\begin{example}\label{ex:msdf-automaton}
    Consider the parametric symbolic automaton~$\Aut$:
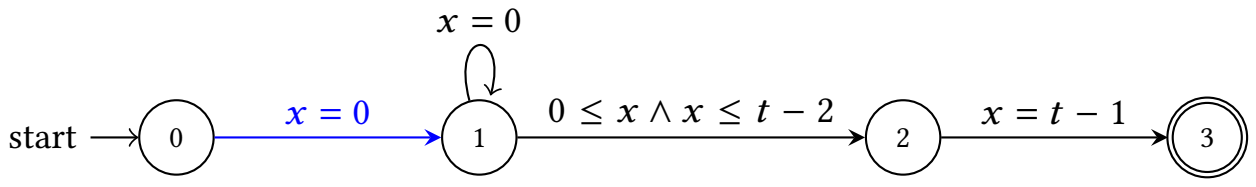
    \begin{center}
        \vspace{-0.2em}
        \begin{tikzpicture}[node distance = 2cm]
            \scriptsize
            \node (0) [state, initial left, scale=0.8] {$0$};
            \node (1) [state, right = 1.5cm of 0, scale=0.8] {$1$};
            \node (2) [state, right = 2.3cm of 1, scale=0.8] {$2$};
            \node (3) [state, accepting, right = 1.5cm of 2, scale=0.8] {$3$};
            \path[-stealth]
            (0) edge[left=0.1, blue] node[above]{$x=0$} (1)
            (1) edge[loop above] node[above]{$x=0$} (1)
            (1) edge[left=0.1] node[above]{$0 \leq x \land x \leq t-2$} (2)
            (2) edge[left=0.1] node[above]{$x = t-1$} (3);
        \end{tikzpicture}%
        \vspace{-0.2em}
    \end{center}
    where missing transitions are implicitly labeled with $\bot$.
    For example, substituting $t$ by $3$ in the transitions,
    we see that the words accepted by the resulting symbolic automaton are of the form $w_{\ell}\dots w_0 \in [X \to \Z]^*$
    where $w_i(x) = 0$ for every $i \in [2..\ell]$, $w_1(x) \in [0..1]$ and $w_0(x) = 2$.
    Observe that $w_{\ell}(x)\dots w_0(x)$ represents an integer in $3$-msdf;
    it is equal to $2$ when $w_1(x) = 0$, or to $5$ when $w_1(x) = 1$.
    Then, for every assignment $m \in  \LangMSD(\Aut)(3)$, we will have that $m(x) = 2$ or $m(x) = 5$.
    More generally, $\LangMSD(\Aut)$ will be defined as the map that, given a $k \in \N_{\geq 2}$, returns the set of all assignments in $[X \to \Z]$ mapping
    $x$ to an integer of the form $a\cdot k + (k-1)$, where $a \in [0..k-2]$.
\end{example}

\pagebreak
\begin{example}\label{ex:example-v-1mod3-par}
    Consider the parametric symbolic automaton~$\mathcal{B}$:
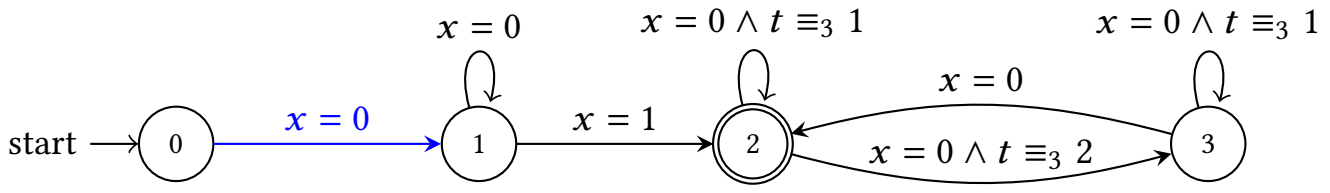
    \begin{center}
        \vspace{-0.2em}
        \hspace{1.7em}
        \begin{tikzpicture}[node distance = 2cm]
            \scriptsize
            \node (0) [state, initial left, scale=0.8] {$0$};
            \node (1) [state, right= 1.5cm of 0, scale=0.8] {$1$};
            \node (2) [state, accepting, right= 1.3cm of 1, scale=0.8] {$2$};
            \node (3) [state, right= 2.5cm of 2, scale=0.8] {$3$};
            \path[-stealth]
            (0) edge[above, blue] node{$x = 0$} (1)
            (1) edge[loop above] node{$x = 0$} (1)
            (1) edge[above] node{$x = 1$} (2)
            (2) edge[loop above] node{$x = 0\land t\equiv_3 1$} (2)
            (2) edge[above, bend right=15] node{$x = 0\land t\equiv_3 2$} (3)
            (3) edge[above, bend right=15] node{$x = 0$} (2)
            (3) edge[loop above] node{$x = 0\land t\equiv_3 1$} (3);
        \end{tikzpicture}
        \vspace{-0.2em}
    \end{center}
    where for $r\in\{0,1,2\}$ the congruence
    $t\equiv_3 r$ is a shorthand for $\exists y\,(t=3\cdot y + r)$.
    This example illustrates a peculiarity of parametric symbolic automata compared to
    symbolic automata under the standard semantics: constraints on the parameter $t$ propagate
    through a path. Taking the transition $2 \to 3$ requires $t \equiv_3 2$,
    which disables the self-loop on state~$3$. Similarly,
    the transition $3 \to 2$ is implicitly only taken
    when $t \equiv_3 2$, despite its label not mentioning $t$.
    The automaton $\mathcal{B}$ defines a map from $k\in\N_{\geq2}$ to the set of all assignments in $[X \to \Z]$ that encodes, following~\Cref{remark:semantic-differences}, the set $\{1\}$ if $k\equiv_3 0$,
    the set $\{k^{n} : n\in\N\}$ if $k\equiv_3 1$, and the set $\{k^{2n} : n\in\N\}$ if $k\equiv_3 2$.
    For instance, $m(x)$ is a power of $4$ for every $m \in  \LangMSD(\mathcal{B})(2)$,
    and it is a power of $10$ for $m \in  \LangMSD(\mathcal{B})(10)$.
\end{example}

To formally define~\LangMSD, start by considering an automaton $\Aut \coloneq (\Phi, Q, q_0, F, \delta)$ from $\AutPAt$, and some integer $k \geq 2$.
By definition, the language $\Lang_{\kPAmodel_k}(\Aut)$, where $\kPAmodel_k$ is the structure of Presburger arithmetic with digits~from~\Cref{subsec:PrA}, is a set of words $w_1\dots w_n$ where each letter $w_i$ is a map from $X$ to $\Z$.
Let us define $\Lang_k(\Aut) \coloneqq \Lang_{\kPAmodel_k}(\Aut) \cap \Mk$ with $\Mk$ from \Cref{subsec:encodings}.
Intuitively, these are the (non-empty) words that respect the constraints on the digits imposed by the msdf representation of integers in $k$'s
complement.
We call these the \emph{words accepted by $\Aut$ in $k$-msdf}.
We remark that the empty word is never in $\Lang_k(\Aut)$, no matter whether $q_0$ belongs to $F$ or not.

As explained in~\Cref{subsec:encodings}, every word accepted by $\Aut$ in $k$-msdf can be seen as a map from $X$ to $\Z$, by applying the \mbox{$k$-msdf}
representation of integers variable-wise:
for $\vec w \coloneqq w_\ell\dots w_0\in\Lang_k(\Aut)$,
this is just the map $\intrep{\vec w}_k \colon X \to \Z$.
Similarly to a formula, which only constrains its free variables, $\Aut$ should only constrain the variables occurring in its labels.
Let $X_\Aut$ be the (finite) set of variables occurring free in the labels $\delta(q,q')$, for $q,q' \in Q$.
We say that a word $\vec w \in \Mk$ \emph{encodes} a map $\mu \colon X \to \Z$, or that $\vec w$ is an \emph{encoding} of~$\mu$, whenever $\intrep{\vec w}_k(x) = \mu(x)$ for every $x \in X_\Aut$.
Finally,~$\LangMSD(\Aut)$ is defined as the map that for $k \in \N_{\geq 2}$ returns the set $\LangMSD(\Aut)(k)$ of all maps encoded by some word in~$\Lang_k(\Aut)$.
We restate~\Cref{theorem:bb-par} using the introduced notation.

\begin{thmstar}[\ref{theorem:bb-par}]
    The set of formulas of parametric B\"uchi arithmetic and the set of parametric symbolic automata represent the same class of functions from $[\N_{\geq 2} \to \powerset[X \to \Z]]$.
    That is, we have the equality $\{ \LangMSD(\Aut) : \Aut \in \AutPAt \} = \{ \sem{\phi} : \phi \in \Formulas_\PBA \}$.
\end{thmstar}

To prove this theorem, we are going to show that for every $\Aut \in \AutPAt$ there is $\phi \in \Formulas_\PBA$ such that
$\LangMSD(\Aut) = \sem{\phi}$, and vice versa.
For example, for the automaton $\mathcal{B}$ from \Cref{ex:example-v-1mod3-par}
we have the equality $\LangMSD(\mathcal{B}) = \sem{V_t(x,x)\land \exists y : x = 3\cdot y + 1}$.
Both directions of the proof are constructive: they yield algorithms to perform the translations.

\subsection{From parametric B\"uchi arithmetic to symbolic automata}\label{subsec:from-formulas-to-automata}

For this direction, we follow the classical proof~\cite[Theorem~6.1]{BruyereHMV94}.
However,
before translating parametric B\"uchi arithmetic formulas into parametric symbolic automata, we must address an encoding issue.
Observe that the automaton from~\Cref{ex:msdf-automaton} accepts, for any given $k \in \N_{\geq 2}$,
the word $w_2w_1w_0$ with $w_2(x) = 0$, $w_1(x) = 0$ and $w_0(x) = k-1$,
while it rejects the word $u_1u_0$ with $u_1(x) = 0$ and $u_0(x) = k-1$.
In both words, $x$ encodes the same integer, since $\intrep{00(k-1)}_k = \intrep{0(k-1)}_k$ by~\Cref{fact:msdf-padding}. This is problematic: consider the automaton $\mathcal{C}$ obtained from $\Aut$
by removing state~$(1)$ and adding a transition from state $(0)$ to state $(2)$ with label $x = 0$:
\begin{center}
    \vspace{-0.2em}
    \begin{tikzpicture}[node distance = 2cm]
        \scriptsize
        \node (0) [state, initial left, scale=0.8] {$0$};
        \node (2) [state, right = 1.5cm of 0, scale=0.8] {$2$};
        \node (3) [state, accepting, right = 1.5cm of 2, scale=0.8] {$3$};
        \path[-stealth]
        (0) edge[left=0.1, blue] node[above]{$x=0$} (2)
        (2) edge[left=0.1] node[above]{$x = t-1$} (3);
    \end{tikzpicture}%
    \vspace{-0.2em}
\end{center}
While the intersection of $\LangMSD(\Aut)(k)$ and $\LangMSD(\mathcal{C})(k)$ contains the map sending $x$ to $k-1$,
the algorithm for intersecting symbolic automata (from~\Cref{fact:sfa-closure}; it just performs the Cartesian product construction)
yields, when applied to
$\Aut$ and $\mathcal{C}$, an automaton~$\mathcal{D}$
with $\Lang_{\kPAmodel_k}(\mathcal{D}) = \emptyset$:
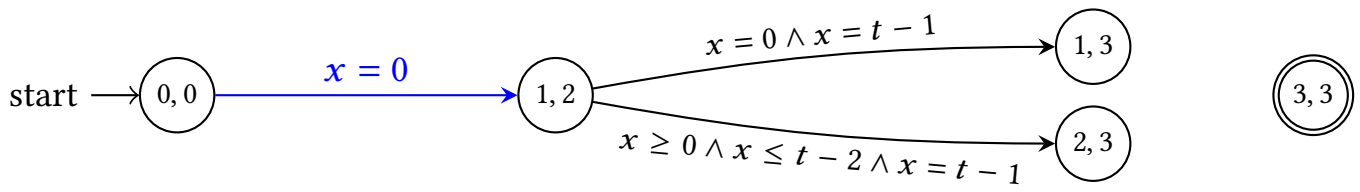
\begin{center}
    \vspace{-0.2em}
    \begin{tikzpicture}[node distance = 2cm]
        \scriptsize
        \node (00) [state, initial left, scale=0.8]                  {$0,0$};
        \node (12) [state, right=of 00, scale=0.8]                  {$1,2$};
        \node (13) [state, above right = -1pt and 3.2cm of 12, scale=0.8]            {$1,3$};
        \node (23) [state, below right = -1pt and 3.2cm of 12, scale=0.8]            {$2,3$};
        \node (33) [state,accepting, right= 4.5cm of 12, scale=0.8]        {$3,3$};
        \path[-stealth]
        (00) edge[blue] node[above]{$x=0$} (12)
        (12) edge[bend left=5] node[sloped,above,scale=0.85]{$x=0 \land x=t-1$} (13)
        (12) edge[bend right=5] node[sloped,below,scale=0.85]{$x \ge 0 \land x \le t-2 \land x=t-1$} (23);
    \end{tikzpicture}
    \vspace{-0.2em}
\end{center}

For the (standard) finite automata for B\"uchi arithmetic on a fixed base, this issue is resolved
by requiring that if a word is accepted, then \emph{all} its paddings by repeating the sign digits are also accepted~\cite{Boigelot21}.
We adapt this notion to symbolic automata as follows:

\begin{definition}[Valid NSA]\label{def:valid-nfa}
    An NSA $\Aut \in \AutPAt$ is \emph{valid} if it satisfies the following property for every $k \in \N_{\geq 2}$:
    for every $\mu \in \LangMSD(\Aut)(k)$
    and word $\vec w \in \Mk$,
    if $\vec w$ encodes $\mu$, then $\vec w \in \Lang_k(\Aut)$.
\end{definition}

\begin{example}\label{example:unvalid-nfa}
    Consider the parametric symbolic automaton
    
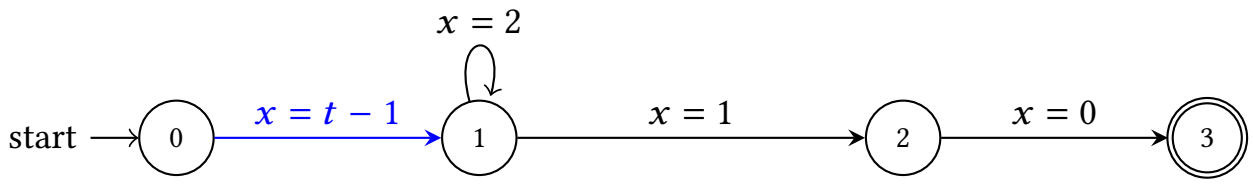
\begin{center}
        \vspace{-0.2em}
        \begin{tikzpicture}[node distance = 2cm]
            \scriptsize
            \node (0) [state, initial left, scale=0.8] {$0$};
            \node (1) [state, right = 1.5cm of 0, scale=0.8] {$1$};
            \node (2) [state, right = 2.3cm of 1, scale=0.8] {$2$};
            \node (3) [state, accepting, right = 1.5cm of 2, scale=0.8] {$3$};
            \path[-stealth]
            (0) edge[left=0.1, blue] node[above]{$x=t-1$} (1)
            (1) edge[loop above] node[above]{$x=2 $} (1)
            (1) edge[left=0.1] node[above]{$x = 1$} (2)
            (2) edge[left=0.1] node[above]{$x = 0$} (3);
        \end{tikzpicture}%
        \vspace{-0.2em}
    \end{center}
    This automaton does not satisfy the property in~\Cref{def:valid-nfa}
    for $k = 2$ nor for $k \geq 4$, and it is thus not a valid NSA.
    Indeed, note that for $k = 2$ the word $110$ is accepted, but not the words $1110$
    or $10$, which encode the same integer. Similarly, for $k \geq 4$, the word $(k-1)210$ is accepted, but not the word $(k-1)(k-1)210$.
    However, the required property is satisfied for $k = 3$:
    in this case we have $\Lang_3(\Aut) = \{ 2^i10 : i \in \N_{\geq 1} \}$,
    which is closed under padding by repeating the sign digit $2$.
\end{example}

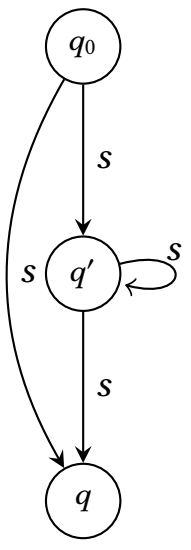
\begin{wrapfigure}{r}{0.09\textwidth}
    \vspace{-13pt}
    \begin{tikzpicture}[node distance = 2cm]
        \scriptsize
        \node (0) [state, scale=0.8] {$q_0$};
        \node (1) [state, below = 1cm of 0, scale=0.8] {$q'$};
        \node (2) [state, below = 1cm of 1, scale=0.8] {$q$};
        \path[-stealth]
        (0) edge[left=0.1] node[right]{$s$} (1)
        (1) edge[loop right] node[above]{$s$} (1)
        (1) edge[left=0.1] node[right]{$s$} (2);
        \path[-stealth]
        (0) edge[bend right] node[right]{$s$} (2);
    \end{tikzpicture}
\end{wrapfigure}
In~\cite{Boigelot21}, finite automata for B\"uchi arithmetic are made valid by a simple
procedure that ``completes'' the automaton by adding transitions to pad accepted words by repeating the sign digits.
This procedure considers any state $q$ reachable from the initial state $q_0$ by reading a word in the language $s^+$,
where $s$ is a tuple of sign digits, and adds to the automaton the gadget on the right,
where $q'$ is a fresh state. In the completed automaton, $q$ is guaranteed to be reachable from $q_0$ by any word in $s^+$, instead of just some of them.
A similar procedure also works for parametric symbolic automata, but we must take into account that, as shown in~\Cref{example:unvalid-nfa}, the property
in~\Cref{def:valid-nfa} can be satisfied in some bases but not in others. The pseudocode is given in~\Cref{algo:nsa-validity}.

\begin{algorithm}[t]
    \begin{algorithmic}[1]
        \Require An NSA $\Aut = (\Phi,Q,q_0,F,\delta)$ from $\AutPAt$
        \Ensure A valid NSA $\Aut' \in \AutPAt$ such that $\LangMSD(\Aut') = \LangMSD(\Aut)$
        \setstretch{1.05}
        \State\textbf{let} $\{x_1,\dots,x_m\} \coloneqq X_\Aut$, i.e., the set of free variables occurring in the labels of~$\Aut$
        \State\textbf{let} $Q' \coloneqq \{q_0'\}$ where $q_0'$ is a fresh state, and set $\delta(q_0',q) \coloneqq \bot$ for every $q \in Q$
        \For{every map~$\sigma \colon \{x_1,\dots,x_m\} \to \{0,t-1\}$}\label{line:sign-maps}
        \For{every path $q_0q_1\dots q_n \in Q^*$ such that $n \geq 1$ and $q_1,\dots,q_n$ are pairwise distinct}\label{line:paths}
        \State\textbf{let} $q''$ be a fresh state, and $\psi$ be the formula $(\bigwedge_{i=1}^m x_i = \sigma(x_i)) \land \bigwedge_{i=0}^{n-1}\delta(q_i, q_{i+1})$\label{line:fresh-state-q}
        \State Set $\delta(q_0',q_n) \coloneqq \delta(q_0',q_n) \lor \psi$
        \State Add $q''$ to $Q'$ and extend $\delta$ with $(q_0', q'') \mapsto \psi$, $(q'',q'') \mapsto \psi$ and $(q'', q_n) \mapsto \psi$\label{line:new-transition-validity}
        \EndFor
        \EndFor
        \State Complete $\delta$ on the domain $(Q \cup Q') \times (Q \cup Q')$ by mapping missing pairs to $\bot$
        \State\textbf{return} $(\Img(\delta), Q \cup Q', q_0', F, \delta)$
    \end{algorithmic}
    \caption{Makes a parametric symbolic automaton valid.}\label{algo:nsa-validity}
\end{algorithm}

\pagebreak
\begin{restatable}{lemma}{LemmaMakeNSAValid}\label{lemma:make-NSA-valid}
    \Cref{algo:nsa-validity} respects its specification.
\end{restatable}

\begin{proof}[Proof idea]
    \vspace{-0.2em}
    Let us informally discuss~\Cref{algo:nsa-validity}.
    Lines~\ref{line:sign-maps} and~\ref{line:paths} of the algorithm enumerate all possible maps of
    sign digits, and all possible paths of the automaton (disregarding loops, except those involving the initial state~$q_0$).
    Consider, for example, the map $x \mapsto t-1$ and the path $0 \to 1 \to 2$ in the automaton from~\Cref{example:unvalid-nfa}. Suppose we were to add a gadget similar to that of~\cite{Boigelot21},
    just with the labels ``$s$'' replaced by $x = t-1$.
    This would fix the automaton in~\Cref{example:unvalid-nfa} for $k=2$ (now $\Lang_2(\Aut) = \{ 1^i0 : i \in \N_{\geq 1} \}$), but inadvertently changes $\Lang_3(\Aut)$ to also include words from $2^+0$. Instead,~\Cref{algo:nsa-validity} gathers all formulas along the considered path, and strengthens the labels of the gadget with their conjunction. In our example, the path $0 \to 1 \to 2$ carries the labels $x=t-1$ (redundant) and $x=1$, so the gadget's transitions are instead labeled with $x = t-1 \land x = 1$. This formula is only satisfiable for $t = 2$, and thus the gadget does not change the language of the automaton in other bases.
\end{proof}

We can thus always assume that parametric symbolic automata are valid.
On these automata, the basic algorithms for symbolic automata work as expected with respect to the semantics of~$\LangMSD$

\begin{restatable}{lemma}{LemmaValidSFAClosure}\label{lemma:valid-sfa-closure}
    There are algorithms satisfying the following specifications for automata in $\AutPAt$:
    \begin{description}
        \item[Determinization:] For a valid NSA $\Aut$, returns a valid DSA $\Aut'$ such that $\LangMSD(\Aut') = \LangMSD(\Aut)$.
        \item[Intersection:] For valid NSAs $\Aut_1$ and $\Aut_2$, return a valid NSA $\Aut'$ such that for every $k \in \N_{\geq 2}$, $\LangMSD(\Aut')(k) = \LangMSD(\Aut_1)(k) \cap
                  \LangMSD(\Aut_2)(k)$.
        \item[Complementation:] For a valid NSA $\Aut$, returns a valid NSA $\Aut'$ such that for every $k \in \N_{\geq 2}$, $\LangMSD(\Aut')(k) = [X \to \Z] \setminus \LangMSD(\Aut)(k)$.
    \end{description}
\end{restatable}

To translate parametric B\"uchi arithmetic formulas into parametric symbolic automata, it now suffices to show closure under variable projection, and to provide translations for the atomic formulas.
Let us start with variable projection, which corresponds to existential quantification over a variable \( x\in X \).
Note that the automaton cannot enforce any relationship between different digits of a number (besides those encoded via the states of the automaton).
It therefore suffices to quantify each digit of \( x \) separately.
We define ${\exists_x \colon \AutPAt \to \AutPAt}$ as the map guarding every formula in the input NSA with the bounded quantifier $(\exists x < t)$.
On the first letter of a word, the quantified digit must also be a sign digit; to enforce this, we add a fresh initial state.
Formally, given NSA $\Aut = (\Phi, Q, q_0, F, \delta)$, we have $\exists_x (\Aut) \coloneqq (\Img(\delta'), Q \cup \{q_\star\}, q_\star, F, \delta')$, where $q_\star \notin Q$ is a fresh state,
$\delta'(q,q') \coloneqq (\exists x < t)\,\delta(q,q')$ and $\delta'(q_\star,q') \coloneqq (\exists x < t)\,\big((x = 0 \lor x = t-1) \land \delta(q_0,q')\big)$ for every $q,q' \in Q$,
$\delta'(q,q_\star) \coloneqq \bot$ for every $q \in Q \cup \{q_\star\}$.
This map ensures closure under projection:

\begin{restatable}{lemma}{LemmaProjectionNSA}\label{lemma:projection-nsa}
    Consider $\Aut \in \AutPAt$ and $x \in X$. Then, for every $k \in \N_{\geq 2}$ and $m \in [X \to \Z]$, $ m \in \LangMSD(\exists_x(\Aut))(k)$ if and only if there is $j \in \Z$
    such that $m[x \mapsto j] \in \LangMSD(\Aut)(k)$.
\end{restatable}

Note that the output of $\exists_x$ need not be a valid parametric symbolic automaton.
Indeed, it only accepts words whose length is also the length of some encoding of the projected variable~\( x \).
As a result, short encodings of solutions may be rejected.
Since this is problematic for subsequent Boolean operations,
we apply~\Cref{algo:nsa-validity} after each variable projection.

Thanks to closure under projection, translating arbitrary atomic formulas reduces to handling the following base cases: $x = 0$, $x = 1$, $x \leq y$, $V_t(x,y)$ and $x + y = z$
(arbitrary atomic formulas can be expressed with existential formulas over these base cases).
We have already encountered the parametric symbolic automaton for $x + y = z$: it is the one obtained from~\Cref{fig:example-adder-par}
by replacing all occurrences of $k$ by the parameter $t$.
Here are the remaining base cases:

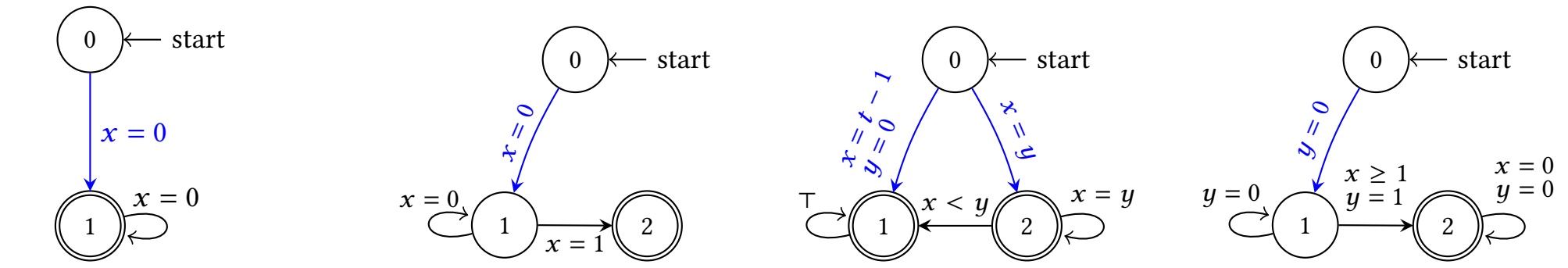
\begin{figure}[H]
    \vspace{-0.9em}
    \begin{subfigure}{.25\textwidth}
        \centering
        \begin{tikzpicture}[node distance = 1cm]
            \scriptsize
            \node (0) [state, initial right, scale=0.9] {$0$};
            \node (1) [state, accepting, below = 1cm of 0, scale=0.9] {$1$};
            \path[-stealth]
            (0) edge[blue, left=0.1] node[right]{$x = 0$} (1)
            (1) edge[loop right] node[above=2pt]{$x = 0$} (1);
        \end{tikzpicture}%
        \caption{NSA for $x = 0$.}
        \label{fig:nsa-zero}
    \end{subfigure}%
    \begin{subfigure}{.25\textwidth}
        \centering
        \begin{tikzpicture}[node distance = 1cm]
            \scriptsize
            \node (0) [state, scale=0.9, initial right] {$0$};
            \node (1) [state, below left = 1cm and 0.2cm of 0, scale=0.9] {$1$};
            \node (2) [state, accepting, below right = 1cm and 0.2cm of 0, scale=0.9] {$2$};
            \path[-stealth]
            (0) edge[blue, left=0.1, bend right=7] node[midway,sloped,above,scale=0.9]{$x = 0$} (1)
            (1) edge[loop left] node[above=2pt, scale=0.9]{$x = 0$} (1)
            (1) edge[left=0.1] node[below, scale=0.9]{$x = 1$} (2);
        \end{tikzpicture}
        \caption{NSA for $x = 1$.}
        \label{fig:nsa-one}
    \end{subfigure}%
    \begin{subfigure}{.25\textwidth}
        \centering
        \begin{tikzpicture}[node distance = 1cm]
            \scriptsize
            \node (0) [state, scale=0.9, initial right] {$0$};
            \node (1) [state, accepting, below left = 1cm and 0.2cm of 0, scale=0.9] {$1$};
            \node (2) [state, accepting, below right = 1cm and 0.2cm of 0, scale=0.9] {$2$};
            \path[-stealth]
            (0) edge[blue, left=0.1, bend right=7] node[midway,sloped,above,scale=0.9]{$\begin{aligned} x &= t - 1\\[-5pt] y &= 0\end{aligned}$} (1)
            (0) edge[blue, left=0.1, bend left=7] node[midway,sloped,above,scale=0.9]{$x = y$} (2)
            (1) edge[loop left] node[above=2pt, scale=0.9]{$\top$} (1)
            (2) edge node[above, scale=0.9]{$x < y$} (1)
            (2) edge[loop right] node[above=2pt, scale=0.9]{$x = y$} (2);
        \end{tikzpicture}
        \caption{NSA for $x \leq y$.}
        \label{fig:nsa-leq}
    \end{subfigure}%
    \begin{subfigure}{.25\textwidth}
        \centering
        \begin{tikzpicture}[node distance = 1cm]
            \scriptsize
            \node (0) [state, scale=0.9, initial right] {$0$};
            \node (1) [state, below left = 1cm and 0.2cm of 0, scale=0.9] {$1$};
            \node (2) [state, accepting, below right = 1cm and 0.2cm of 0, scale=0.9] {$2$};
            \path[-stealth]
            (0) edge[blue, left=0.1, bend right=7] node[midway,sloped,above,scale=0.9]{$y = 0$} (1)
            (1) edge[loop left] node[above=2pt, scale=0.9]{$y = 0$} (1)
            (1) edge[left=0.1] node[above, scale=0.9]{$\begin{aligned} x &\geq 1\\[-5pt] y &= 1\end{aligned}$} (2)
            (2) edge[loop right] node[above=2pt, scale=0.9]{$\begin{aligned} x &= 0\\[-5pt] y &= 0\end{aligned}$} (2);
        \end{tikzpicture}
        \caption{NSA for $V_t(x,y)$.}
        \label{fig:nsa-Vt}
    \end{subfigure}
\end{figure}
\vspace{-1em}

It is routine to verify that all the NSAs for the base cases are correct. The right-to-left direction of~\Cref{theorem:bb-par} then follows by structural induction on $\phi \in \Formulas_\PBA$, using~\Cref{lemma:valid-sfa-closure,lemma:projection-nsa,lemma:make-NSA-valid}.

\begin{restatable}{lemma}{LemmaFromFormulasToAutomata}\label{lemma:from-formulas-to-automata}
    For every formula $\phi$ in $\Formulas_\PBA$
    there is an NSA $\Aut \in \AutPAt$
    with~$\LangMSD(\Aut) = \sem{\phi}$.
\end{restatable}

\begin{remark}
  B\"uchi arithmetic is~\tower-complete~\cite{Buechi1960,Meyer75}:
  it captures all problems solvable in time bounded by a tower of exponentials
  of elementary height in the size of the input~\cite{Schmitz16}.
  Since, for a fixed base, the emptiness of an automaton in~$\AutPAt$ can be checked in elementary time,
  no translation from formulas to automata produces automata of elementary size.
  The translation of~\Cref{lemma:from-formulas-to-automata} matches this lower bound.
  Each complementation (done via determinization)
  and each projection (followed by~\Cref{algo:nsa-validity})
  adds at most one exponential to the size of the automaton.
  So, the automaton for a formula~$\phi$ has size bounded by a tower of exponentials
  of height linear in the size of~$\phi$.
\end{remark}

\subsection{Arithmetization of parametric symbolic automata}\label{subsec:from-automata-to-formulas}

We now prove the other direction of~\Cref{theorem:bb-par}: computations of parametric symbolic automata can be represented by formulas of parametric B\"uchi arithmetic.
In B\"uchi-Bruy\`ere theorem~\cite{Villemaire92,BruyereHMV94}, the analogous statement is obtained via a translation yielding a formula $\phi\in\Formulas_{\kBA}$ with a bounded number of quantifier alternations.
A slight modification shows that $\phi$ can be taken
from the $\exists^{*}\forall^{*}$ fragment of $\Formulas_{\kBA}$~\cite{HaaseR2021}.

As in~\cite{BruyereHMV94}, we encode a run of the automaton using one variable per state,
whose digit is~$1$ exactly at the positions where the run visits that state.
A universal quantification over positions then checks that consecutive states agree with the transition function.
Adapting this encoding to the parametric setting raises two challenges that lead to deviations from the original proof.
The first challenge concerns labels.
When the base is fixed, labels come from a finite set, hence Boolean connectives can be used to enumerate the digits that occur at a certain position of a number.
In the parametric setting, the alphabet $[0..t-1]$ is not fixed, so enumeration is impossible.
Moreover, the digit~$d$ at a given position $i$ cannot be extracted directly,
only the monomial $d \cdot t^i$ can.
Our solution is to ``shift the labels'': a label~$\phi(x_1,\dots,x_n)$ is not evaluated on the digits $d_1,\dots,d_n$ read at position~$i$, but, after a suitable transformation, on the monomials $d_1 \cdot t^i, \dots, d_n \cdot t^i$.
\Cref{lemma:shifted-labels} (which has no counterpart in~\cite{BruyereHMV94}) makes this precise.

The second challenge is comparing digits from consecutive positions.
For a fixed base $k$, this can be easily achieved thanks to the
equation $y = k \cdot x$. In the parametric case, $y = t \cdot x$ is however not definable, as we will see in~\Cref{sec:multiplication-effects}.
We instead rely on the following formula $S_t(x,y)$ that
holds whenever $x$ and $y$ are subsequent powers of $t$:
\[
  S_t(x,y) \coloneqq P_t(x) \land P_t(y) \land x < y \land \forall z \,\bigl(P_t(z) \implies z \leq x \lor y \leq z\bigr).
\]
Above, $P_t(x) \coloneqq V_t(x, x)$ characterises the powers of $t$.
Observe that $S_t(x,y)$ is universal;
this will ultimately limit our translation to the $\exists^{*}\forall^{*}\exists^*$ fragment  instead of $\exists^{*}\forall^{*}$.

To start with the reduction, let us define a predicate $\mathrm{digit}_t(x, p, p', y)$ with the following semantics:
for every $k \in \N_{\geq 2}$ and $\vec w \coloneqq w_\ell \dots w_0 \in \Mk$
such that $\intrep{\vec w}_k(p) = k^i$ and $\intrep{\vec w}_k(p') = k^{i+1}$ for some $i \in \N$,
we have
$\intrep{\vec w}_k \models \mathrm{digit}_t(x, p, p', y)$
if and only if $\intrep{\vec w}_k(y) = d \cdot k^i$ and $w_{\min(\ell,i)}(x) = d$
for some $d \in [0..k-1]$.
In other words, $\mathrm{digit}_t(x, p, p', y)$ is true whenever $y = d \cdot t^i$
(shorthand for ``the integer assigned to $y$ is $d \cdot t^i$'')
and $d$ is the $i$-th digit of $x$ (assuming $p = t^i$ and $p' = t^{i+1}$).
This is, in some sense, a ``parametric generalization'' of the predicate~${\in_{d, k}(x, y)}$  from~\cite{BruyereHMV94}.

\begin{restatable}{lemma}{LemmaDigitT}\label{lemma:digit-t}
    The predicate \( \mathrm{digit}_t(x, p, p', y) \) is definable in both \( \exists^{*} \) and \( \forall^{*} \) fragments of \( \Formulas_\PBA \).
\end{restatable}
\begin{proof}
    For simplicity, we start by considering non-negative integers only.
    Assuming $p = t^i$ and $p' = t^{i+1}$,
    the following quantifier-free formula $\mathrm{split}_t(x,p,p',p'',x_1,x_2,x_3)$
    decomposes $x$ into three parts $x_1$, $x_2$ and $x_3$,
    where $x_1$ contains
    digits less significant than $i$, $x_3$ contains digits more significant than $i$,
    and $x_2$ is exactly the ``monomial'' $d \cdot t^i$ we are after.
    The auxiliary variable $p''$ is a large power of $t$ used to capture the digits more significant than $i$.
    \begin{align*}
        \mathrm{split}_t(x,p,p',p'',x_1,x_2,x_3) \coloneqq {} & (x =  x_1 + x_2 + x_3) \land\textstyle\bigwedge_{j\in\{1,2,3\}} (0\leq x_j) \,\land (x_1 < p)
        \,\land
        \\
                                                              & (x_2 = 0 \lor (V_t(x_2, p)\land x_2 < p')) \land (x_3=0 \lor (p' \leq p'' \land V_t(x_3,p''))).
    \end{align*}
    There is a unique decomposition of $x$ into the sum of $x_1,x_2$ and $x_3$.
    Defining $\mathrm{digit}_t$ on the non-negative integers, denoted $\mathrm{digit}_t^+$, can be done in
    both the $\exists^*$ and $\forall^*$ fragment of $\Lang_{\PBA}$ as follows:
    \begin{align*}
        \mathrm{digit}_t^+(x, p, p', y)             & \coloneq \hspace{1.2em}\exists p'',x_1,x_3 \hspace{0.5em}\mathrm{split}_t(x,p,p',p'',x_1,y,x_3),
        \\
        \text{or \ }\mathrm{digit}_t^+(x, p, p', y) & \coloneq \forall p'',x_1,x_2,x_3\left(\mathrm{split}_t(x,p,p',p'',x_1,x_2,x_3)\implies x_2=y\right).
    \end{align*}
    One then relies on~\Cref{fact:msdf-inverses} to extend the definition to the negative numbers:
    \[
        \mathrm{digit}_t(x, p, p', y)\coloneqq \big((x<0)\land\mathrm{digit}_t^+(-x-1, p, p', p' - p - y)\big) \lor \big((0\leq x)\land\mathrm{digit}_t^+(x, p, p', y)\big). \qedhere
    \]
\end{proof}

The last predicate we need before describing the translation,
denoted $\mathrm{local}_{\phi}$, merits an informal explanation.
The formulas from~$\Formulas_{\PAt}$ labeling transitions
of the automata read only one digit at a time.
However, as discussed above, parametric B\"uchi arithmetic cannot extract
a digit: we can only extract monomials $d \cdot t^i$,
as done by the formula~$\mathrm{digit}_t$.
The formula $\mathrm{local}_{\phi}$ resolves this mismatch by
``shifting'' the solutions of $\phi$ by an arbitrary power of $t$.
Its formal semantics is given below.

\begin{restatable}{lemma}{LemmaShiftedLabels}\label{lemma:shifted-labels}
    Let $\phi(x_1,\ldots,x_n)$ be from $\Formulas_{\PAt}$.
    We can construct a formula~$\mathrm{local}_{\phi}(p,p',x_1,\ldots,x_n)$,
    in the $\exists^*$ fragment of $\Formulas_{\PBA}$ with the property:
    for every $k \in \N_{\geq 2}$,
    and $\vec w \coloneqq w_\ell \dots w_0 \in \Mk$, if
    $\intrep{\vec w}_k(p) = k^i$ and $\intrep{\vec w}_k(p') = k^{i+1}$ for some $i \in \N$, then
    $w_i \models_k \phi
        \Longleftrightarrow
        \intrep{\vec w}_k \models_k \mathrm{local}_{\phi}$.
\end{restatable}

\begin{proof}
    To obtain an existential formula, we rely on the fact
    that Presburger arithmetic has quantifier elimination
    in the extended structure $(\Z,0,1,+,(\equiv_d)_{d \in \N_{\geq 2}},\leq)$,
    where $\equiv_d$ stands for the congruence modulo $d$ relation~\cite{Presburger29}:

    \begin{fact}\label{fact:presburger-quantifier-elimination}
        The structure $(\Z,0,1,+,(\equiv_d)_{d \in \N_{\geq 2}},\leq)$ admits quantifier elimination:
        there is an algorithm that, given a first-order formula $\phi$ over this structure, returns an equivalent quantifier-free formula $\psi$.
    \end{fact}
    Seeing the parameter $t$ as a standard integer variable,
    we compute a quantifier-free formula $\psi$ equivalent to $\phi$.
    Without loss of generality, we assume $\psi$ to be a \emph{positive} (i.e., negation-free) Boolean combination of constraints
    $\tau \sim 0$,
    where $\tau$ is a linear term of the form $a_1 \cdot x_1+\dots+a_n \cdot x_n + b\cdot t + c$,
    with $a_i$, $b$ and $c$ integers,
    and $\sim$ is either $\leq$ or $\equiv_d$ for some $d \in \N_{\geq 2}$.
    Positivity can be forced thanks to the equivalences $\lnot(\tau\leq 0)\iff (-\tau+1\leq 0)$ and \( \lnot\left(\tau\equiv_d 0\right) \iff \bigvee_{r\in[1,d-1]}\tau - r\equiv_d 0\).

    We now ``shift'' the solutions to $\psi$,
    constructing a formula $\psi^*(y_1,\dots,y_n)$ such that,
    when each $y_j$ is set to a monomial $d_j \cdot t^i$,
    $\psi^*$ is satisfied if and only if so is $\phi(d_1,\dots,d_n)$.
    Recall that $\intrep{\vec w}_k(p) = k^i$ and $\intrep{\vec w}_k(p') = k^{i+1}$.
    The formula $\psi^*$ is obtained by manipulating $\psi$ as explained below.

    Inequalities $a_1 \cdot x_1 + \dots + a_n \cdot x_n + b\cdot t + c \leq 0$
    are rewritten to $a_1 \cdot y_1 + \dots + a_n \cdot y_n + b\cdot p' + c \cdot p \leq 0$.
    Intuitively, every monomial in the original inequality is multiplied by $p$,
    after which each term $x_j \cdot p$ is replaced by $y_j$, and $t \cdot p$ is replaced by $p'$.

    For a congruence $\tau \equiv_{d} 0$, we can similarly start by multiplying everything by $p$,
    obtaining $\tau \cdot p \equiv_{d \cdot p} 0$. This is not, however, a formula
    in $\Formulas_{\PBA}$, since the modulus is no longer a constant.
    Let us nonetheless proceed as for inequalities, replacing $x_j \cdot p$ by $y_j$, and $t \cdot p$ by $p'$. We obtain a congruence of the form
    $a_1\cdot y_1 +\dots+a_n \cdot y_n + b\cdot p' + c\cdot p \equiv_{d\cdot p} 0$.
    This congruence is equivalent to $\exists y\, (d \cdot p \cdot y = a_1\cdot y_1 +\dots+a_n \cdot y_n + b\cdot p' + c\cdot p)$, where $y$ ranges over $\Z$.
    But $p \cdot y$ is nothing more than an integer that is divisible by $p$,
    and so this formula is equivalent to the $\Formulas_{\PBA}$ formula
    \[
        \exists y \exists p''\, (d \cdot y = a_1\cdot y_1+\dots+a_n\cdot y_n +b\cdot p' + c) \land V_t(y, p'') \land p \leq p''.
    \]

    Once the formula $\psi^*$ has been constructed following the transformations above,
    we define:
    \[
        \textstyle\mathrm{local}_\phi(p,p',x_1,\dots,x_n) \coloneqq \exists y_1 \dots \exists y_n\, \psi^{*}(p, p', y_1,\dots,y_n)\land
        \bigwedge_{i=1}^n \mathrm{digit}_t(x_i, p, p', y_i)\text{.}
    \]
    The fact that $\mathrm{local}_\phi$ is correct and is an existential formula then
    follows from~\Cref{lemma:digit-t}.
\end{proof}

With these preparation steps done, we can proceed to translate automata.

\begin{restatable}{lemma}{LemmaSigmaTwoWithNt}\label{lemma:sigma-2-with-nt}
    For every \( \mathcal{A} \in \AutPAt \), we can compute a formula \( \phi \) in the \( \exists^{*}\forall^{*} \) fragment of parametric B\"uchi arithmetic extended with the
    predicate \( S_t \), such that \( \LangMSD(\mathcal{A})=\sem{\phi} \).
\end{restatable}

\begin{proof}
    Let \( \mathcal{A}=(\Phi, Q, q_0, F, \delta) \) be a parametric symbolic automaton
    from $\AutPAt$, and let $\vec{x}=\left(x_1,\dots,x_n\right)$ be the variables in~$X_\Aut$.
    By~\Cref{lemma:make-NSA-valid,lemma:valid-sfa-closure}, we can assume $\Aut$ to be a valid DSA.
    As in~\cite{BruyereHMV94}, we also assume $\Aut$ to be \emph{complete}: for every $k \in \N_{\geq 2}$, $q \in Q$ and $w \colon X \to [0..k-1]$,
    there is one $q' \in Q$ such that $w \models_k \delta(q,q')$.
    (It suffices to add a non-accepting state~$q_\bot \notin Q$, setting $\delta(q_\bot,q_\bot) \coloneqq \top$, $\delta(q_\bot,q) \coloneqq \bot$
    and $\delta(q,q_\bot) \coloneqq \lnot\bigvee_{q' \in Q}\delta(q,q')$ for every $q \in Q$, and extending~$\Phi$ accordingly.)
    Hence, $\Aut$ has exactly one run on every word over~$[X \to [0..k-1]]$ and, as $\Aut$ is valid, a map~$\mu$ belongs to~$\LangMSD(\Aut)(k)$
    if and only if this run ends in~$F$ for some (equivalently, every) $k$-msdf encoding of~$\mu$.

    For every state $q \in Q$, we introduce an integer variable with the same name,
    and denote these variables collectively as~$\vec q$. The intended reading is that,
    for a fixed base $t$, the variables $\vec q$ jointly describe a single run of $\mathcal{A}$:
    at every position~$i$, exactly one variable $q \in Q$ has digit $1$ in its base-$t$
    expansion at position $i$, other variables from $Q$ have digit $0$, and this $q$ is the state that the run occupies just after
    consuming the digit at position $i$ from the input.
    To make this more precise, we need a way to identify the length of the run.
    Fix a base $t$ and consider
    a run $q_0 \xrightarrow{w_m} q_1 \xrightarrow{w_{m-1}} \dots \xrightarrow{w_0} q_{m+1}$ of $\mathcal{A}$,
    on a word $\vec w \coloneqq w_m \dots w_0 \in \Mk$ accepted by $\mathcal{A}$; here $q_j$, for
    $j \in [0..m+1]$, is the state reached after reading the $j$ symbols $w_m,\dots,w_{m-j+1}$,
    so that $q_j$ is exactly the state we want active at digit position $m+1-j$. We want to encode
    this run using the variables $\vec q$, for an arbitrary base $t$ and an arbitrary (unknown)
    length $m+1$. Since $m$ cannot be referred to directly in our formula, we instead introduce
    an auxiliary variable $u$ standing for $t^m$, together with two further variables $u'$ and
    $u''$ standing for $t^{m+1}$ and $t^{m+2}$, respectively.

    For $q \in Q$, denote by $q[0],q[1],\dots$ the sequence of digits in $[0..t-1]$ such that
    $q = \sum_{i \in \N} q[i] \cdot t^i$.
    Formalizing the intuition above, we want the variables $\vec q$ to satisfy:
    \begin{enumerate}
        \item\label{it:pbb:states:p1} for every $i \in [0..m+1]$ exactly one $q \in Q$ is such that $q[i] = 1$, the others have a zero digit;
        \item\label{it:pbb:states:p2} $q_0[m+1] = 1$, and $q[0] = 1$ for some $q \in F$.
    \end{enumerate}
    Item~\ref{it:pbb:states:p1} forces $\vec q$ to describe, at every position from $0$ to
    $m+1$, a single well-defined ``currently active'' state.
    Item~\ref{it:pbb:states:p2} fixes the two endpoints of the sequence.
    Assuming $u' = t^{m+1}$ and $u'' = t^{m+2}$, we capture \Cref{it:pbb:states:p1} with the formula $\phi_{\ref{it:pbb:states:p1}}(\vec q, u', u'')$ defined as
    \[
        \bigwedge_{q\in Q} \left(0\leq q< u''\right) \land \forall s \forall s' \Big(S_t(s,s')
        \land s \leq u'\implies
        \bigvee_{q\in Q} \!\big(\mathrm{digit}_t(q, s, s', s)\land \!\bigwedge_{q'\in Q \setminus \{q\}} \!\!\mathrm{digit}_t(q', s, s', 0)\big)\Big).
    \]
    The first part states that the variables $q\in Q$ may have only non-negative values and that their most significant bit
    is at most $(m+1)$. The second part uses the predicate $\mathrm{digit}_t$ to
    assert that at every position $i\in[0..m+1]$ exactly one state is active (with digit one), and the remaining ones are inactive (with digit zero).
    We capture Item~\ref{it:pbb:states:p2} with the formula $\phi_{\ref{it:pbb:states:p2}}(\vec q, u', u'')$ defined as
    \[
        \mathrm{digit}_t(q_0,u',u'',u')\,\land\,\forall v\, \big(S_t(1, v)\implies \textstyle\bigvee_{q\in F} \mathrm{digit}_t(q,1,v,1)\big).
    \]
    Note that, in this formula, the premise of the implication can be true only when the variable \( v \) is equal to \( t \). Both formulas $\phi_{\ref{it:pbb:states:p1}}(\vec q, u', u'')$ and $\phi_{\ref{it:pbb:states:p2}}(\vec q, u', u'')$ only feature universal quantifiers (when using the $\forall^*$ definition of $\mathrm{digit}_t$ from \Cref{lemma:digit-t}). Below, let $\mathrm{run}_{Q,F}(\vec q,u',u'') \coloneqq(\phi_{\ref{it:pbb:states:p1}}\land\phi_{\ref{it:pbb:states:p2}})$.

    The variables $\vec{x}=\left(x_1,\dots,x_n\right)$
    are the free variables of our final formula. So far, $\mathrm{run}_{Q,F}(\vec q,u',u'')$
    only constrains $\vec q$ to encode \emph{some} sequence of states, without tying this
    sequence to the transition relation $\delta$ of $\mathcal{A}$ or to the values of $\vec x$.
    To capture a valid run and not just an arbitrary sequence of states, we conjoin to
    $\mathrm{run}_{Q,F}(\vec q,u',u'')$ a further formula, denoted $\Delta(\vec q,u',\vec x)$,
    which asserts that consecutive active states are indeed connected by a transition of
    $\delta$ labelled consistently with the digits of $\vec x$. This formula is defined as
    follows,
    \begin{multline*}
        \Delta(\vec q,u',\vec x)\coloneqq \forall s\forall s'\forall s''\, \left(S_t(s, s')\land S_t(s',s'')\land s\leq u'\right)\implies \\
        \bigwedge_{q,q'\in Q} \left(\mathrm{local}_{\delta(q,q')}(s,s',\vec x)\land\mathrm{digit}_t(q,s',s'',s')\implies \mathrm{digit}_t({q'},s,s',s)\right).
    \end{multline*}
    Here $s,s',s''$ range, via $S_t(s,s')\land S_t(s',s'')$, over triples of consecutive powers
    of $t$, i.e.\ $s=t^i$, $s'=t^{i+1}$, $s''=t^{i+2}$ for some position $i\in[0..m]$; the bound
    $s\leq u'$ restricts~$i$ to the range covered by the run. Recall that the run visits the
    state active at position $i+1$ \emph{before} the state active at position~$i$, since
    positions are read from most to least significant. Accordingly, $\Delta$ asserts that
    whenever a state $q$ is active at the more significant position $i+1$
    ($\mathrm{digit}_t(q,s',s'',s')$) and the transition label $\delta(q,q')$ is satisfied
    locally at position $i$ by the digits of $\vec x$ ($\mathrm{local}_{\delta(q,q')}(s,s',\vec x)$),
    the successor state $q'$ must indeed be the state active at position $i$
    ($\mathrm{digit}_t(q',s,s',s)$).

    The formula $\Delta(\vec q,u',\vec x)$ lies in the $\forall^*$ fragment of parametric
    Büchi arithmetic with $S_t$: note that $\mathrm{local}_{\delta(q,q')}$ and the first
    $\mathrm{digit}_t$ conjunct occur negated (as antecedents of the inner implication), so
    their $\exists^*$ definitions in~\Cref{lemma:shifted-labels,lemma:digit-t}
    turn into $\forall^*$ definitions. The remaining, positive occurrence of
    $\mathrm{digit}_t$ is instantiated with its $\forall^*$ definition from
    \Cref{lemma:digit-t}.

    We can now give the final $\exists^*\forall^*$ formula~$\phi(\vec x)$:
    \[
        \exists \vec q,u, u', u''\, S_t(u,u')\land S_t(u',u'') \land \textstyle\bigwedge_{x\in\vec x}\big(u > \norm{x}\big) \,\land\, \mathrm{run}_{Q,F}(\vec q,u',u'') \,\land\,
        \Delta(\vec q,u',\vec x),
    \]
    where $(u >\norm{x}) \coloneqq (x > 0 \land u > x) \lor (x = 0 \land u > 1) \lor (x < 0 \land u > -x-1)$
    ensures that the most significant non-sign digit of \( x \) occurs on position at most \( m-1 \),
    with $u = t^{m}$. This guarantees that the most significant digit of \( x \)
    that the formula sees is, in msdf, either a $0$ (and $x$ is non-negative) or a $t-1$ (and $x$ is negative),
    which is exactly what the semantics of $\LangMSD(\mathcal{A})$ requires.
\end{proof}

Recalling that $S_t$ can be defined in the $\forall^*$ fragment of parametric B\"uchi arithmetic, we obtain the following corollary, which completes the proof of~\Cref{theorem:bb-par}.
\begin{corollary}\label{lemma:sigma-3-without-nt}
    For every \( \mathcal{A} \in \AutPAt \), we can compute a formula \( \phi \) in the \( \exists^{*}\forall^{*}\exists^{*} \) fragment of parametric B\"uchi arithmetic
    such that \( \LangMSD(\mathcal{A})=\sem{\phi} \).
\end{corollary}

\subsection{Decision problems on parametric B\"uchi arithmetic}\label{subsec:corollaries}

Starting from~\Cref{theorem:bb-par}, and relying on the fact that Presburger arithmetic has quantifier elimination, we now show that the set of bases for which a parametric B\"uchi arithmetic formula is satisfiable is ultimately periodic.
A subset $S \subseteq \N$ is \emph{ultimately periodic} if there are integers $N$ and $p \geq 1$ such that for every $n \geq N$, $n \in S$ if and only if $n + p \in S$.
Observe that such a set can be represented finitely by a triple $(B,N,p)$ with $N$ and $p$ as in the definition, and $B \subseteq [0..N+p-1]$ satisfying $B = S \cap [0..N+p-1]$.
Ultimately periodic sets are closed under union, intersection, and complementation, and these operations can be performed effectively on their representations (see,~e.g.,~\cite[Proposition~9]{HaaseM21}).

\begin{algorithm}[t]
    \begin{algorithmic}[1]
        \Require A formula $\phi$ in $\Formulas_\PBA$
        \Ensure A representation of $\realbases(\phi)$ as an ultimately periodic set
        \setstretch{1.05}
        \State\textbf{let} $\Aut \coloneqq (\Phi,Q,q_0,F,\delta)$ be an NSA in $\AutPAt$ such that $\LangMSD(\Aut) = \sem{\phi}$.
        \State add a fresh state $\widetilde q_0$ to $Q$, and make it the initial state of $\Aut$\footnotemark\label{algo:realizable-bases:line:fresh-initial}
        \State extend $\delta$, adding $(\widetilde q_0,q) \mapsto \delta(q_0,q) \land \bigwedge_{x \in X_\Aut} (x = 0 \lor x = t-1)$ and $(q,\widetilde q_0) \mapsto \bot$ for all $q \in Q$\label{algo:realizable-bases:line:sign-digit}
        \State\textbf{let} $b$ be an empty map from $Q \times Q$ to ultimately periodic sets\label{algo:realizable-bases:map-b}
        \For{every pair of states $q,q' \in Q$}\label{algo:realizable-bases:line:compute-b}
        \State\textbf{let}
        \begin{minipage}[t]{0.85\textwidth}
            $\psi(t)$ be the quantifier-free formula of Presburger arithmetic obtained by performing quantifier elimination on $t \geq 2 \land (\exists \vec x < t)\delta(q,q')$, where $\vec x$ lists the variables in~$X_\Aut$, seeing $t$ as a (free) integer variable
        \end{minipage}\vspace{2pt}
        \State $b(q,q') \gets {}$ultimately periodic set representation of the solutions of $\psi(t)$\label{algo:realizable-bases:line:end-for}
        \EndFor
        \State\textbf{let}
        $m$ be the map from $Q$ to ultimately periodic sets given by
        $m(q) \coloneqq {}$%
        \scalebox{0.9}{
            $\begin{cases}
                    (\{2\},2,1)     & \text{if } q \in F             \\
                    (\emptyset,0,1) & \text{if } q \in Q \setminus F
                \end{cases}$}
        \vspace{-5pt}
        \While{\textbf{true}}\label{algo:realizable-bases:line:while}
        \State\textbf{let} $m'$ be a copy of $m$
        \For{every state $q \in Q$}
        \ $m(q) \gets m(q) \cup \bigcup_{q' \in Q} \left(b(q,q') \cap m(q')\right)$\label{algo:realizable-bases:line:update-m}
        \EndFor
        \If{$m = m'$}\label{algo:realizable-bases:line:return}
        \textbf{return} $m(\widetilde q_0)$
        \EndIf
        \EndWhile
    \end{algorithmic}
    \caption{Computes the realizable bases of a parametric B\"uchi arithmetic formula.}\label{algo:realizable-bases}
\end{algorithm}

We describe an algorithm that, given as input a parametric B\"uchi arithmetic formula $\phi$, computes an ultimately periodic representation of the set
$\realbases(\phi)$ of realizable bases of $\phi$. The algorithm is based
on \Cref{fact:presburger-quantifier-elimination} and the following classic fact from~\cite{GinsburgS66}.

\begin{fact}\label{fact:ultimately-periodic}
    The set of solutions of a univariate quantifier-free formula over $(\Z,0,1,+,(\equiv_d)_{d \in \N_{\geq 2}},\leq)$ is ultimately periodic, and a representation of this set can be computed effectively from the formula.
\end{fact}

The pseudocode of the procedure is given in~\Cref{algo:realizable-bases}.\footnotetext{Lines~\ref{algo:realizable-bases:line:fresh-initial} and~\ref{algo:realizable-bases:line:sign-digit} make the algorithm correct for every NSA $\Aut$ with $\LangMSD(\Aut) = \sem{\phi}$. However, they are not needed when $\Aut$ is constructed as described in~\Cref{subsec:from-formulas-to-automata}.}
The algorithm starts by translating the input formula $\phi$ into a parametric symbolic automaton $\Aut$, following~\Cref{lemma:from-formulas-to-automata}. Since in the $k$-msdf encoding
the first digit is a sign digit in $[X \to \{0,k-1\}]$, the algorithm changes the initial state to a fresh state $\widetilde q_0$, and modifies the transition relation $\delta$ to make sure that the automaton only reads words whose first letter is a sign digit; see the constraints $x = 0 \lor x = t-1$, for $x \in X_\Aut$. The updates to $\delta$ also ensure that the initial state is not a final state, since in the $k$-msdf encoding the empty word is never accepted. Next (lines~\ref{algo:realizable-bases:map-b}--\ref{algo:realizable-bases:line:end-for}), the algorithm computes for each pair of states $q,q' \in Q$ an ultimately periodic set $b(q,q')$ characterizing the bases $k \geq 2$ such that $\delta(q,q')$ is satisfiable over~$\kPAmodel_k$.
This set is computed by relying on~\Cref{fact:presburger-quantifier-elimination,fact:ultimately-periodic}, seeing $t$ as a free integer variable instead of the constant symbol from the language of Presburger arithmetic with digits.
The last part of the algorithm (lines~\ref{algo:realizable-bases:line:while}--\ref{algo:realizable-bases:line:return}) performs a fixpoint computation to populate a map $m$.
Once stabilized, this map associates to each state $q \in Q$ an ultimately periodic set $m(q)$ characterizing the bases $k \geq 2$ for which there is a run from $q$ to some final state in $\Aut$ over a word with letters in $[X \to [0..k-1]]$;
moreover, for $\widetilde q_0$, these words must start with a sign digit.
The map is initialized so that $m(q)$ represents $\N_{\geq 2}$ if $q$ is a final state, and $\emptyset$ otherwise. Line~\ref{algo:realizable-bases:line:update-m} updates $m(q)$ by adding to it all bases $k \in b(q,q') \cap m(q')$, for all $q' \in Q$.
Upon reaching a fixpoint, $m(\widetilde q_0) = \realbases(\phi)$.

\pagebreak
\begin{restatable}{lemma}{LemmaAlgoRealbasesCorrectness}\label{lemma:algo-realbases-correctness}
    \Cref{algo:realizable-bases} respects its specification and terminates.
\end{restatable}

\begin{proof}[Proof sketch]
    \looseness=-1 The correctness of the algorithm follows from the discussion above.
    For termination, consider the set~$U$ of ultimately periodic sets generated by taking
    unions and intersections of the sets $b(q,q')$, $(\{2\},2,1)$, and $(\emptyset,0,1)$, for all $q,q' \in Q$. The set $U$ is a partially ordered set with respect to the subset relation~$\subseteq$.
    Also define $\dot\subseteq$ as the point-wise order on $[Q \to U]$ given by $m \mathbin{\dot\subseteq} m'$ if and only if $m(q) \subseteq m'(q)$ for every $q \in Q$.
    Note that $[Q \to U]$ is a partially ordered set with respect to $\dot\subseteq$.

    The body of the while loop corresponds to an operator $\Phi \colon [Q  \to U] \to [Q \to U]$
    that updates the current map $m$. It is defined as
    $\Phi(m)(q) \coloneqq m(q) \cup \bigcup\nolimits_{q' \in Q} \left(b(q,q') \cap m(q')\right)$.
    This operator is inflationary, i.e., $m \mathbin{\dot\subseteq} \Phi(m)$ for every $m \in [Q \to U]$. Hence, regardless of how $m$ is initialized, iterates of $\Phi$ form an ascending chain in $[Q \to U]$. Since $[Q \to U]$ is finite, this chain must reach a fixpoint after finitely many iterations, and the while loop terminates.
\end{proof}

\TheoremUltimatelyPeriodic*

\Cref{lemma:algo-realbases-correctness} implies~\Cref{theorem:ultimately-periodic} from~\Cref{subsec:contributions}, and shows that the decision problems of \textsc{Existence} and \textsc{Universality} for parametric B\"uchi arithmetic are decidable.

\begin{theorem}\label{theorem:bb-par-decidability}
    The \textsc{Existence} and \textsc{Universality}
    problems for $\Formulas_\PBA$ are decidable.
\end{theorem}

\begin{proof}
    For both problems, we first run~\Cref{algo:realizable-bases} on the input formula $\phi$ to compute a representation~$(B,N,p)$ of the ultimately periodic set $\realbases(\phi)$. The formula $\phi$ is a positive
    instance of \textsc{Existence} if and only if $B$ is non-empty.
    And it is a positive instance of \textsc{Universality} if and only if $(B,N,p)$ represents $\N_{\geq 2}$, which holds whenever $N \geq 2$ and $B = [2..N+p-1]$.
\end{proof}

\section{The effects of adding (full) multiplication by $t$}\label{sec:multiplication-effects}

The reader may notice that in the proof of \Cref{theorem:bb-par} we compared two subsequent configurations of parametric symbolic automata using a version of multiplication $(x \mapsto t \cdot x)$, restricted to inputs that are powers of $t$.
In B\"uchi arithmetic on base $k$, the function $(x\mapsto k\cdot x)$ is clearly definable by a quantifier-free formula, and it is used extensively in the
proof of~\Cref{theorem:bb}, as well as in~\cite{Starchak23}, where recognizable languages are characterized using existential formulas.

The latter paper shows that a set $\mathcal{S}\subseteq\N^n$, being encoded in base $k$,
is recognizable by a finite automaton if and only if it is existentially definable in the structure $(\N; 0,1,+, \&_k,\leq)$,
where $\&_k$ is the bitwise minimum operation on base $k$ on non-negative integers:
\begin{equation}\label{eq:bitwise-minimum}
    \intrep{0a_\ell\dots a_0}_k \&_k \intrep{0b_\ell\dots b_0}_k = \intrep{0\min(a_\ell,b_\ell)\dots \min(a_0,b_0)}_k.
\end{equation}
Informally, the roles of $(x\mapsto k\cdot x)$ and $\&_k$ can be
described as follows: the former is used to shift a representation of a sequence of
configurations by one step, while the latter serves as a comparator of each pair of
subsequent configurations.
For~\Cref{theorem:bb}, this means that in converting finite automata to B\"uchi arithmetic on base $k$, quantifier alternation is only needed to define $\&_k$.

In this section we study computational properties of parametric B\"uchi arithmetic enriched with $(x\mapsto t\cdot x)$, which we denote by \PBAM.
We start by considering a different signature, denoted~\PBAMA, which is obtained from~\PBAM by replacing the predicate $V_t$ with the binary function $\&_t$ to be interpreted as the bitwise minimum operation.
We show that, in contrast to the case when the parameter $t$ is replaced with a fixed integer $k\geq2$, existential formulas over~\PBAMA are expressive enough to encode computations of
counter machines.
This implies that the \textsc{Existence} problem for the existential fragment of $\Formulas_{\PBAMA}$ is already undecidable.
Since the function~$\&_t$ is expressible in $\Formulas_{\PBAM}$ using a formula with only universal quantifiers, this implies that the \textsc{Existence} problem for the $\exists^*\forall^*$ fragment of $\Formulas_{\PBAM}$ is undecidable, which in turn means (due to~\Cref{theorem:bb-par-decidability}) that the function $(x\mapsto t\cdot x)$ is not definable in parametric B\"uchi arithmetic.

For simplicity, throughout the section we work with interpretations over the non-negative integers. Since \PBA features the order~$\leq$, our undecidability
results trivially transfer to the integers.

\subsection{The semantics of multiplication and bitwise minimum, and counter machines}\label{subsec:counter-machines}

The satisfaction relation $\models_k$ and semantics $\sem{\_}_{-}$ of parametric B\"uchi arithmetic can be naturally updated to the signatures \PBAM and \PBAMA and to work over $\N$.
Given~${k \in \N_{\geq 2}}$, one can consider a single structure
$\kBAmodel_k^+ \coloneqq (\N;I)$, where $0,1,+$ and $\leq$
are interpreted over $\N$ in the usual way, the symbol $V_t$ with the
set $\{(n,d) \in \N \times \N : \text{$d$ is the largest power of $k$ dividing $n$} \}$, and
the binary function $\&_t \colon \N \times \N \to \N$ with the bitwise minimum operation on base $k$,
following the equation in~\eqref{eq:bitwise-minimum}.
For $\Formulas = \Formulas_{\PBAM}$ or $\Formulas = \Formulas_{\PBAMA}$,
this structure naturally yields a satisfaction relation $\_ \models_k \_ \colon [X \to \N] \to \Formulas \to \{\top, \bot\}$,
from which one can then define the (parametric) semantics $\sem{\_}_{-} \colon \Formulas \to \N_{\geq 2} \to \powerset[X \to \N]$,
exactly as done for parametric B\"uchi arithmetic.
One can then define the set of realizable bases $\realbases(\phi)$ and the \textsc{Existence}
and \textsc{Universality} problems for $\phi$ in $\Formulas_{\PBAM}$ or $\Formulas_{\PBAMA}$, exactly as in~\Cref{sec:preliminaries}.

Since we are now working over $\N$, some care is needed regarding subtraction.
In this section, we solely consider the binary minus operator $\_ - \_ \colon \N \times \N \to \N$ given by $x - y = z$ if and only if $(z = 0 \land x < y) \lor (x = z + y)$.
We will work as if this operator is available in the signature.
It is easy to see that any formula featuring binary minus can be rewritten into an equivalent one
without it, at the expenses of adding further existentially quantified variables.

\paragraph*{Counter machines.}
A counter machine~\cite{Minsky67,Kozen97} is a tuple $\mathcal{M}\coloneqq(Q,\mathcal{C},q_0, F, \Delta)$, where, $Q$ is a finite set of states,
$\mathcal{C}=\{c_1,\dots,c_n\}$ is a finite set of $n \geq 1$ counters,
each storing a non-negative integer, $q_0\in Q$ is the initial state, $F\subseteq Q$ are final states,
and $\Delta\subseteq {Q\times \{\id,\inc,\dec\}^n\times Q}$ is the transition relation,
where $\id$, $\inc$, and $\dec$ are the identity, successor, and predecessor operations on integers, respectively.
A~\emph{configuration of $\mathcal{M}$} is a tuple $(p,c_1,\dots,c_n)\in Q\times\N^n$.
We write $\leadsto$ for the smallest binary relation on configurations such that, for every $(p,c_1,\dots,c_n) \in Q\times\N^n$,
\begin{equation*}
    (p,c_1,\dots,c_n)\leadsto(q,f_1(c_1),\dots,f_n(c_n)) \ \Longleftrightarrow \ \exists (p,f_1,\dots,f_n,q) \in \Delta \text{ s.t. } f_i(c_i) \geq 0 \text{ for all } i\in[1..n].
\end{equation*}
A~\emph{run in $\mathcal{M}$} is a sequence of configurations $(p,c_1,\dots,c_n)\leadsto\dots\leadsto(p',c_1',\dots,c_n')$.
We say that a counter machine $\mathcal{M}$ \emph{accepts} $x\in\N$ whenever
for some $q\in F$ there is a run in~$\mathcal{M}$ of the form $(q_0,x,0,\dots,0)\leadsto\dots\leadsto(q,0,0,\dots,0)$.
The following classical result (see, e.g.,~\cite{Kozen97}), implies that the non-emptiness problem is undecidable for counter machines with $3$ counters.

\begin{fact}\label{fact:counter-machines}
    A set $S\subseteq\N$ is recursively enumerable if and only if there is a counter machine with $3$ counters that accepts exactly the numbers in $S$.
\end{fact}

\subsection{Encoding runs with existential formulas}\label{subsec:encoding-runs}

Using counter machines, we now show how to encode any recursively-enumerable set $S\subseteq\N$
as an existential formula in $\Formulas_{\PBAMA}$.
Other encoding of counter/register machines into purely existential formulas
can be found, for instance, in~\cite{JonesM84} for the $\exists^*$ fragment of $(\N;0,1,+,\cdot,x^y,=)$ (i.e., the structure of exponential Diophantine equations) and in~\cite{Starchak23} for the $\exists^*$ fragment of $(\N;0,1,+,\&_k,\frown_k,\leq)$,
where $\frown_k$ stands for concatenation of $k$-ary representations of non-negative integers.
Though these two theories and $\Formulas_{\PBAMA}$ can represent all recursively-enumerable sets,
we do not see any simple reduction between them.

\begin{theorem}\label{theorem:counters}
    Let $S \subseteq \N$ be a recursively-enumerable set. There is a univariate existential formula $R(x)$
    in $\Formulas_{\PBAMA}$ such that $S=\{a\in\N : \text{there is } k \geq 2 \text{ such that } [x\mapsto a]\models_k R(x)\}$.
\end{theorem}

\begin{proof}

    We construct $\phi$ from a counter machine $\mathcal{M}\coloneqq(Q,\{c_1,c_2,c_3\},q_0, F, \Delta)$ accepting the numbers in $S$. Let us start by giving an intuition on how to encode accepting runs of $\mathcal{M}$ via variables over the non-negative integers.
    In an accepting run $(q_0,x,0,0) \leadsto \dots \leadsto(q,0,0,0)$
    with $q \in F$ and $x \in \N$,
    the values of the three counters are necessarily bounded by some non-negative integer, say $M$. If we force the parameter $t$ to be
    greater than $M$ (in fact $M+2$ for technical reasons), then every value of the counters can be encoded with a single digit in base $t$.
    The idea is then to use variables $y_1,y_2,y_3$ to encode the values of the three counters in the run: the $i$-th digit of $y_j$ in base $t$
    will be equal to the value of counter $c_j$ in the $i$-th configuration of the run (plus $1$, again for technical reasons).
    As in the proof of~\Cref{lemma:sigma-2-with-nt},
    to encode the sequence of states in the run,
    for each state $q \in Q$ we will introduce a non-negative integer variable with the same name, and require that the base $t$ encoding of $q$ belongs to $\{0,1\}^*$, with that $i$-th digit equal to $1$ if and only if the state in the $i$-th configuration of the run is $q$.
    An additional variable $w$ is used to encode a power $t^{m}$, where $m+1$ is the length of the run.
    Most of the work will then be to express the fact that the sequence of configurations encoded by these variables is indeed a valid and accepting run of $\mathcal{M}$.

    Let us formalize the above intuition. To do so, we need a few auxiliary predicates, the first set of which is defined in the next table.
    \begin{center}
        {\arrayrulecolor{gray!85}\renewcommand{\arraystretch}{1.15}%
        \begin{tabular}{l|l|>{\raggedright\arraybackslash}p{0.5\linewidth}}
            \rowcolor{gray!15}
            {\it Predicate:}      & {\it Definition:}                        & {\it Meaning:}                                                                              \\ \hline
            $P_t(x)$              & $1\leq x \land x \,\&_t\, (x-1) = 0$     & $x$ is a power of $t$.                                                                      \\[2pt]
            $x \pleq_t y$         & $x \,\&_t\, y = x$                       & Every digit of $x$ is smaller than or equal to the corresponding digit in $y$.              \\[2pt]
            $x = \textbf{1}_t(w)$ & $P_t(w)\land (t-1)\cdot x = t\cdot w-1$  & $w$ is of the form $t^m$ and the base $t$ encoding of $x$ is a sequence of $m+1$ many $1$s.
        \end{tabular}\arrayrulecolor{black}}
    \end{center}
    \noindent
    For example, we have $15 = \textbf{1}_2(8)$ and $111=\textbf{1}_{10}(100)$.
    We will often use $x = \textbf{1}_t(w)$ as a function $\textbf{1}_t(w)$. Occurrences of this function
    can be expressed with the predicate by introducing fresh existential variables.
    We will also need to extract from two non-negative integers $x$ and $y$
    the unique integer $z$ filtering exactly the digits of $x$ and $y$ that equal: for every position $i\in\N$ we have $z[i]= \textbf{if } x[i] = y[i] \textbf{ then } x[i] \textbf{ else } 0$.
    This ternary predicate is denoted by $z = \textit{eq}_t(x,y)$,
    and can be defined as the following $\exists^*$ formula of $\Formulas_{\PBAMA}$:
    \[
        \exists v \colon P_t(v) \land (x < v) \land z = x \,\&_t\, \big((t-1)\cdot
        \blue{\textbf{(}}\textbf{1}_t(v) -
        \green{\textbf{(}}\textbf{1}_t(v)\,\&_t\,\orange{\textbf{(}}(x -  x\,\&_t\,y) + (y -  y\,\&_t\,x)\orange{\textbf{)}}\green{\textbf{)}}\blue{\textbf{)}}\big).
    \]
    This definition executes a sequence of digit-wise transformations.
    The subformula $x -  x\,\&_t\,y$ defines an integer $z_x$ such that $z_x[i] = 0$ if and only if $x[i] \leq y[i]$;
    while $y - y \&_t x$ symmetrically yields an analogous integer $z_y$.
    Their sum $z_x + z_y$, corresponding to the expression within the~\orange{orange} parentheses,
    is an integer satisfying $(z_x + z_y)[i]=0$ if and only if
    $x[i]=y[i]$. The next step applies a digit-wise indicator function to this sum, mapping non-zero digits to $1$ and zeros to $0$.
    This is achieved by using a sufficiently large power of $t$ (the variable $v$) to generate a mask of $1$s via $\textbf{1}_t(v)$,
    which is then intersected with $z_x + z_y$ via the bitwise minimum (expression within the parentheses in \green{light blue}).
    Subtracting this sequence from $\textbf{1}_t(v)$ (expression within the parentheses in \blue{blue})
    thus yields the unique integer $r$ such that, for every position $i \in \N$,
    $r[i] = \textbf{if } x[i] = y[i] \textbf{ then } 1 \textbf{ else } 0$.
    Then, the target integer $z$ is obtained by scaling $r$ by $(t-1)$ and taking its bitwise minimum with~$x$.

    Let us now consider the counter machine $\mathcal{M}\coloneqq(Q,\{c_1,c_2,c_3\},q_0, F, \Delta)$. As done in the proof of~\Cref{lemma:sigma-2-with-nt}, for every state $q \in Q$
    we introduce a variable with the same name, and denote these variables collectively as~$\vec q$.
    To capture the sequence of states in an accepting run of length $m+1$, we impose the following conditions:
    \begin{enumerate}
        \item\label{it:states:p1} for every $i \in [0..m]$ exactly one $q \in Q$ is such that $q[i] = 1$, the others have a zero digit;
        \item\label{it:states:p2} $q_0[0] = 1$, and $q[j] = 1$ for some $j \in [0..m]$ and $q \in F$.
    \end{enumerate}
    Considering a further variable $w$, used to encode the power $t^m$,
    we can capture these two properties with the following formula:
    \begin{equation*}
        \mathrm{run}_{Q,F}(w, \vec q) \coloneqq \underbrace{\textstyle\sum_{q \in Q} q = \textbf{1}_t(w) \land \bigwedge_{q \neq q' \in Q}
            q \,\&_t\, q' = 0}_{\text{Item~\ref{it:states:p1}}} \land
        \underbrace{\textstyle 1 \pleq_t q_0 \land w \pleq_t\sum_{q \in F} q}_{\text{Item~\ref{it:states:p2}}}.
    \end{equation*}

    The values of the counters $c_1,c_2,c_3$ during the run are encoded
    by three variables $y_1,y_2,y_3$.
    We will restrict the digits of these variables in base $t$ to $[1..t-2]$.
    This is done for simplicity and technical reasons: avoiding the digit $t-1$
    ensures the additions we will perform do not produce a carry, while starting form $1$ allows
    us to use $0$ as a ``missing value''.
    For every $j \in \{1,2,3\}$ and $i \in [0..m]$, the $i$-th digit of $y_j$ in base $t$ is equal to \emph{$1$ plus} the value of counter $c_j$ in the $i$-th configuration of the run.
    Furthermore,
    since $y_1,y_2,y_3$ are used to encode the values of counters that untouched, increased by~$1$, or decreased by~$1$ at each step of the run, we also want to impose that the base $t$ representation of $y \in \{y_1,y_2,y_3\}$ satisfies $y[i+1]\in \{y[i]-1,y[i],y[i]+1\}$
    for every position such that for every $i\geq 0$.
    With this in mind, consider the formula:
    \begin{align}
        C(w,y,v,u_{\id}, u_{\inc}, u_{\dec})\coloneqq (\textbf{1}_t(w)\subseteq_t y) \land (y\subseteq_t (t-2)\cdot\textbf{1}_t(w)) \land
        (y = u_{\id} + u_{\inc} + u_{\dec} + v) \,\land\label{eq:undec-counter-1}
        \\
        (v \leq t - 2) \,\land (u_{\id} \,\&_t\, u_{\inc} = 0) \,\land
        (u_{\id} \,\&_t\, u_{\dec} = 0) \,\land
        (u_{\inc} \,\&_t\, u_{\dec} = 0) \,\land
        \label{eq:undec-counter-2}
        \\
        u_{\id}  = \mathrm{eq}_t (y , (t\cdot y)) \,\land
        u_{\inc} = \mathrm{eq}_t (y , (t\cdot (y + \textbf{1}_t(w)))) \,\land
        u_{\dec} = \mathrm{eq}_t (y , (t\cdot (y - \textbf{1}_t(w)))).\:\:
        \label{eq:undec-counter-4}
    \end{align}
    In this formula, $w$ again encodes the power $t^m$.
    The first two conjuncts in~\eqref{eq:undec-counter-1} ensure that the digits of $y$ are in $[1..t-2]$, while the last conjunct decomposes $y$ into four distinct parts, stored in the free variables $v$, $u_{\id}$, $u_{\inc}$, and $u_{\dec}$. Because of~\eqref{eq:undec-counter-2}
    and~\eqref{eq:undec-counter-4}, this decomposition is disjoint: for every~$i\in[0..m]$, exactly one of the four variables has a non-zero digit in position $i$.
    Indeed, observe that (among other things) the conjuncts in~\eqref{eq:undec-counter-4}
    ensure that the least-significant digits of $u_{\id}$, $u_{\inc}$, and $u_{\dec}$ are all equal to zero. Then, the conjunctions in~\eqref{eq:undec-counter-2} ensure that $v$
    has a single digit in base $t$ (it is in fact the least significant digit of $y$),
    and that the other three variables have no overlapping non-zero digits.
    Looking at $u_{\id}  = \mathrm{eq}_t (y , (t\cdot y))$,
    we see that $u_{\id}$ has a non-zero digit in position $i \geq 1$ if and only if $y[i] = y[i-1]$, and in this case $u_{\id}[i] = y[i]$.
    Similarly, $u_{\inc} = \mathrm{eq}_t (y , (t\cdot (y + \textbf{1}_t(w))))$
    and $u_{\dec} = \mathrm{eq}_t (y , (t\cdot (y - \textbf{1}_t(w))))$ ensure that $u_{\inc}$ and $u_{\dec}$ have a non-zero digit in position $i \geq 1$ if and only if $y[i] = y[i-1]+1$ and $y[i] = y[i-1]-1$, respectively, and in these cases $u_{\inc}[i] = y[i]$ and $u_{\dec}[i] = y[i]$.
    Note that, as each digit of $y$ are in $[1..t-2]$,
    the addition and subtraction of $\textbf{1}_t(w)$ in these formulas do not produce any carry.

    \begin{example}
        Suppose that the values of a counter $c$ follows the following progression during a run:
        $2,1,0,0,0,1,2,1,1,0$. This sequence will be represented
        in the base $t\geq 5$ encoding using the non-negative integer $y=(\green{1}\orange{2}\green{2}\blue{32}\orange{11}
            \green{1}\green{2}3)_t$.
        The formula $C(w,y,v,u_{\id}, u_{\inc}, u_{\dec})$ forces $v$ to be equal to the least significant digit of $y$, which is $3$ in this example. This corresponds to the initial value of the counter~$c$, plus $1$. The subformula $\textbf{1}_t(w) \pleq_t y$ imposes
        $\textbf{1}_t(w)= (1111111111)_t$, so $w=t^{9}$ (the run has length $10$).
        Together with $v$, the remaining integers in the formula decompose $y$ as follows: $u_{\id}=(\orange{200011000})_t$, $u_{\inc}=(\blue{3200000})_t$, and $u_{\dec}=(\green{1020000120})_t$.
    \end{example}

    Lastly, we define the formula $R(x)$ that expresses the existence of an accepting run of $\mathcal{M}$ on input $x$. Denoting $\vec y \coloneqq (y_1,y_2,y_3)$ and $\vec u \coloneqq (u^1_{\id}, u^1_{\inc}, u^1_{\dec}, u^2_{\id}, u^2_{\inc}, u^2_{\dec}, u^3_{\id}, u^3_{\inc}, u^3_{\dec})$, we define
    \begin{align*}
        R(x) \coloneqq \exists w,\vec y,\vec u,\vec q \,\Big( \mathrm{run}_{Q,F}(w, \vec q)
        \land C(w,y_1,x + 1,u^1_{\id}, u^1_{\inc}, u^1_{\dec}) \land
        \bigwedge_{i\in\{2,3\}}C(w,y_i,1,u^i_{\id}, u^i_{\inc}, u^i_{\dec})
        \\
        {} \land \bigwedge_{i\in\{1,2,3\}}(y_i < 2\cdot w) \land \hspace{-0.5cm} \bigwedge_{(q,f_1,f_2,f_3)\in Q\times \{\id,\inc,\dec\}^3} \hspace{-0.5cm} (t\cdot q) \,\&_t\, u^1_{f_1} \,\&_t\, u^2_{f_2} \,\&_t\,u^3_{f_3}
        \subseteq_t \hspace{-5pt}\sum_{\substack{q' \in Q \colon {} \\ (q,f_1,f_2,f_3,q') \in \Delta}}  \hspace{-5pt} q' \Big).
    \end{align*}

    We now verify that $R(x)$ correctly encodes an accepting run.

    Both the $\mathrm{run}_{Q,F}(w, \vec q)$ and $C(\dots)$ formulas use $w$ within the $\textbf{1}_t(\cdot)$ function,
    specifically in the expressions $\sum_{q \in Q} q = \textbf{1}_t(w)$
    and $(\textbf{1}_t(w)\subseteq_t y) \,\land\, (y\subseteq_t (t-2)\cdot\textbf{1}_t(w))$.
    These constraints force $y_1,y_2,y_3$ and $\sum_{q \in Q} q$ to all have the same length
    $m+1$ in base $t$, which determines that $w = t^m$.
    The length $m+1$ thus represents the number of configurations in the encoded run.

    Recall that in $C(w,y,v,u_{\id}, u_{\inc}, u_{\dec})$, the parameter $v$ captures the least significant digit of $y$.
    In $R(x)$, we instantiate $C$ with $v = x+1$ for the first counter and $v = 1$ for the remaining counters,
    thereby initializing the three counters to $(x,0,0)$ for the initial configuration of the run.
    The formula $C$ also decomposes each subsequent digit of $y_i$ based on whether the counter
    remains unchanged ($u^i_{\id}$), increments ($u^i_{\inc}$), or decrements ($u^i_{\dec}$).
    For the run to be accepting, the final configuration must have all counters at zero.
    Since counter values are shifted by $+1$ in our encoding, this means each $y_i$
    must have most significant digit equal to $1$, which is equivalent to requiring $y_i < 2\cdot w$.

    The formula $\mathrm{run}_{Q,F}(w, \vec q)$ ensures that the encoded run visits exactly one state at each step,
    starting from the initial state $q_0$ and ending in a final state $q \in F$.
    The remaining task is to encode the transition relation~$\Delta$, which is done by the final conjunction in $R(x)$.

    At each step $i \in [0..m]$, exactly one state variable $q \in Q$ satisfies $q[i] = 1$.
    Similarly, for the transition from step $i$ to step $i+1$ (where $i \in [0..m-1]$), exactly one triple
    $(f_1,f_2,f_3) \in \{\id,\inc,\dec\}^3$ of counter operations is applied, meaning
    $u^1_{f_1}[i+1]$, $u^2_{f_2}[i+1]$, and $u^3_{f_3}[i+1]$ are all non-zero.
    To verify that the transition from step $i$ to step $i+1$ respects $\Delta$,
    we use multiplication by $t$ to shift the state information: since $(t \cdot q)[i+1] = q[i]$,
    the term $(t \cdot q) \,\&_t\, u^1_{f_1} \,\&_t\, u^2_{f_2} \,\&_t\,u^3_{f_3}$
    isolates exactly those positions $i+1$ where state $q$ is active at step $i$
    and operations $(f_1,f_2,f_3)$ are applied from step $i$ to step $i+1$.
    For the run to be valid, the successor state at step $i+1$ must be reachable via this quadruple.
    The condition $(t \cdot q) \,\&_t\, u^1_{f_1} \,\&_t\, u^2_{f_2} \,\&_t\,u^3_{f_3}
        {}\subseteq_t{} \sum_{q' \in Q \colon (q,f_1,f_2,f_3,q') \in \Delta}  q'$
    enforces this: wherever the left-hand side has a non-zero digit (indicating that
    state $q$ and operations $(f_1,f_2,f_3)$ are active), the right-hand side must also have
    a non-zero digit, indicating that at least one valid successor state $q' \in Q$
    exists according to $\Delta$.
\end{proof}

\Cref{theorem:counters} and the undecidability of the non-emptiness problem for counter machines imply:
\begin{restatable}{corollary}{CorrollaryExistenceUndecidable}\label{corollary:existence-undecidable}
    The \textsc{Existence} problem for the $\exists^*$ fragment of $\Formulas_{\PBAMA}$ is undecidable.
\end{restatable}

With~\Cref{theorem:bb-par-decidability}, the following corollary
implies~\Cref{theorem:mul} from~\Cref{subsec:contributions}.

\begin{corollary}\label{corollary:ea-undec}%
    The \textsc{Existence} problem for the $\exists^*\forall^*$ fragment of $\Formulas_{\PBAM}$ is undecidable.
\end{corollary}
\begin{proof}
    Because of~\Cref{corollary:existence-undecidable},
    it suffices to show how to express any existential formula of $\Formulas_{\PBAMA}$ as a formula in the $\exists^*\forall^*$ fragment of $\Formulas_{\PBAM}$.

    Let $\exists \vec x \, \phi(\vec x)$ be an existential formula in $\Formulas_{\PBAMA}$,
    with $\phi$ quantifier-free.
    Applying the function
    \[
        \Psi \coloneqq \psi \mapsto
        \begin{cases}
            \psi                                                                           & \!\text{if $\psi$ does not contain $\&_t$,}                                       \\
            \exists z \, \Big( z = \tau \&_t \rho \land \Psi(\psi[z/\tau \&_t \rho]) \Big) & \!\text{if $\psi$ contains $\tau \&_t \rho$, with $\tau,\rho$ terms from~$\PBAM$}
        \end{cases}
    \]
    on $\phi$, we translate $\exists \vec x \, \phi(\vec x)$ into an equivalent formula $\exists \vec x \Psi(\phi)$
    in which $\&_t$ occurs only in atomic formulas of the form $z = \tau \&_t \rho$, and never in the scope of a negation.
    To prove the corollary it thus suffices
    to express $z = x \&_t y$ using a formula in~$\Formulas_{\PBAM}$ with only universal quantifiers.

    In our definition, we use the quantifier-free formula $\mathrm{split}_t(x,p,p',p'',x_1,x_2,x_3)$ from \Cref{lemma:digit-t}.
    Recall that, assuming that $p$ and $p'$ are powers of $t$ such that $p'=t\cdot p$, the variable
    $x$ is uniquely decomposed into $x_1 < p$, $x_2=d\cdot p$ for some digit $d\in[0..t-1]$, and $x_3$ is divisible by $p''\geq p'$.
    It remains to compare the digits in $x,y$, and $z$ using universal quantifiers.
    \begin{align*}
        z=x\&_t y \iff \forall p,\vec p, \vec x,\vec y,\vec z\,\Big( \mathrm{split}_t(x,p,t\cdot p,p_x,\vec x)
        \land \mathrm{split}_t(y,p,t\cdot p,p_y,\vec y) \land \mathrm{split}_t(z,p,t\cdot p,p_z,\vec z)
        \\
        \implies (z_2 = x_2 \land x_2 \leq y_2) \lor (z_2 = y_2 \land y_2 \leq x_2)
        \Big),
    \end{align*}
    where $\vec p\coloneqq(p_x,p_y,p_z)$, and the triples of variables in the decomposition of $x,y,z$ are gathered
    into $\vec x \coloneqq (x_1,x_2,x_3),\vec y\coloneqq (y_1,y_2,y_3)$, and $\vec z\coloneqq (z_1,z_2,z_3)$.
    The second variable in each triple is responsible for the digit base $t$ on the
    $i$-th position for $p=t^i$, and we take the minimum of $x_2$~and~$y_2$. \qedhere

\end{proof}

\section{Practical considerations}\label{sec:tool}
\suppressfloats[t]

Much effort has been put into designing and implementing optimized translations to automata
(see, e.g.,~\cite{Boigelot21,OeiMSH21,HabermehlHHHL24,NicolF26}), and into
understanding which arithmetical and combinatorial problems can be solved with B\"uchi
arithmetic~\cite{Shallit22,HieronymiMOSSS24,PopoliSS24}. This section is similar in spirit.
In~\Cref{subsec:better-inequalities} we optimize
the translation of linear inequalities from~\Cref{subsec:from-formulas-to-automata}.
In~\Cref{subsec:smt-formalization,subsec:experiments} we express arithmetical problems from
the literature as bounded existence and universality problems
for quantifier-free \PBA, and describe our experiments with a prototype tool on these problems.
The goal of~\Cref{subsec:experiments} is not to put forward a new tool,
but to give practical information that helps developers of established tools assess whether our techniques can be useful to them.

\subsection{Translating inequalities}\label{subsec:better-inequalities}

Translating inequalities ${\textstyle\sum_{i=1}^n a_i \cdot x_i} \leq b$ by combining the ``base cases'' as in~\Cref{subsec:from-formulas-to-automata} is extremely inefficient in practice.
For B\"uchi arithmetic in base $k$, Algorithms~1 and~2 from~\cite{Boigelot21} translate equations and inequalities into automata with $O(k \cdot \log \abs{b} \cdot \norm{\vec a}_1)$ states, where $\vec a = (a_1,\dots,a_n)$ and $\norm{\vec a}_1 \coloneqq \sum_{i=1}^n \abs{a_i}$.
For equations over $\N$,~\cite[Proposition~5]{GuepinH019} gives a base-independent translation with $O(\max(\norm{\vec a}_1,\abs{b}))$ states.

We present a variant of the algorithm from~\cite{Boigelot21} for parametric symbolic automata,
together with a novel complexity analysis, showing that inequalities can be translated
into parametric symbolic automata with $O(\sqrt{|b| \cdot \norm{\vec a}_1} + \norm{\vec a}_1)$ states.
Asymptotically, this bound is better than in~\cite{GuepinH019},
and (in addition to being base-independent) it improves on~\cite{Boigelot21} by a logarithmic factor whenever $\abs{b} \leq \norm{\vec a}_1$.
The pseudocode is given in~\Cref{algo:nsa-inequality}.%

As a preliminary step, the algorithm normalizes the coefficients by dividing them by their gcd~$g$ (lines~\ref{algo:ineq:line:g}--\ref{algo:ineq:line:divide-g}), using the equivalence
$\sum_{i=1}^n a_i \cdot x_i \leq b \iff \sum_{i=1}^n (a_i/g) \cdot x_i \leq \floor{b/g}$.
This step, absent from~\cite{Boigelot21}, simplifies both the definition of the state set~$Q$ and the complexity analysis.

\begin{algorithm}[t]
    \begin{algorithmic}[1]
        \Require An inequality $\phi \coloneqq \sum_{i=1}^n a_i \cdot x_i \leq b$ with $a_i, b \in \Z$
        \Ensure A valid NSA $\Aut \in \AutPAt$ such that $\LangMSD(\Aut) = \sem{\phi}$
        \setstretch{1.05}
        \State\textbf{let} $g \coloneqq \gcd(a_1,\dots,a_n)$\label{algo:ineq:line:g}
        \State $(a_1,\dots,a_n,b) \gets (a_1/g,\dots,a_n/g,\floor{b/g})$\label{algo:ineq:line:divide-g}
        \State\textbf{let} $a_+ \coloneqq \sum_{a_i > 0} a_i$ and $a_- \coloneqq \sum_{a_i < 0} a_i$
        \State\textbf{let} $Q \coloneqq \{\floor{b/j} + \ell : j \in [1..\abs{b}+1], \ell \in [-a_+..{-a_-}] \}$ and $F \coloneqq \{ b\}$\label{algo:ineq:line:states}
        \State\textbf{let} $q_0$ be a fresh state, and let the transition function $\delta$ be defined as follows:\label{algo:ineq:line:delta}
        \Statex\vspace{3pt}
        \begin{center}
            $\begin{aligned}
                    (v_1,v_2) & \mapsto (v_1 \cdot t \leq (v_2 - \textstyle\sum_{i=1}^n a_i \cdot x_i) < (v_1 + 1) \cdot t) &  & \text{for all } v_1 \in Q, v_2 \in Q \\
                    (q_0,v)   & \mapsto \textstyle\sum_{i=1}^n a_i \cdot x_i \geq v \cdot (1-t)                             &  & \text{for all } v \in Q              \\
                    (q,q_0)   & \mapsto \bot                                                                                &  & \text{for all } q \in Q \cup \{q_0\}
                \end{aligned}$
        \end{center}\vspace{3pt}
        \State\textbf{return} $(\Img(\delta), Q \cup \{q_0\}, q_0, F, \delta)$
    \end{algorithmic}
    \caption{Builds the parametric symbolic automaton of an inequality.}\label{algo:nsa-inequality}
\end{algorithm}

The idea behind the subsequent steps is to identify each non-initial state of the resulting automaton with an integer~$v$ corresponding to the right-hand side
of the inequality $\sum_{i=1}^n a_i \cdot \intrep{\vec w}_k(x_i) \leq v$.
More precisely, we seek for the following semantics for $v$:
for any base $k \in \N_{\geq 2}$, a word $\vec w\in \Mk$ (recall: $\Mk \coloneqq {[X \to \{0,k-1\}] \cdot [X \to [0..k-1]]^*}$)
reaches~$v$ from the initial state
if and only if $\sum_{i=1}^n a_i \cdot \intrep{\vec w}_k(x_i) \leq v$.
A priori, it is not clear why such a set of integers $v$ should be finite and independent of the base $k$.
Both properties stem from the following observation.

Assume for simplicity $b \geq 0$, and define the integers
$a_+ \coloneqq \sum_{a_i > 0} a_i$ and $a_- \coloneqq \sum_{a_i < 0} a_i$.
Take a word $\vec w = w_\ell \dots w_0 \in \Mk$ with $\ell \geq 1$ and let $\vec w' = w_{\ell} \dots w_1$.
By definition of $\intrep{\cdot}_k$, we have
\begin{equation}\label{eq:automaton-step}
    \textstyle\sum_{i=1}^n a_i \cdot \intrep{\vec w}_k(x_i) \leq b
    \ \text{ if and only if } \
    \sum_{i=1}^n a_i \cdot \intrep{\vec w'}_k(x_i) \leq \floor{\frac{b - \sum_{i=1}^n a_i \cdot w_0(x_i)}{k}}.
\end{equation}
Since the image $\Img(w_0)$ is included in $[0..k-1]$, we see that
\[
    a_- \cdot (k-1) \leq \textstyle\sum_{i=1}^n a_i \cdot w_{0}(x_i) \leq a_+ \cdot (k-1).
\]
Together with $0 \leq \frac{b}{k} \leq b$, this means that the
floor division in the right-hand side $\floor{\frac{b - \sum_{i=1}^n a_i \cdot w_0(x_i)}{k}}$
``transforms'' $b$ into an integer in the interval $[-a_+..b-a_-]$.
This interval does not depend on the base~$k$, on the length $\ell$, and on the map $w_\ell$.
Hence, to recognize the inequality $\sum_{i=1}^n a_i \cdot \intrep{\vec w}_k(x_i) \leq b$, it is sufficient for the automaton to track only
integers from this finite set, and take $F = \{b\}$ as the only accepting state.
The set of states $Q$ defined in line~\ref{algo:ineq:line:states} of~\Cref{algo:nsa-inequality} follows this idea, but is slightly more refined
in order to obtain a better bound on the number of states of the automaton.

\begin{restatable}{lemma}{LemmaClosurePropertiesQ}\label{lemma:closure-properties-Q}
    Let~$Q$ be the set defined in line~\ref{algo:ineq:line:states} of~\Cref{algo:nsa-inequality},
    with respect to the integers~$a_1,\dots,a_n,b$ in line~\ref{algo:ineq:line:divide-g}.
    Then,
    (1) \ $b \in Q$, and
    (2) \ $\floor{\frac{v-\sum_{i=1}^n a_i \cdot w(x_i)}{k}} \in Q$
    for all~$v \in Q$, $k \in \N_{\geq 2}$ and $w \colon X \to [0..k-1]$.
\end{restatable}

Let us now examine the transitions.
Since states in~$Q$ encode right-hand sides of inequalities,
the transitions between them follow directly from~\eqref{eq:automaton-step}:
a pair $(v_1,v_2) \in Q \times Q$ must be labelled by a formula equivalent to
$v_1 = \floor{\frac{v_2 - \sum_{i=1}^n a_i \cdot x_i}{t}}$.
Line~\ref{algo:ineq:line:delta} provides exactly this formula,
expressed as the pair of linear inequalities
$v_1 \cdot t \leq v_2 - \sum_{i=1}^n a_i \cdot x_i < (v_1 + 1) \cdot t$.

It remains to discuss the transitions from the initial state~$q_0 \notin Q$
(the reverse transitions, to~$q_0$, are all labelled~$\bot$).
These handle single-letter words $w \in [X \to \{0,k-1\}]$; recall that our goal in this algorithm is to
express the fact that $w$ reach $v \in Q$ from $q_0$ if and only if $\sum_{i=1}^n a_i \cdot \intrep{w}_k(x_i) \leq v$.
By definition of the map $\intrep{\cdot}_k$,
$\intrep{w}_k(x) = 0$ when $w(x)=0$ and $\intrep{w}_k(x) = -1$ when $w(x)=k-1$,
hence $\intrep{w}_k(x) = \frac{w(x)}{1-k}$ for all $x \in X$.
Thus, $\sum_{i=1}^n a_i \cdot \intrep{w}_k(x_i) \leq v$
rewrites as $\sum_{i=1}^n a_i \cdot w(x_i) \geq v \cdot (1-k)$,
which is precisely the formula labelling the pair $(q_0,v)$ in line~\ref{algo:ineq:line:delta}.

\begin{restatable}{lemma}{LemmaNSAInequalityCorrectness}\label{lemma:nsa-inequality-correctness}
    \Cref{algo:nsa-inequality} respects its specification.
\end{restatable}

\begin{proof}[Proof idea]
    Consider an inequality $\phi \coloneqq \sum_{i=1}^n a_i \cdot x_i \leq b$,
    assumed to satisfy~$\gcd(a_1,\dots,a_n) = 1$
    thanks to lines~\ref{algo:ineq:line:g} and~\ref{algo:ineq:line:divide-g} of~\Cref{algo:nsa-inequality}.
    Let $\Aut = (\Phi, Q \cup \{q_0\}, q_0, F, \delta)$ be the automaton for~$\phi$ constructed by~\Cref{algo:nsa-inequality}.
    To prove the lemma, one shows that for every $k \in \N_{\geq 2}$ we have $\LangMSD(\Aut)(k) = \sem{\phi}_k$,
    and that $\Aut$ is a valid NSA.
    For this, it suffices to consider $k \in \N_{\geq 2}$,
    a word $\vec w \coloneqq w_\ell \dots w_0 \in \Mk$,
    and $v \in Q \subseteq \Z$.
    A simple induction on $\ell$ shows that $\sum_{i=1}^n a_i \cdot \intrep{\vec w}_k(x_i) \leq v$
    if and only if
    there is a path $q_0\dots q_{\ell+1}$ such that $q_{\ell+1} = v$ and $w_{\ell+1-i} \models_k \delta(q_{i-1},q_{i})$ for every $i \in [1..\ell+1]$.
    As $b \in F$ (line~\ref{algo:ineq:line:states}), taking $v = b$ proves the lemma.
\end{proof}

Complexity-wise, to show that~\Cref{algo:nsa-inequality} computes an NSA with $O(\sqrt{|b| \cdot \norm{\vec a}_1} + \norm{\vec a}_1)$ states, it suffices to bound
the size of the set $Q$ from line~\ref{algo:ineq:line:states}. The following lemma provides such a bound.

\begin{lemma}\label{lemma:inequality-bound-Q}
    For any integers $a,b,c$ with $a \leq c$, the number of integers in the Minkowski sum
    $[a..c] + \{\floor{b/k} : k \in [1..\abs{b}+1]\}$ is at most~$2 \cdot \sqrt{\abs{b} \cdot (c-a+1)} + c - a + 1$.
\end{lemma}

\begin{proof}
    Recall that Minkowski sum $A+B$ of two sets $A,B\subseteq\Z$ is just $\{a+b \,:\, a\in A \text{ and } b\in B \}$.
    For simplicity, assume $b \geq 0$; the case $b < 0$ is symmetric.
    The key observation is that as $k$ grows, the intervals $I_k \coloneqq [a..c] + \floor{b/k}$
    begin to overlap. This is already visible at the tail:
    $I_{b} \cup I_{b+1} = ([a..c] + 1) \cup [a..c] = [a..c+1]$.
    The idea is then to find an integer $k_0$ such that
    $\bigcup_{k=k_0}^{b+1} I_k$ is, informally, nearly a single interval, whereas $I_1,\dots,I_{k_0-1}$ are distinct enough.
    We take $k_0 \coloneqq \sqrt{\frac{b}{c-a+1}}$ as a ``good enough'' cut-off point and finish the proof as follows.

    First, consider $k \geq k_0$. The set $\{\floor{b/k} : k \in [\ceil{k_0}..b+1]\}$ is included in the real interval $[0,b/k_0] = [0,\sqrt{b \cdot (c-a+1)}]$.
    Therefore, $[a..c] + \{\floor{b/k} : k \in [\ceil{k_0}..b+1]\} \subseteq [a,c+\sqrt{b \cdot (c-a+1)}]$,
    and thus this part of the sum contributes at most $\sqrt{b \cdot (c-a+1)} + c - a + 1$ integer points.

    Consider now $k < k_0$. The integers in $\{\floor{b/k} : k \in [1..\ceil{k_0}-1]\}$ are all ``sparse'' but very few, and so a crude estimate suffices.
    The set $[a..c] + \{\floor{b/k} : k \in [1..\ceil{k_0}-1]\}$ has at most $k_0 \cdot (c-a+1) \leq \sqrt{b \cdot (c-a+1)}$ integer points.

    Hence, $[a..c] + \{\floor{b/k} : k \in [1..b+1]\}$ has size at most~$2 \cdot \sqrt{b \cdot (c-a+1)} + c - a + 1$.
\end{proof}

\begin{remark}
    Modifying~\Cref{algo:nsa-inequality} so that it translates equalities $\sum_{i=1}^n a_i \cdot x_i = b$
    into valid NSAs is straightforward.
    First, if $b$ is not divisible by $\gcd(a_1,\dots,a_n)$, we have the algorithm return an automaton for the empty language.
    Otherwise, it suffices to update the transition function as follows:
    \begin{center}
        $\begin{aligned}
                (v_1,v_2) & \mapsto v_1 \cdot t  = v_2 - \textstyle\sum_{i=1}^n a_i \cdot x_i &  & \text{for all } v_1 \in Q, v_2 \in Q \\
                (q_0,v)   & \mapsto \textstyle\sum_{i=1}^n a_i \cdot x_i = v \cdot (1-t)      &  & \text{for all } v \in Q
            \end{aligned}$
    \end{center}
    Correctness follows by minor adjustments to the proof of~\Cref{lemma:nsa-inequality-correctness},
    since the equality analogue of the double implication in~\eqref{eq:automaton-step} is simply
    $\sum_{i=1}^n a_i \cdot \intrep{\vec w}_k(x_i) = b \Longleftrightarrow
        \sum_{i=1}^n a_i \cdot \intrep{\vec w'}_k(x_i) = \frac{b - \sum_{i=1}^n a_i \cdot w_0(x_i)}{k}$.
\end{remark}

\subsection{SMT formalization of bounded arithmetical problems}\label{subsec:smt-formalization}

Parametric B\"uchi arithmetic enables simultaneous reasoning about multiple bases, allowing us to
express many natural arithmetical problems. In this section, we present two examples of such
problems
and formalize them as the \textsc{Bounded Existence} and
\textsc{Bounded Universality} problems. For an input formula $\phi$ and $\bmin, \bmax \in \mathbb{N}_{\geq 2}$,
the question is whether $[\bmin..\bmax] \cap \realbases(\phi) \neq \emptyset$ (resp.~$[\bmin..\bmax] \subseteq \realbases(\phi)$).

Towards our implementation (\Cref{subsec:experiments}), we formalize these problems
in the syntax of our tool.
This is the SMT-LIB syntax~\cite{BarrettFT16} for quantifier-free formulas over integers and strings,
enriched with ${\texttt{int.v}: \textbf{Int} \times \textbf{Int} \to \textbf{Bool}}$
(corresponding to~$V_t$).
Variables are restricted to the sorts \textbf{Int} and \textbf{String},
and one can use the function $\texttt{str.to\_int}: \textbf{String} \to \textbf{Int}$
for string-to-integer conversions (in base~$t$).
The logic features the standard arithmetic operations and comparisons on integers
from QF\_LIA, plus operations to build regular expressions (\textbf{RegEx}) over the alphabet $\{0,1\}$,
as for instance {\tt str.to\_re} to construct a regular expression from a 0-1 string, {\tt re.++} for concatenation, and {\tt re.*} for Kleene star.
Membership of a variable of sort \textbf{String} in a regular expression can be expressed using the predicate $\texttt{str.in\_re}: \textbf{String} \times \textbf{RegEx} \to \textbf{Bool}$.

\paragraph{Bounded local divisibilities.}
Our first example, already mentioned in~\Cref{subsec:limitations},
comes from the algorithms to decide the existential fragment of LIA with divisibility constraints~\cite{Beltyukov76,Lipshitz78}. Without entering into the details of these algorithms,
it suffices to say that they perform several manipulations of the original formula,
until reaching a system of inequalities
\begin{equation}\label{eq:bounded-local-divisibility}
    \bigwedge\nolimits_{i=1}^n v_p(f_i(\vec x)) \leq v_p(g_i(\vec x)) \land f_i(\vec x) \neq 0 \land g_i(\vec x) \neq 0,
\end{equation}
where $p$ is a prime number, $f_i$ and $g_i$ are linear polynomials with integer coefficients, $\vec x$ are integer variables,
and $v_p$ is the $p$-adic valuation $v_p(n) \coloneqq \max\{k \in \mathbb{N} : p^k \text{ divides } n\}$
(with ${v_p(0) \coloneqq +\infty}$).
The algorithm guarantees that the original formula from LIA with divisibility constraints is satisfiable if and only if this system has a solution for all primes $p$ in some range $[\bmin..\bmax]$,
and that one can equivalently look at all integers in this range instead of just the primes.
This is a \textsc{Bounded Universality} problem,
which we call the \emph{bounded local divisibilities problem}.

Since $V_p(n,d) \iff d = p^{v_p(n)}$ for all non-zero $n,d \in \mathbb{Z}$,
the inequality $v_p(f_i(\vec x)) \leq v_p(g_i(\vec x))$ in System~\eqref{eq:bounded-local-divisibility} can alternatively be expressed using $V_p$.
It is then straightforward to express System~\eqref{eq:bounded-local-divisibility} in the syntax of our tool. For example, for $f(x,y) \coloneqq 3 \cdot x - 6 \cdot y + 1$ and $g(x,y) \coloneqq x - 1$, the formula $v_p(f) \leq v_p(g) \land f \neq 0 \land g \neq 0$ can be expressed as
\begin{alltt}
    \textcolor{vscBracketA}{(}and \textcolor{vscBracketB}{(}int.v \textcolor{vscBracketC}{(}+ \textcolor{vscBracketA}{(}* 3 x\textcolor{vscBracketA}{)} \textcolor{vscBracketA}{(}- \textcolor{vscBracketB}{(}* 6 y\textcolor{vscBracketB}{)}\textcolor{vscBracketA}{)} 1\textcolor{vscBracketC}{)} z\textcolor{vscBracketB}{)} \textcolor{vscBracketB}{(}int.v \textcolor{vscBracketC}{(}- x 1\textcolor{vscBracketC}{)} w\textcolor{vscBracketB}{)} \textcolor{vscBracketB}{(}<= z w\textcolor{vscBracketB}{)}\textcolor{vscBracketA}{)}
\end{alltt}
where {\tt z} and {\tt w} are auxiliary integer variables.
The constraints $f \neq 0$ and $g \neq 0$ are implied by the fact that $V_t(n,d)$ does not hold for $n = 0$.

\paragraph{Bounded modular polynomial roots.}
Our second example is a classic problem from number theory: given a polynomial $f(t)$ with integer coefficients, a modulus $m \in \mathbb{N}_{\geq 2}$, and two integers $\bmin, \bmax \in \mathbb{Z}$,
does there exist an integer $t \in [\bmin..\bmax]$ such that $f(t) \equiv 0 \mod m$? Since we are working modulo $m$, $\bmin$ and $\bmax$ can be shifted so that they are both positive, and $\bmin \geq 2$.
This problem is \np-complete already for $f$ being a quadratic polynomial~\cite{MandersA76}.

For example, for $f(t) = 31 \cdot t^2 + 17 \cdot t + 48$ and $m = 96$, we write $f(t) \equiv 0 \mod m$ as follows:
\begin{alltt}
    \textcolor{vscBracketA}{(}and \textcolor{vscBracketB}{(}str.in_re t1 \textcolor{vscBracketC}{(}re.++ \textcolor{vscBracketA}{(}re.* \textcolor{vscBracketB}{(}str.to_re "0"\textcolor{vscBracketB}{)}\textcolor{vscBracketA}{)} \textcolor{vscBracketA}{(}str.to_re "10"\textcolor{vscBracketA}{)}\textcolor{vscBracketC}{)}\textcolor{vscBracketB}{)}
         \textcolor{vscBracketB}{(}str.in_re t2 \textcolor{vscBracketC}{(}re.++ \textcolor{vscBracketA}{(}re.* \textcolor{vscBracketB}{(}str.to_re "0"\textcolor{vscBracketB}{)}\textcolor{vscBracketA}{)} \textcolor{vscBracketA}{(}str.to_re "100"\textcolor{vscBracketA}{)}\textcolor{vscBracketC}{)}\textcolor{vscBracketB}{)}
         \textcolor{vscBracketB}{(}= \textcolor{vscBracketC}{(}+ \textcolor{vscBracketA}{(}* 31 \textcolor{vscBracketB}{(}str.to_int t2\textcolor{vscBracketB}{)}\textcolor{vscBracketA}{)} \textcolor{vscBracketA}{(}+ \textcolor{vscBracketB}{(}* 17 \textcolor{vscBracketC}{(}str.to_int t1\textcolor{vscBracketC}{)}\textcolor{vscBracketB}{)} 48\textcolor{vscBracketA}{)}\textcolor{vscBracketC}{)} \textcolor{vscBracketC}{(}* 96 q\textcolor{vscBracketC}{)}\textcolor{vscBracketB}{)}\textcolor{vscBracketA}{)}
\end{alltt}
where {\tt t1}, {\tt t2} and {\tt q} are auxiliary variables.
Notably, we use the base $t$ of parametric B\"uchi arithmetic as the variable of the polynomial $f$.
The formula encodes this by requiring {\tt t1} and {\tt t2} to have base $t$ representations $0^*10$ and $0^*100$ respectively, which correspond to $t$ and $t^2$.
The variable {\tt q} is used to express the quotient of the division of $f(t)$ by $m$.

\begin{remark}
    Of course, the algorithms tailored to the problems above would perform much better than reducing the problems
    to \PBA and using a solver for it.
    The point of this section is to show that parametric B\"uchi arithmetic gives a general framework
    to express natural arithmetical problems.
\end{remark}

\subsection{Experiments on bounded existence and universality problems}\label{subsec:experiments}

Solving \textsc{Bounded Existence} and \textsc{Bounded Universality} problems does not strictly require parametric B\"uchi arithmetic; standard B\"uchi arithmetic
suffices by simply iterating through the bases in the range $[\bmin..\bmax]$. This raises a pragmatic question: \emph{Is there any benefit to using the parametric approach for bounded problems?}
It is a priori unclear whether the answer is positive or negative.
On the one hand, a parametric automaton must encode the behavior of the formula across all bases,
making it larger than any single-base finite automaton. On the other hand, parametric automaton
is constructed only once for all bases in $[\bmin..\bmax]$. We investigate this question with our tool.

\vspace{1em}
\textit{Our tool} is a prototype implementation of the algorithms presented in this paper. Written in OCaml, it relies on the {\tt smt.ml} (version 0.30.0)
backend~\cite{PereiraMAAACCMSS26} to call the SMT solver {\tt Z3}~\cite{MouraB08} (version 4.16.0)
for solving the satisfiability of Presburger arithmetic formulas with digits, and to extract the realizable bases.
Input formulas follow the syntax described in~\Cref{subsec:smt-formalization}.
The formula is preprocessed in a few ways, including putting it in negation normal form and replacing negated inequalities
$\lnot (f \leq g)$ by $g +1 \leq f$, to avoid performing complementation on non-atomic formulas.
The solver has two operating modes:
\begin{itemize}
    \item \parmode (short for \emph{parametric mode}): This mode uses the parametric symbolic automata described in the paper.
          It constructs a single automaton bottom-up for a given formula.
    \item \symmode (short for \emph{symbolic automata mode}): This mode uses symbolic automata for fixed bases:
          edge labels are Presburger formulas, but they feature the fixed base instead of the parameter~$t$.
          On bounded problems, this mode iteratively constructs automata for each
          fixed $k\in[\bmin..\bmax]$, e.g.~using a symbolic variant of~\cite[Algorithms~1 and~2]{Boigelot21} for (in)equalities.
\end{itemize}

In both modes, to prevent excessive memory consumption, we employ a heuristic that estimates the number of states in the bottom-up construction of the resulting
automaton for a Boolean combination of atomic constraints.
If it exceeds 500\,000 states (this threshold is configurable via the {\tt -active} option), instead of flattening these combinations into a single automaton,
the solver perform an on-the-fly / lazy product construction similar to~\cite[Algorithm~9]{Esparza2023}. In both active and lazy modes, we use depth-first search, and
for every current state $q\in Q$, combine all labels of the outgoing transitions into a disjunction of \PAt formulas, supplement the disjunction with the digit
semantics for each digit variable, and employ {\tt Z3} to extract a successor of the state~$q$, if the formula is satisfiable. The process terminates as soon
as a path to an accepting state is found (or enough paths are found, in the case of \textsc{Bounded Universality}). We omit further details.

\paragraph{The benchmarks.} We generated 120 instances: 30 for bounded local divisibilities, 30 for bounded modular polynomial roots,
and 60 random ones. Apart from the ``shape'' of the formulas in the first two categories, all instances are random.
Generative AI (Anthropic's Claude Opus~4.7 and~5.5) was used to help prepare these benchmarks, but not to develop the prototype.
Here are some statistics:
\begin{itemize}
    \item Bounded local divisibilities: 1--3 comparisons of $V_t$, 3--6 variables, integers in $[-15..15]$.
    \item Bounded modular polynomial roots: degree 1--4, modulus ${m \in [19..1024]}$ (other integers are reduced modulo $m$ by the tool).
    \item Random instances: 3--6 variables, 1--3 arithmetic constraints, 1--3 regular constraints, 1--2 occurrences of the {\tt int.v} predicate, integers in $[-20..20]$.
\end{itemize}
As for the bases, we consider three intervals: $[2..20]$, $[21..200]$ and $[201..220]$.
These intervals are chosen to perform three tests.
(i) The first interval tests whether \symmode mode outperforms \parmode mode on small intervals, where constructing a few small automata could be faster than constructing a single parametric automaton.
(ii) The second interval tests whether \parmode mode becomes faster for larger ranges.
(iii) The third interval examines how base size affects performance in small ranges.

\begin{figure}[t]
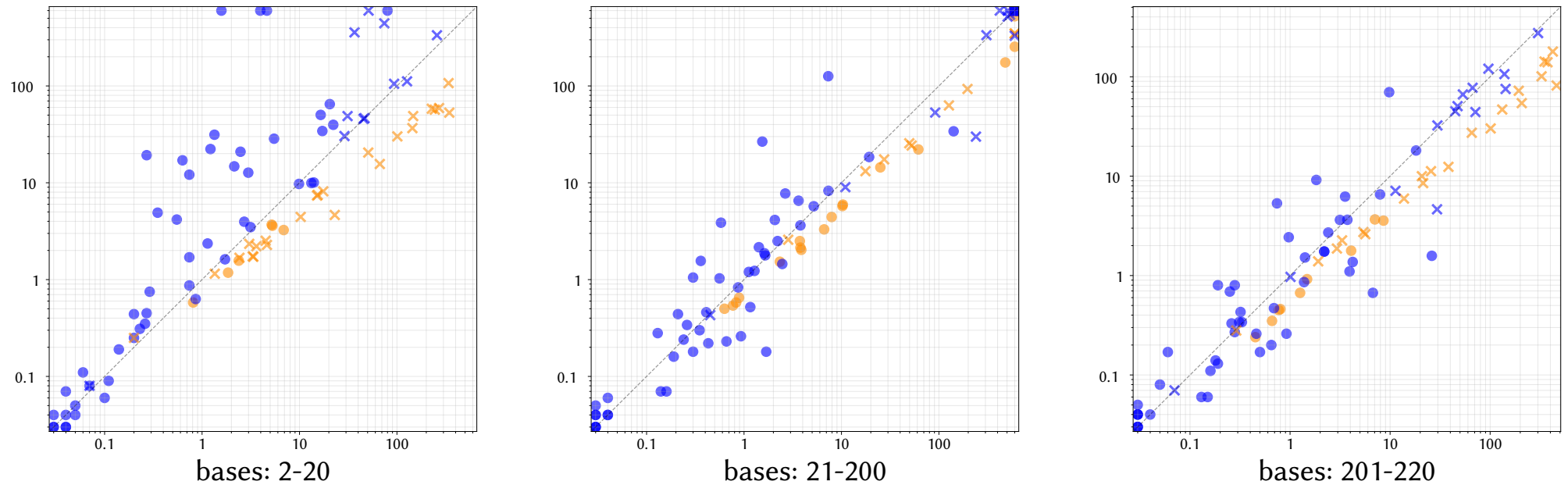
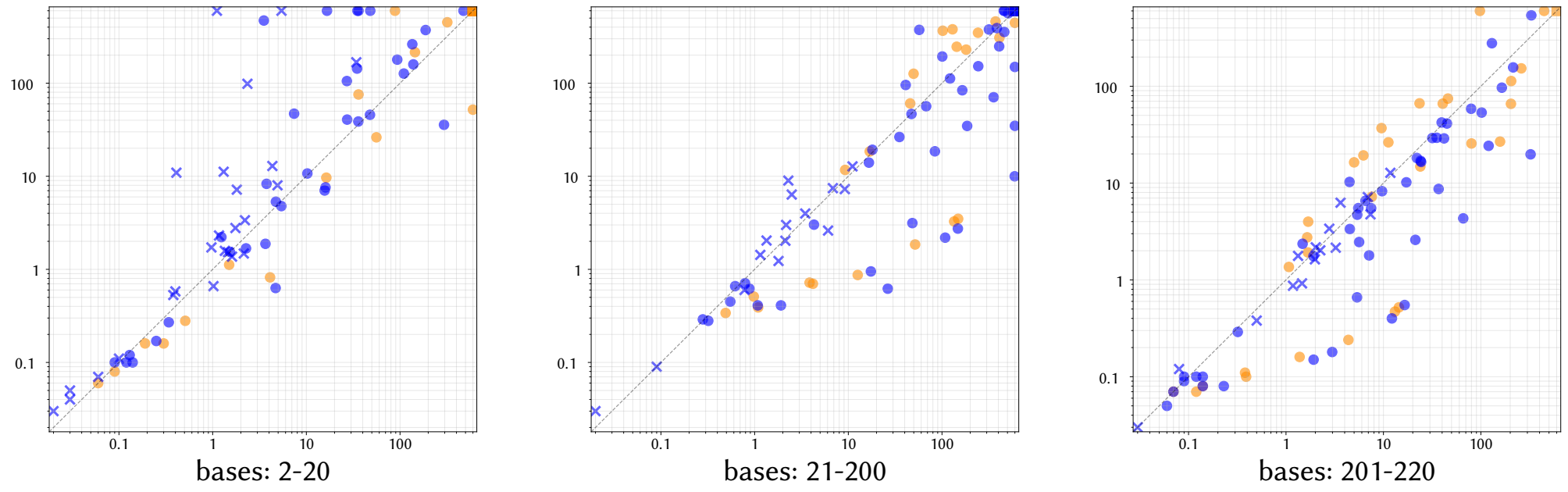

    \begin{subfigure}[b]{\textwidth}
        \begin{minipage}{0.33\textwidth}\centering
            \scalebox{0.23}{\input{plots/scatter-2-20-existence.pgf}}
        \end{minipage}%
        \begin{minipage}{0.33\textwidth}\centering
            \scalebox{0.23}{\input{plots/scatter-21-200-existence.pgf}}
        \end{minipage}%
        \begin{minipage}{0.33\textwidth}\centering
            \scalebox{0.23}{\input{plots/scatter-201-220-existence.pgf}}
        \end{minipage}%
        \caption{\textsc{Bounded Existence}: {\color{blue}random instances} and {\color{orange}instances of bounded modular polynomial roots}.}\label{figure:scatter-existence}
    \end{subfigure}
    \vspace{-0.5em}

    \begin{subfigure}[b]{\textwidth}
        \begin{minipage}{0.33\textwidth}\centering
            \scalebox{0.23}{\input{plots/scatter-2-20-universality.pgf}}
        \end{minipage}%
        \begin{minipage}{0.33\textwidth}\centering
            \scalebox{0.23}{\input{plots/scatter-21-200-universality.pgf}}
        \end{minipage}%
        \begin{minipage}{0.33\textwidth}\centering
            \scalebox{0.23}{\input{plots/scatter-201-220-universality.pgf}}
        \end{minipage}%
        \caption{\textsc{Bounded Universality}: {\color{blue}random instances} and {\color{orange}instances of bounded local divisibilities}.}\label{figure:scatter-universality}
    \end{subfigure}
    \vspace{-1.5em}
    \caption{%
        Scatter plots for \textsc{Bounded Existence} and \textsc{Bounded Universality} problems.
        Circles represent SAT instances, while crosses represent UNSAT instances.
        The $x$- and $y$-axes show the execution time in seconds for the \symmode and \parmode modes of the solver, respectively.
        Plots are in logscale; the limits are $600$\,s and 16\,GiB.}\label{figure:scatters}
\end{figure}

\paragraph{Results.}
\Cref{figure:scatters} shows the results of our benchmarks. Regarding the three tests above, they seem to indicate that
(i) \symmode mode is faster on intervals $[\bmin..\bmax]$ where $\bmax$ is small,
(ii) \parmode mode is faster on large intervals, and
(iii) the performance of \symmode mode degrades for large bases, with \parmode becoming faster on $[\bmin..\bmax]$ with $\bmin$ large.
For example, on \textsc{Bounded Existence} with bases in $[201..220]$,
\parmode solves the 90 instances in 1\,997 seconds in total, against 3\,942 seconds for \symmode mode. Moreover, \parmode mode is consistently faster than \symmode mode on the bounded modular polynomial root problem, also for bases in $[2..20]$, although most instances were UNSAT, which in retrospect is expected when the range of bases is small relative to the modulus. This might indicate that for instances that require exploring (almost) the whole range of bases, i.e.~negative instances for \textsc{Bounded Existence} and positive ones for \textsc{Bounded Universality}, \parmode mode is more efficient.

\section{Conclusion}

We have introduced parametric symbolic automata that allow reasoning about integers simultaneously in different, potentially infinitely many
positional encodings. Its arithmetical counterpart $\Formulas_\PBA$ provides a language that completely captures the expressive power of
parametric symbolic automata. Using this language, we have proved decidability of the existence and universality problems for the standard B\"uchi arithmetic,
and have shown that natural arithmetical problems featuring non-linear constraints
can be modeled as parametric symbolic automata.

Several avenues towards applications remain open.
On the practical side, \Cref{sec:tool} suggests that automata-based provers for B\"uchi arithmetic,
such as Walnut~\cite{Mousavi16,Shallit22} and Pecan~\cite{OeiMSH21},
could benefit from parametric symbolic automata on problems that must be solved for many bases, or for a single large one.
On the theoretical side, our results show that, for regular languages, one can quantify over the size of the alphabet,
and that parametric symbolic automata capture this added expressive power.
For applications such as program verification, WS1S is often a more natural viewpoint than B\"uchi arithmetic (see~\Cref{sec:introduction}).
A parametric extension of WS1S, capturing the same sets as parametric B\"uchi arithmetic,
is a natural question to explore.

\begin{acks}
    We thank Georgii Belianin for contributing to the implementation of the prototype of~\Cref{sec:tool}.
    Alessio Mansutti and Jakub Kaszycki are co-funded by the European Union (GA 101154447) and by MCIN/AEI (GA CEX2024-001471-M). Mikhail Starchak is funded by the European Research Council (ERC) under the HORIZON ERC Synergy Grant DynAMiCs (GA 101167561).
    Views and opinions expressed are those of the authors only and do not necessarily reflect those of the European Union, the European Commission,
    or the European Research Council Executive Agency. Neither the European Union nor the granting authority can be held responsible for them.
\end{acks}

\subsection*{Disclosure of Use of AI in Research}
All results, proofs and exposition in this manuscript are entirely the work of the authors.
Generative AI was used only for the preparation and packaging of the research artifact.
Specifically, Anthropic's Claude Opus~4.7 and, later, Claude Opus~5.5 were used to help prepare the benchmarks of~\Cref{subsec:experiments}, but not to develop the prototype.
These models were also used to structure and refine the wording of the repository documentation (the \texttt{README.md} file) archived on Zenodo.
The authors reviewed and verified all AI-generated benchmarks and documentation edits, and they take full responsibility for the correctness and integrity of the software and data.

\subsection*{Data Availability Statement}
The artifact of this paper~\cite{KaszyckiMS26artifact} is archived on Zenodo.
It contains the prototype of~\Cref{sec:tool} (release v0.2.0, as an executable and as source code), the benchmarks of~\Cref{subsec:experiments},
the logs of all runs of~\Cref{subsec:experiments}, a script that repeats these runs, and the Jupyter notebook that produces~\Cref{figure:scatters}.
To reproduce the results of~\Cref{subsec:experiments}, use the version of the artifact cited above (POPL-2027-AEC-final).
The version evaluated by the POPL 2027 Artifact Evaluation Committee
(POPL-2027-AEC-submission, DOI \href{https://doi.org/10.5281/zenodo.23312247}{10.5281/zenodo.23312247})
contains the same prototype, but the logs of an earlier run of the experiments, made with release v0.1.0.
For reuse, the prototype is developed at \href{https://github.com/mklstarch/Parsym}{github.com/mklstarch/Parsym}.

\bibliographystyle{ACM-Reference-Format}
\bibliography{bibliography}

\newpage
\appendix
\section{Proofs of statements in Section~\ref{subsec:from-formulas-to-automata}}\label{app:proofs-NSA}

We recall the statements:

\LemmaMakeNSAValid*

\begin{proof}
	Let \( \mathcal{A}=(\Phi,Q,q_0,F,\delta) \), let \( \mathcal{A}' \) be the output of~\Cref{algo:nsa-validity} on \( \mathcal{A} \), and fix \( k\in\N_{\geq 2} \).
	Recall that \( X_{\mathcal{A}}=\{x_1,\dots,x_m\} \) (line~1 of the algorithm).
	The free variables of the formulas \( \psi \) defined in line~\ref{line:fresh-state-q} are \( x_1,\dots,x_m \), so \( X_{\mathcal{A}'}=X_{\mathcal{A}} \).
	Hence, encodings for \( \mathcal{A} \) and for \( \mathcal{A}' \) are the same.
	We show that \( \mathcal{A}' \) accepts every encoding of every map in \( \LangMSD(\mathcal{A})(k) \), and that every map encoded by a word in \( \Lang_k(\mathcal{A}') \) belongs to \( \LangMSD(\mathcal{A})(k) \).
	These two properties imply that \( \LangMSD(\mathcal{A}')(k)=\LangMSD(\mathcal{A})(k) \), and that \( \mathcal{A}' \) is valid.

	Below, two letters \( a,b\colon X\to[0..k-1] \) \emph{agree} if \( a(x)=b(x) \) for every \( x\in X_{\mathcal{A}} \).
	Letters that agree satisfy the same labels of \( \mathcal{A} \) and of \( \mathcal{A}' \).

	\emph{Encodings of the same map.}
	Let \( \vec w\in\Mk \), and let \( s \) be its first letter.
	Write \( \vec w=\vec p\,\vec w' \), where \( \vec p \) is the longest prefix of \( \vec w \) whose letters all agree with~\( s \).
	(The word \( \vec w' \) may be empty.)
	Consider \( \vec v\in\Mk \) such that \( \intrep{\vec v}_k(x)=\intrep{\vec w}_k(x) \) for every \( x\in X_{\mathcal{A}} \).
	We claim that \( \vec v=\vec p''\vec w'' \), where all letters of the non-empty word \( \vec p'' \) agree with \( s \).
	Moreover, \( \vec w'' \) has length \( \abs{\vec w'} \), and its \( i \)-th letter agrees with the \( i \)-th letter of \( \vec w' \), for every \( i \).
	Indeed, consider \( x\in X_{\mathcal{A}} \).
	The \( x \)-tracks of \( \vec v \) and \( \vec w \) encode the same integer, and every integer has at most one encoding of each length.
	So, by~\Cref{fact:msdf-padding}, the longer of the two tracks is obtained from the shorter one by adding \( \abs{\abs{\vec v}-\abs{\vec w}} \) leading copies of the sign digit \( s(x) \).
	Since this number of copies does not depend on \( x \), the claim follows.

	\emph{\( \mathcal{A}' \) accepts every encoding of a map in \( \LangMSD(\mathcal{A})(k) \).}
	Let \( \vec w\in\Lang_k(\mathcal{A}) \), with first letter \( s \), and write \( \vec w=\vec p\,\vec w' \) as above.
	Fix an accepting run of \( \mathcal{A} \) on \( \vec w \), and let \( q_0q_1\dots q_N \) be the states it visits while reading \( \vec p \), so that \( N=\abs{\vec p}\geq 1 \).
	While \( q_1\dots q_N \) contains a cycle of states \( q\dots q \), we replace this cycle with \( q \).
	We obtain a path \( q_0q'_1\dots q'_n \) with \( n\geq 1 \), \( q'_n=q_N \), and \( q'_1,\dots,q'_n \) pairwise distinct.
	Every two consecutive states of this path are also consecutive in the run.
	So, the label of the transition between them is satisfied by a letter of \( \vec p \), and thus by every letter that agrees with \( s \).
	Let \( \sigma \) map \( x_i \) to \( 0 \) if \( s(x_i)=0 \), and to \( t-1 \) otherwise.
	Consider the formula \( \psi \) and the state \( q'' \) defined in line~\ref{line:fresh-state-q} for \( \sigma \) and the path \( q_0q'_1\dots q'_n \).
	For \( t=k \), every letter that agrees with \( s \) satisfies \( \psi \).
	Now, let \( \vec v \) be an encoding of a map encoded by \( \vec w \), and write \( \vec v=\vec p''\vec w'' \) as above.
	The automaton \( \mathcal{A}' \) reads \( \vec p'' \) along the states \( q_0'q'_n \) if \( \abs{\vec p''}=1 \), and along \( q_0'q''\dots q''q'_n \) otherwise.
	Indeed, the labels of these transitions are \( \psi \) or, for the pair \( (q_0',q'_n) \), a disjunction containing~\( \psi \).
	From \( q'_n=q_N \), the automaton \( \mathcal{A}' \) reads \( \vec w'' \) by following the run of \( \mathcal{A} \) on \( \vec w' \), which ends in~\( F \).
	Therefore, \( \vec v\in\Lang_k(\mathcal{A}') \).

	\emph{Every map encoded by a word in \( \Lang_k(\mathcal{A}') \) is in \( \LangMSD(\mathcal{A})(k) \).}
	Let \( \vec v\in\Lang_k(\mathcal{A}') \), with first letter \( s \), and fix an accepting run of \( \mathcal{A}' \) on \( \vec v \).
	The state \( q_0' \) has no incoming transitions, and no state of \( Q \) has a transition towards a state of \( Q' \).
	Hence, the run first reads a non-empty prefix \( \vec p'' \) of \( \vec v \) through states of \( Q' \), and then reaches a state \( q_n\in Q \).
	By construction, there are a map \( \sigma \) and a path \( q_0q_1\dots q_n \) with \( n\geq 1 \) such that all letters of \( \vec p'' \) satisfy the formula \( \psi \) defined in line~\ref{line:fresh-state-q} for \( \sigma \) and this path.
	Write \( \vec v=\vec p''\vec w'' \).
	Since all letters of \( \vec p'' \) satisfy \( \psi \), they agree with \( s \), and \( s \) satisfies \( \delta(q_i,q_{i+1}) \) for every \( i\in[0..n-1] \).
	Therefore, \( \mathcal{A} \) accepts the word \( \vec u\coloneqq s^n\,\vec w'' \): it reads \( s^n \) along \( q_0q_1\dots q_n \), and then \( \vec w'' \) as \( \mathcal{A}' \) does from~\( q_n \).
	Since \( n\geq 1 \), the word \( \vec u \) starts with \( s \), and so \( \vec u\in\Mk \).
	(This is where the condition \( n\geq 1 \) of line~\ref{line:paths} is needed.)
	For every \( x\in X_{\mathcal{A}} \), the \( x \)-tracks of \( \vec u \) and \( \vec v \) only differ in the number (at least one) of their leading copies of~\( s(x) \).
	By~\Cref{fact:msdf-padding}, they encode the same integer.
	We conclude that every map encoded by \( \vec v \) is encoded by \( \vec u \in \Lang_k(\mathcal{A}) \), and thus belongs to \( \LangMSD(\mathcal{A})(k) \).
\end{proof}

\LemmaValidSFAClosure*

\begin{proof}[Proof of determinization.]
	Suppose that \( \mathcal{A}=\left(\Phi, Q, q_0, F, \delta\right) \) is an NSA.
	To find an equivalent DSA, we will use the powerset construction, with special handling for transitions.
	For a set of formulae \( \Phi \), by \( \mathrm{BC}(\Phi) \) denote the set of Boolean combinations of \( \Phi \), which is infinite when \( \Phi \) is nonempty.
	Namely, let \( \mathcal{A}'=\left(\Phi', Q', q_0', F', \delta'\right) \), where
	\begin{itemize}
		\item \( \Phi' := \delta'(Q'\times Q') \subseteq_{\mathrm{fin}} \mathrm{BC}(\Phi) \),
		\item \( Q' := \powerset(Q) \),
		\item \( q_0' := \left\{q_0\right\} \),
		\item \( F' := \left\{A\subseteq Q\mid A\cap F\ne \varnothing\right\} \), and
		\item \( \delta'(A, B) := \bigwedge_{q'\in Q}\left(q'\in B\iff \bigvee_{q\in A} \delta(q, q')\right) \).
	\end{itemize}
	Note that, since \( B \) is known, the \( \iff \) operator used in the definition of \( \delta' \) can be elided, yielding a~formula in \( \Formulas_{\PAt} \).
	The elements \( Q' \), \( q_0' \) and \( F' \) are exactly the same as in the powerset construction for finite automata.
	The function \( \delta' \) follows these semantics by establishing the conditions when the set \( A\subseteq Q \) of states transitions exactly to the set \( B\subseteq Q \) of states.
	It must be the case that \( B \) is exactly the set of states, to which at least one state in \( A \) transitions under the currently read symbol.

	To prove that \( \mathcal{A}' \) is indeed deterministic, we proceed by induction on run length.
	The inductive thesis is that \( \mathcal{A}' \) has exactly one run for any word.
	For the empty word, this is trivially true.
	To see the inductive step, suppose that \( \mathcal{A}' \) is currently in state \( A\in Q' \), and the symbol \( a\in\sigma \) is currently being read.
	To see that there is at least one state, note that, for each state \( q' \), each term of form \( \bigvee_{q\in A} \delta(q, q') \) is either true or false for \( a \).
	We can pick a subset \( B\subseteq Q \) such that \( q'\in B \) exactly when \( a\models_k \bigvee_{q\in A} \delta(q, q') \).

	To see that there is at most one state, let \( B_1,B_2\in Q' \) be two states it transitions into.
	Therefore, the current symbol \( a\in\sigma \) satisfies both \( \delta'(A, B_1) \) and \( \delta'(A, B_2) \).
	We can equivalently express this as a conjunction
	\[
		a\models_k \delta'(A, B_1)\land\delta'(A, B_2)\text{.}
	\]
	Because conjunction is associative and commutative, we can substitute the definition of \( \delta' \) and rearrange to obtain
	\[
		a\models_k \bigwedge_{q'\in Q}\left[\left(q'\in B_1 \iff \bigvee_{q\in A}\delta(q, q')\right) \land \left(q'\in B_2 \iff \bigvee_{q\in A}\delta(q, q')\right)\right]\text{.}
	\]
	We can then decompose this conjunction, obtaining, for every \( q'\in Q \),
	\[
		a\models_k \left(q'\in B_1 \iff \bigvee_{q\in A}\delta(q, q')\right) \land \left(q'\in B_2 \iff \bigvee_{q\in A}\delta(q, q')\right)\text{.}
	\]
	It follows that
	\[
		a\models_k \left(q'\in B_1 \iff q'\in B_2\right)\text{,}
	\]
	and therefore the sets \( B_1 \) and \( B_2 \) are equal by extensionality.

	The language equivalence of \( \mathcal{A} \) and \( \mathcal{A}' \) follows by another induction on run length.
	The inductive thesis is as follows: a run of \( \mathcal{A} \) on a word to state \( q\in Q \) exists if and only if the only run of \( \mathcal{A}' \) on that word ends up in a state \( A \) that contains the state \( q \) when understood as a subset of \( Q \)
	It is easy to see that for the empty word this is true.
	Suppose \( \mathcal{A}' \) is in the state \( A\in Q' \) and transitions to \( B\in Q' \) by reading \( a\in\sigma \) after having previously read \( w\in\sigma^* \).
	By the inductive assumption, \( A \) is exactly the set of states which runs on the word without the currently read symbol can reach.
	The very definition of \( \delta' \) requires that the set of states after a transition is exactly the set of states that have a transition from at least one of states in \( A \).
	Because every run over \( wa \) to \( q \) splits into a~run over \( w \) to some \( q' \) and a transition from \( q' \) to \( q \) over \( a \), the inductive step is established.
	From the inductive thesis and the definition of \( F' \), \( \mathcal{A}' \) accepts a word exactly when \( \mathcal{A} \) does.
	Note that \( X_{\mathcal{A}'}=X_{\mathcal{A}} \), since every label \( \delta(q,q') \) occurs in the labels \( \delta'(\{q\},B) \).
	Validity of \( \mathcal{A}' \) follows from validity of \( \mathcal{A} \), since these automata accept the same \emph{encodings}, not just number tuples.
\end{proof}

\begin{proof}[Proof of intersection.]
	To establish intersection, we again proceed analogously to finite automata, this time using the product construction.
	Given \( \mathcal{A}_1 \) and \( \mathcal{A}_2 \), where \( \mathcal{A}_i=\left(\Phi_i, Q_i, q_0^i, F_i, \delta_i\right) \), we construct \( \mathcal{A}'=\left(\Phi', Q', q_0', F', \delta'\right) \) as
	\begin{itemize}
		\item \( \Phi' := \Phi_1 \land \Phi_2 \),
		\item \( Q' := Q_1\times Q_2 \),
		\item \( q_0' := \left<q_0^1, q_0^2\right>\in Q' \),
		\item \( F' := F_1 \times F_2 \subseteq Q' \), and
		\item \( \delta'\!\left(\left<q_1, q_2\right>, \left<q_1', q_2'\right>\right) := \delta_1(q_1, q_1') \land \delta_2(q_2, q_2') \).
	\end{itemize}

	The correctness of this construction again relies on induction on run length.
	This time, the inductive thesis is simpler: a run of \( \mathcal{A}' \) to the state \( \left<q_1, q_2\right> \) exists if and only if a run of \( \mathcal{A}_1 \) to the state \( q_1 \) and a run of \( \mathcal{A}_2 \) to the state \( q_2 \) both exist.
	The base case is immediate.
	Similarly, a run can be extended by a letter if and only if the runs of the two automata can be extended by that letter, by the definition of \( \delta' \).
	By the inductive thesis and the definition of \( F' \), \( \mathcal{A}' \) accepts a word exactly when both \( \mathcal{A}_1 \) and \( \mathcal{A}_2 \) do.
	Every label of \( \mathcal{A}' \) is the conjunction of a label of \( \mathcal{A}_1 \) and a label of \( \mathcal{A}_2 \), and every label of \( \mathcal{A}_i \) occurs in such a conjunction.
	So, \( X_{\mathcal{A}'}=X_{\mathcal{A}_1}\cup X_{\mathcal{A}_2} \), and an encoding of a map \( \mu \) for \( \mathcal{A}' \) is also an encoding of \( \mu \) for \( \mathcal{A}_1 \) and for \( \mathcal{A}_2 \).
	It follows that \( \LangMSD(\mathcal{A}')(k)\subseteq\LangMSD(\mathcal{A}_1)(k)\cap\LangMSD(\mathcal{A}_2)(k) \).
	Conversely, let \( \mu \) be in this intersection.
	Since \( \mathcal{A}_1 \) and \( \mathcal{A}_2 \) are valid, they both accept every encoding of \( \mu \) for \( \mathcal{A}' \), and so does \( \mathcal{A}' \).
	This proves the reverse inclusion, as well as the validity of \( \mathcal{A}' \).
\end{proof}

\begin{proof}[Proof of complement.]
	Recall that \( \Mk \), defined in~\Cref{subsec:encodings}, is the language of valid encodings of number tuples.
	Let \( \mathcal{B}=\left(\Phi_{\mathcal{B}}, Q_{\mathcal{B}}, q_0^{\mathcal{B}}, F_{\mathcal{B}}, \delta_{\mathcal{B}}\right) \) be the DSA obtained from \( \mathcal{A} \) by the determinization construction above.
	Recall that \( \mathcal{B} \) has exactly one run on every word.
	(Determinism alone does not suffice: it only guarantees at most one run, and swapping the accepting states would then not complement the language.)
	Now let \( \mathcal{B}':=\left(\Phi_{\mathcal{B}}, Q_{\mathcal{B}}, q_0^{\mathcal{B}}, Q_{\mathcal{B}}\setminus F_{\mathcal{B}}, \delta_{\mathcal{B}}\right) \).
	Since \( \mathcal{B} \) has exactly one run on every word, it follows that
	\[
		\Lang_k(\mathcal{B}')=\Mk\setminus \Lang_k(\mathcal{B}) = \Mk\setminus \Lang_k(\mathcal{A})\text{.}
	\]
	Moreover, \( X_{\mathcal{B}'}=X_{\mathcal{A}} \), and since \( \mathcal{A} \) is valid, either all encodings of a map \( \mu \) belong to \( \Lang_k(\mathcal{A}) \) (when \( \mu\in\LangMSD(\mathcal{A})(k) \)) or none of them does.
	As every map has an encoding (see~\Cref{subsec:encodings}), \( \LangMSD(\mathcal{B}')(k)=[X\to\Z]\setminus\LangMSD(\mathcal{A})(k) \), and \( \mathcal{B}' \) is valid.
	The automaton \( \mathcal{B}' \) is also a DSA, therefore an NSA.
	Let \( \mathcal{C} \) be the NSA recognizing the language \( \Mk \), shown in \Cref{fig:nsa-valid}.
	We can take the automaton recognizing the intersection of the languages recognized by \( \mathcal{B}' \) and \( \mathcal{C} \) as the requested NSA \( \mathcal{A}' \).
	\begin{figure}[t]
		\centering
		\begin{tikzpicture}[node distance = 1cm]
			\scriptsize
			\node (0) [state, initial left, scale=0.8] {$0$};
			\node (1) [state, accepting, right = 3cm of 0, scale=0.8] {$1$};
			\path[-stealth]
			(0) edge[blue] node[above]{$\bigwedge_{x\in X_{\mathcal{A}}}\left[x = 0 \lor x = t - 1\right]$} (1)
			(1) edge[loop right] node[above=2pt]{$\top$} (1);
		\end{tikzpicture}%
		\caption{NSA for the language \Mk .}
		\label{fig:nsa-valid}
	\end{figure}
\end{proof}

\LemmaProjectionNSA*

\begin{proof}
	Note that \( X_{\exists_x(\mathcal{A})}=X_{\mathcal{A}}\setminus\{x\} \), by definition of \( \exists_x \).
	Suppose that \( m\in\LangMSD(\exists_x(\mathcal{A}))(k) \).
	Therefore, there exists some encoding \( w \) of \( m \) accepted by \( \exists_x(\mathcal{A}) \).
	Fix an accepting run of the automaton.
	For each \( i\in\left\{1,\dots,\abs{w}\right\} \), we can look at the acceptance condition on the transition taken when reading \( w_i \).
	There is some digit \( j_i \in [0..k-1] \) found by the existential quantifier.
	Since the first transition of the run leaves the initial state \( q_\star \), we have \( j_1 \in \{0,k-1\} \).
	Let \( w'_i=w_i[x\mapsto j_i] \).
	Then \( w'=w'_1\dots w'_{\abs{w}} \) belongs to \( \Mk \), and it is accepted by \( \mathcal{A} \) (replace \( q_\star \) with \( q_0 \) in the run).
	This word encodes the map \( m[x\mapsto j] \), where \( j=\intrep{j_1\dots j_{\abs{w}}}_k \).

	Suppose that \( m[x\mapsto j]\in\LangMSD(\mathcal{A})(k) \), and let \( w\in\Lang_k(\mathcal{A}) \) be an encoding of \( m[x\mapsto j] \).
	Fix an accepting run of \( \mathcal{A} \) on \( w \).
	A run of \( \exists_x(\mathcal{A}) \) on \( w \) follows this run, starting in \( q_\star \) instead of \( q_0 \), and guesses the digit of \( x \) in the \( i \)-th letter to be \( w_i(x) \).
	For \( i = 1 \), this digit is a sign digit, since \( w \in \Mk \).
	Since \( x\notin X_{\exists_x(\mathcal{A})} \), the word \( w \) is an encoding of \( m \) for \( \exists_x(\mathcal{A}) \).
	Therefore \( m\in\LangMSD(\exists_x(\mathcal{A}))(k) \).
\end{proof}

\LemmaFromFormulasToAutomata*

\begin{proof}
	The proof proceeds by structural induction on \( \phi \).
	First, as described in~\Cref{subsec:from-formulas-to-automata}, one can represent any atomic formula with a formula only containing atoms of the form:
	\begin{itemize}
		\item \( x = 0 \),
		\item \( x = 1 \),
		\item \( x \leq y \),
		\item \( V_t(x, y) \), and
		\item \( x + y = z \).
	\end{itemize}
	This is at the expense of adding new quantified variables.
	However, because the newly added variables are obtained using functional relations, one may introduce them using either existential or universal quantifier (i.e. they are definitional quantifiers), as not to increase the alternation rank of the formula.
	For example, the formula \( x \leq 2 \) is equivalent to
	\[
		\exists y \exists z \colon \left(x\leq z\right) \land \left(y + y = z\right) \land \left(y = 1\right)\text{,}
	\]
	as well as
	\[
		\forall y \forall z \colon \left[\left(y + y = z\right) \land \left(y = 1\right)\right]\implies \left(x\leq z\right)\text{.}
	\]

	With that in mind, we can proceed with the induction.
	The base cases have already been outlined above, so we can proceed with the inductive steps.
	Note that the base cases are all valid NSA, and the induction shall preserve that property.
	\begin{itemize}
		\item
		      For a formula of form \( \phi \land \psi \), let \( \mathcal{A}_1 \) and \( \mathcal{A}_2 \) be NSA such that, by the inductive assumption, \( \LangMSD(\mathcal{A}_1)=\sem{\phi} \) and \( \LangMSD(\mathcal{A}_2)=\sem{\psi} \).
		      By~\Cref{lemma:valid-sfa-closure}, we can construct a valid intersection NSA \( \mathcal{A}' \) such that
		      \[
			      \LangMSD(\mathcal{A}')=\LangMSD(\mathcal{A}_1)\cap\LangMSD(\mathcal{A}_2)=\sem{\phi}\cap\sem{\psi}=\sem{\phi\land\psi}\text{.}
		      \]
		\item
		      For a formula of form \( \lnot \phi \), let \( \mathcal{A} \) be the (valid) NSA such that \( \LangMSD(\mathcal{A})=\sem{\phi} \).
		      By~\Cref{lemma:valid-sfa-closure}, we can construct a valid complement NSA \( \mathcal{A}' \).
		      Then, for every \( k\in\N_{\geq 2} \),
		      \[
			      \LangMSD(\mathcal{A}')(k)=[X\to\Z]\setminus\LangMSD(\mathcal{A})(k)=[X\to\Z]\setminus\sem{\phi}_k=\sem{\lnot\phi}_k\text{.}
		      \]
		\item
		      For a formula of form \( \phi \lor \psi \), use De Morgan's Laws,
		      \[
			      \phi \lor \psi \equiv \lnot\left[\left(\lnot \phi\right) \land \left(\lnot \psi\right)\right]\text{.}
		      \]
		\item
		      For a formula of form \( \exists x\colon \phi \), let \( \mathcal{A} \) be the NSA such that \( \LangMSD(\mathcal{A})=\sem{\phi} \).
		      By~\Cref{lemma:projection-nsa}, we can construct a projection NSA \( \mathcal{A}' \) with respect to the variable \( x \).
		      Let \( \mathcal{A}'' \) be a valid NSA recognizing all valid msdf encodings of the number tuples that have any encoding recognized by \( \mathcal{A}' \), constructed using \Cref{algo:nsa-validity}.
		      Then, for every \( k\in\N_{\geq 2} \),
		      \[
			      \LangMSD(\mathcal{A}'')(k)=\pi_x\left(\LangMSD(\mathcal{A})(k)\right)=\pi_x\left(\sem{\phi}_k\right)=\sem{\exists x\colon \phi}_k\text{,}
		      \]
		      where \( \pi_x(S)\coloneqq\{m\in[X\to\Z] : m[x\mapsto j]\in S \text{ for some } j\in\Z\} \) for \( S\subseteq[X\to\Z] \).
		      The first equality follows from \( \LangMSD(\mathcal{A}'')=\LangMSD(\mathcal{A}') \) (\Cref{lemma:make-NSA-valid}) and from~\Cref{lemma:projection-nsa}.
		      Therefore
		      \[
			      \LangMSD(\mathcal{A}'')=\sem{\exists x\colon \phi}\text{.}
		      \]
		\item
		      For a formula of form \( \forall x\colon \phi \), use De Morgan's Laws,
		      \[
			      \forall x\colon \phi \equiv \lnot \exists x\colon \lnot \phi\text{.}
		      \]
	\end{itemize}
	This concludes the proof, and establishes the effectiveness of the procedure.
\end{proof}

\begin{restatable}{proposition}{PropositionFToACxity}
	The running time of this procedure can be bounded from above by a function related to \( \mathrm{TOWER}(n) \), defined as
	\[
		\mathrm{TOWER}(n)=\underbrace{2^{2^{\iddots^{2}}}}_{\clap{n\text{ copies of }2}}\text{,}
	\]
	with \( n \) being the representation length of the input formula.

	For a formula in prenex form, this running time is strictly related to the alternation rank.
	Restricted to formulae with a fixed number of alternations, a variant of this procedure works in elementary time.

	However, there is no substantially faster procedure for such conversion.
	In particular, it cannot be done in elementary time for arbitrary formulae.
\end{restatable}

\begin{proof}
	For the upper bound, we measure the size of an NSA as its number of states plus the total size of its labels.
	The steps of the procedure affect the size as follows.
	Projection (\Cref{lemma:projection-nsa}) adds one state and one bounded quantifier per label, and intersection is a product construction; both are polynomial.
	Determinization, and hence complementation, is exponential.
	\Cref{algo:nsa-validity} is also exponential: it adds one state for every map \( \sigma \colon X_{\mathcal{A}} \to \{0,t-1\} \) and every path \( q_0 q_1 \dots q_n \) with \( q_1,\dots,q_n \) pairwise distinct, hence at most \( 2^{\abs{X_{\mathcal{A}}}} \cdot \abs{Q}^{\abs{Q}+1} \) states, and each new label is a conjunction of at most \( \abs{Q}+1 \) formulas.
	The procedure performs a number of such steps that is linear in \( n \), which yields the bound in terms of \( \mathrm{TOWER} \).

	For formulas in prenex form with a fixed number \( \ell \) of quantifier alternations, we modify the procedure in two ways, without changing the semantics of the resulting NSA.
	First, we apply \Cref{algo:nsa-validity} once after each maximal block of quantifiers, instead of after each projection.
	This is correct because \Cref{lemma:projection-nsa} does not require its input to be valid, so projecting the variables of a block one after the other yields an NSA for the quantified formula; \Cref{lemma:make-NSA-valid} then makes this NSA valid before the next Boolean operation, which is the only kind of step that requires validity.
	Second, we treat the quantifier-free matrix as follows.
	Since the variables introduced for the atomic formulas are defined functionally, the matrix can be written as \( \exists \vec y \, (D \land M) \), and also as \( \forall \vec y \, (D \implies M) \), where \( D \) defines \( \vec y \) and \( D \) and \( M \) are Boolean combinations of the base cases.
	We add \( \vec y \) to the innermost block of quantifiers, choosing the form matching its type.
	We then put the matrix in disjunctive normal form \( \psi_1 \lor \dots \lor \psi_r \), with \( r \leq 2^{O(n)} \), and translate it as \( \lnot(\lnot\psi_1 \land \dots \land \lnot\psi_r) \).
	Each \( \psi_i \) is a conjunction of \( O(n) \) base cases or their negations, whose NSAs have constant size, so its NSA has \( 2^{O(n)} \) states.
	Hence, the NSA for the matrix is obtained with a constant number of exponential steps.
	Each of the \( \ell+1 \) blocks then adds a constant number of exponential steps (one for \Cref{algo:nsa-validity}, and two complementations for a universal block), so the resulting NSA, and the running time of the procedure, are bounded by a tower of exponentials of height \( O(\ell) \).
	This is elementary for every fixed \( \ell \).

	For the lower bound, suppose there is some procedure \( P \) that turns formulae of \( \PBA \) into NSA.
	Let us take any sentence \( \phi \) of \( 2\BA \).
	We replace all occurrences of \( V_2 \) (of \( 2\BA \)) with \( V_t \) (of \( \PBA \)) in the sentence \( \phi \), obtaining \( \phi'\in\Formulas_\PBA \).
	We run \( P \) on the sentence \( t=2\land\phi' \), obtaining some NSA \( \mathcal{A} \).
	Next, we make the label formulae quantifier-free by~\Cref{fact:presburger-quantifier-elimination}, which can be done in triply-exponential time by~\cite{Oppen73}.
	We remove all edges that are false for \( t=2 \), and erase the labels of those that are true, obtaining an unlabelled NFA \( \mathcal{A}' \).
	This can be done in polynomial time.
	Language nonemptiness of \( \Lang(\mathcal{A}') \) is equivalent to the truth of \( \phi \).
	Checking if an NFA accepts any word is equivalent to checking reachability in a directed graph, which is decidable in nondeterministic logarithmic space due to~\cite[Theorem 21]{Jones75}, and therefore in polynomial time.

	However, the theory of \( 2\BA \) is known to be \tower-hard \cite{Meyer75,BruyereHMV94}, for the class \tower as understood in~\cite[Section 2.3]{Schmitz16}.
	Substitution of \( V_2 \), quantifier elimination, substitution of \( t=2 \) and the subsequent NFA nonemptiness check are all computable in elementary time.
	Therefore, a correct procedure \( P \) cannot have its worst-case running time bounded by an~elementary function of \( n \), or significantly better than \( \mathrm{TOWER}(n) \) at all.
\end{proof}

\section{Proofs of statements in Section~\ref{subsec:corollaries}}\label{app:proofs-decision-problems}

\LemmaAlgoRealbasesCorrectness*

\begin{proof}[Proof of correctness.]
	We will show that the following are equivalent for a base \( k\in\N_{k\geq 2} \):
	\begin{enumerate}
		\item\label{item:asdfgfa-1} \( \phi \) is satisfiable for \( t=k \),
		\item\label{item:asdfgfa-2} \( \LangMSD(\mathcal{A})(k) \) is nonempty,
		\item\label{item:asdfgfa-3} there exists a sequence \( r_0,\dots,r_n\in Q \) (including the newly created state \( \widetilde q_0 \)) such that
			\begin{itemize}
				\item \( r_0=\widetilde q_0 \),
				\item \( r_n\in F \), and
				\item for each \( i\in[1..n] \), \( k\in b(r_{i-1}, r_i) \);
			\end{itemize}
		\item\label{item:asdfgfa-4} \( k\in m(\widetilde q_0) \) after the algorithm stabilizes.
	\end{enumerate}

  \medskip
	\noindent \( \text{\eqref{item:asdfgfa-1}} \Longleftrightarrow \text{\eqref{item:asdfgfa-2}} \)\textbf{:}\ \
	Follows directly from \Cref{lemma:from-formulas-to-automata}.
	Note that the addition of \( \widetilde q_0 \) and subsequent change in transitions does not change the acceptance for valid encodings.

  \medskip
	\noindent \( \text{\eqref{item:asdfgfa-2}} \Longrightarrow \text{\eqref{item:asdfgfa-3}} \)\textbf{:}\ \
	Suppose that \( \intrep{w_l\dots w_0}_k\in\LangMSD(\mathcal{A})(k) \).
	Fix an accepting run over \( w \).
	Let \( n:=l+1 \), and let \( r_i \) for \( i\in[0..n] \) be the respective states of the accepting run after reading \( i \) symbols.
	Because it is an accepting run, it must satisfy the first two conditions.
	Furthermore, the transition from \( r_{i-1} \) to \( r_i \) for \( i\in[1..n] \) can be taken for \( w_{i-1} \) and therefore is satisfiable, so \( k\in b(r_{i-1}, r_i) \).

  \medskip
	\noindent \( \text{\eqref{item:asdfgfa-3}} \Longrightarrow \text{\eqref{item:asdfgfa-2}} \)\textbf{:}\ \
	We know that, for all \( i\in[1..n] \), \( k\in b(r_{i-1}, r_i) \), so \( \delta(r_{i-1},r_i) \) is satisfiable for \( t=k \).
	Let \( l=n-1 \).
	We choose, for each \( i\in[1..n] \), a letter \( w_{n-i}\colon X\to[0..k-1] \) that satisfies \( \delta(r_{i-1}, r_i) \) and maps every variable outside \( X_{\mathcal{A}} \) to~\( 0 \).
	This is possible since \( \delta(r_{i-1}, r_i) \) only mentions variables in \( X_{\mathcal{A}} \).
	The initial transformation of \( \mathcal{A} \) ensures that \( w_l \) maps every variable to \( 0 \) or \( k-1 \), and thus \( w\in\Mk \).
	Then \( \intrep{w_l\dots w_0}_k\in\LangMSD(\mathcal{A})(k) \).

  \medskip
	\noindent \( \text{\eqref{item:asdfgfa-3}} \Longrightarrow \text{\eqref{item:asdfgfa-4}} \)\textbf{:}\ \
	Because \( r_n \in F \), initially \( k\in m(r_n) \).
	While \( k\in m(r_i) \), and \( k\notin m(r_{i-1}) \) for any \( i\in[1..n] \), the algorithm has outstanding work, so it cannot stabilize.
	Then \( k \) becomes a member of \( m(r_i) \) by the update rule.
	In~\Cref{subsec:corollaries}, we show that the algorithm always stabilizes.
	Therefore, in the end, \( k\in m(r_0)=m(\widetilde q_0) \).

  \medskip
	\noindent \( \text{\eqref{item:asdfgfa-4}} \Longrightarrow \text{\eqref{item:asdfgfa-3}} \)\textbf{:}\ \
	We build the path \( r_0\dots r_n \) inductively, starting from \( r_0:=\widetilde q_0 \).
	We assume that \( k\in m(r_i) \), and it was first the case after \( l_i\in\N \) iterations.
	If \( l_i=0 \), it means that \( r_i\in F \), and we set \( n:=i \).
	Note that this particular case can only happen for \( i>0 \).
	Otherwise, because it was introduced in iteration \( l_i>0 \), it must be the case that \( k\in m(r_{i+1}) \) and \( k\in b(r_i, r_{i+1}) \) for some state which we can call \( r_{i+1} \).
	Also, because \( k \) already was a member of \( m(r_{i+1}) \) at that point, \( l_{i+1}<l_i \).
	We continue the construction in \( r_{i+1} \).
	Because \( l_i \) strictly decreases and takes only natural values, this procedure must terminate for some \( n \).
	The resulting sequence \( r_i \) satisfies the requirements of \eqref{item:asdfgfa-3}.
\end{proof}

\section{Proofs of statements in Section~\ref{sec:tool}}\label{app:tool}

\LemmaClosurePropertiesQ*

\begin{proof}
    The fact that $b$ belongs to $Q$ is straightforward (simply take $j = 1$ and $\ell = 0$ in line~\ref{algo:ineq:line:states}).

    Let us now prove the second item. Take $v \in Q$, $k \in \N_{\geq 2}$ and $w \colon X \to [0..k-1]$, then,
    by definition of the set $Q$, we have
    $v = \floor{\frac{b}{j}} + \ell$ for some $j \in [1..\abs{b}+1]$ and $\ell \in [-a_+..{-a_-}]$
    for the integers $a_+ \coloneqq \sum_{a_i > 0} a_i$ and $a_- \coloneqq \sum_{a_i < 0} a_i$.

    We will now establish the inequality $\floor{\frac{b}{j \cdot k}} -a_+ \leq \floor{\frac{v-\sum_{i=1}^n a_i \cdot w(x_i)}{k}} \leq \floor{\frac{b}{j \cdot k}} -a_-$,
    which implies that $\floor{\frac{v-\sum_{i=1}^n a_i \cdot w(x_i)}{k}} \in Q$
    (note: for $j \cdot k > \abs{b}+1$, we have $\floor{\frac{b}{j \cdot k}} = \floor{\frac{b}{\abs{b}+1}}$).
    Let us only prove the lower bound, because the upper bound is analogous:
    \begin{equation*}\let\displaystyle\textstyle
        \begin{aligned}
            \floor{\frac{v-\sum_{i=1}^n a_i \cdot w(x_i)}{k}}
             & = \floor{\frac{\floor{\frac{b}{j}} + \ell-\sum_{i=1}^n a_i \cdot w(x_i)}{k}}
             & \mathcmt{definition~of~$v$}                                                                                                  \\
             & = \floor{\frac{\frac{b}{j} + \ell -\sum_{i=1}^n a_i \cdot w(x_i)}{k}}
             & \mathcmt{$\floor{\frac{\floor{c}+d}{e}} = \floor{\frac{c+d}{e}}$ for $c \in \R$, $d \in \Z$, $e \in \N_{\geq 1}$}            \\
             & \geq \floor{\frac{\frac{b}{j} -a_+ - a_+ \cdot (k-1)}{k}}
             & \mathcmt{$\ell \geq -a_+$ and $\sum_{i=1}^n a_i \cdot w(x_i) \leq a_+ \cdot (k-1)$}                                          \\
             & = \floor{\frac{b}{j \cdot k}-a_+} = \floor{\frac{b}{j \cdot k}}-a_+
             & \mathcmt{$\floor{c+d} = \floor{c} + d$ for $c \in \R$ and $d \in \Z$}
             &                                                                                                                   & \qedhere
        \end{aligned}
    \end{equation*}
\end{proof}

\LemmaNSAInequalityCorrectness*

\begin{proof}
    Consider an inequality $\phi \coloneqq \sum_{i=1}^n a_i \cdot x_i \leq b$.
    Following lines~\ref{algo:ineq:line:g} and~\ref{algo:ineq:line:divide-g} of~\Cref{algo:nsa-inequality},
    we can assume~that $\gcd(a_1,\dots,a_n) = 1$.
    Let $\Aut = (\Phi, Q \cup \{q_0\}, q_0, F, \delta)$ be the automaton for~$\phi$ constructed by~\Cref{algo:nsa-inequality}.
    To prove the lemma, we are going to show that for every $k \in \N_{\geq 2}$ we have $\LangMSD(\Aut)(k) = \sem{\phi}_k$,
    and that $\Aut$ is a valid NSA.
    For this, it suffices to consider $k \in \N_{\geq 2}$,
    a~word $\vec w \coloneqq w_\ell \dots w_0 \in \Mk$,
    and $v \in Q \subseteq \Z$.
    By induction on $\ell$, we show that $\sum_{i=1}^n a_i \cdot \intrep{\vec w}_k(x_i) \leq v$
    if and only if
    there is a path $q_0\dots q_{\ell+1}$ such that $q_{\ell+1} = v$ and $w_{\ell+1-i} \models_k \delta(q_{i-1},q_{i})$ for every $i \in [1..\ell+1]$.
    As $b \in F$ (line~\ref{algo:ineq:line:states}), taking $v = b$ proves the lemma.

    \begin{description}
        \item[base case: $\ell = 0$.]
              For the base case $\vec w = w_0$, it suffices to show that
              $w_0 \models_k \delta(q_0,v)$ if and only if $\sum_{i=1}^n a_i \cdot \intrep{w_0}_k(x_i) \leq v$.
              Recall that, for all $x \in X$, either $\intrep{w_0(x)}_k = 0$ and $w_0(x) = 0$,
              or $\intrep{w_0(x)}_k = -1$ and $w_0(x) = k-1$.
              In other words, $\intrep{w_0}_k(x) = \intrep{w_0(x)}_k = \frac{w_0(x)}{1-k}$.
              Then,
              \begin{align*}
                  w_0 \models_k \delta(q_0,v)
                   & \Leftrightarrow \textstyle\sum_{i=1}^n a_i \cdot w_0(x_i) \geq v \cdot (1-k) &  & \mathcmt{definition of $\delta$ in line~\ref{algo:ineq:line:delta}} \\
                   & \Leftrightarrow \textstyle\sum_{i=1}^n a_i \cdot \frac{w_0(x_i)}{1-k} \leq v &  & \mathcmt{divide by $1-k < 0$}                                       \\
                   & \Leftrightarrow \textstyle\sum_{i=1}^n a_i \cdot \intrep{w_0}_k(x_i) \leq v  &  & \mathcmt{$\textstyle\intrep{w_0}_k(x) = \frac{w_0(x)}{1-k}$}
              \end{align*}
        \item[induction step: $\ell \geq 1$.]
              Let $\vec w' = w_{\ell} \dots w_1$. So, $\intrep{\vec w}_k(x) = \intrep{\vec w'}_k(x) \cdot k + w_0(x)$ for all~$x \in X$.

              For the right-to-left direction, suppose there is a path $q_0\dots q_{\ell+1}$ such that $q_{\ell+1} = v$ and $w_{\ell+1-i} \models_k \delta(q_{i-1},q_{i})$ for every $i \in [1..\ell+1]$. Note that $q_{\ell}$ is an integer from $Q$,
              and that $w_0 \models_k \delta(q_{\ell},q_{\ell+1})$ implies
              $q_{\ell} \cdot k \leq q_{\ell+1} - \sum_{i=1}^n a_i \cdot w_0(x_i)$, from the definition of $\delta$ in~line~\ref{algo:ineq:line:delta}. Therefore,
              \begin{align*}
                  \textstyle\sum_{i=1}^n a_i \cdot \intrep{\vec w}(x_i)_k
                  ={}    & (\textstyle\sum_{i=1}^n a_i \cdot \intrep{\vec w'}_k(x_i)) \cdot k + (\textstyle\sum_{i=1}^n a_i \cdot w_0(x_i))                                                                         \\
                  \leq{} & q_{\ell} \cdot k + (\textstyle\sum_{i=1}^n a_i \cdot w_0(x_i))
                         & \mathcmt{I.H.}                                                                                                                                                                           \\
                  \leq{} & q_{\ell+1} = v                                                                                                   & \hspace{-2cm}\mathcmt{by $w_0 \models_k \delta(q_{\ell},q_{\ell+1})$}
              \end{align*}

              For the left-to-right direction, suppose that $\sum_{i=1}^n a_i \cdot \intrep{\vec w}_k(x_i) \leq v$.
              By definition of $\vec w'$ together with the fact that $\textstyle\sum_{i=1}^n (a_i \cdot \intrep{\vec w'}_k(x_i))$ is an integer, we have
              \[
                  (\textstyle\sum_{i=1}^n a_i \cdot \intrep{\vec w'}_k(x_i)) \leq \floor{\frac{v - \sum_{i=1}^n a_i \cdot w_0(x_i)}{k}}.
              \]
              Let $v' \coloneqq \floor{\frac{v - \sum_{i=1}^n a_i \cdot w_0(x_i)}{k}}$,
              which belongs to $Q$ by~\Cref{lemma:closure-properties-Q}.
              By induction hypothesis,
              there is a path $q_0\dots q_{\ell}$ such that $q_{\ell} = v'$ and $w_{\ell + 1 - i} \models_k \delta(q_{i-1},q_{i})$ for every $i \in [1..\ell]$. By definition of $v'$ and of the transition function~$\delta$,
              we get $w_0 \models_k \delta(v',v)$.
              So, this path can be extended to $q_0\dots q_{\ell}q_{\ell+1}$ such that $q_{\ell+1} = v$ and $w_{\ell+1-i} \models_k \delta(q_{i-1},q_{i})$ for all $i \in [1..\ell+1]$.
              \qedhere
    \end{description}
\end{proof}

\end{document}